\documentclass[pre,amssymb,reprint,superscriptaddress,showpacs]{revtex4-1}
\usepackage{amsmath}
\usepackage{amsfonts}
\usepackage{amssymb}
\usepackage{epsfig}
\usepackage{graphicx}

\renewcommand{\phi}{\varphi}

\newcommand{\be}{\begin{equation}}
\newcommand{\ee}{\end{equation}}
\newcommand{\bea}{\begin{equnaray}}
\newcommand{\eea}{\end{equnaray}}
\newcommand{\ba}{\begin{align}}
\newcommand{\ea}{\end{align}}

\usepackage{color}
\def\le{\left}
\def\ri{\right}

\def\qc{q_{\rm c}}
\definecolor{green}{rgb}{0.0, 0.44, 0.0}
\definecolor{red}{rgb}{1.0, 0.13, 0.32}
\definecolor{blue}{rgb}{0.06, 0.2, 0.65}
\definecolor{magenta}{rgb}{1.0, 0.0, 1.00}
\definecolor{purple}{rgb}{0.7, 0.0, 0.7}
\definecolor{cyan}{rgb}{0.0, 1.0, 1.0}

\begin{document}

\title{Zero-temperature glass transition in two dimensions}

\author{Ludovic Berthier}

\affiliation{Laboratoire Charles Coulomb (L2C), University of Montpellier, CNRS, Montpellier, France}

\author{Patrick Charbonneau}

\affiliation{Department of Chemistry, Duke University,
Durham, North Carolina 27708, USA}

\affiliation{Department of Physics, Duke University,
Durham, North Carolina 27708, USA}

\author{Andrea Ninarello}

\affiliation{Laboratoire Charles Coulomb (L2C), University of Montpellier, CNRS, Montpellier, France}

\author{Misaki Ozawa}

\affiliation{Laboratoire Charles Coulomb (L2C), University of Montpellier, CNRS, Montpellier, France}

\author{Sho Yaida}

\affiliation{Department of Chemistry, Duke University,
Durham, North Carolina 27708, USA}

\date{\today}

\begin{abstract}
The nature of the glass transition is theoretically understood in the mean-field limit of infinite spatial dimensions, but the problem remains totally open in physical dimensions. Nontrivial finite-dimensional fluctuations are hard to control analytically, and experiments fail to provide conclusive evidence regarding the nature of the glass transition. Here, we use Monte Carlo simulations that fully bypass the glassy slowdown, and access equilibrium states in two-dimensional glass-forming liquids at low enough temperatures to directly probe the transition. We find that the liquid state terminates at a thermodynamic glass transition at zero temperature, which is associated with an entropy crisis and a diverging static correlation length.
\end{abstract}

\maketitle

Difficult scientific problems can drastically simplify in some unphysical limits. For instance, very large dimensions ($d \to \infty$) give relevant fluctuations a simple mean-field character~\cite{book1}, and one-dimensional ($d=1$) models can often be treated exactly~\cite{1dbook}. Yet these two solvable limits are  crude idealizations of our three-dimensional reality. The rich theoretical arsenal developed to interpolate between them has revealed the highly nontrivial role of spatial fluctuations in all areas of science. In particular, as the number of spatial dimensions decreases, a phase transition may change nature or even disappear. Dimensionality thus provides a key tool for understanding the essence of many natural phenomena~\cite{book2}.

The glass transition from a viscous fluid to an amorphous solid is no exception~\cite{rmp11}. Its mean-field description, which becomes exact as $d \to \infty$, explains the dramatic slowdown of glass-forming liquids through the rarefaction of the number of glassy metastable states upon approaching a critical temperature, $T_\mathrm{K}$~\cite{mf1,mf2}. The configurational entropy, $s_{\rm conf}$, which is the logarithm of the number of such states, becomes subextensive when $T\leq T_\mathrm{K}$. The equilibrium glass transition thus corresponds to an entropy crisis, a hypothesis first suggested by Kauzmann in his visionary analysis of experimental data~\cite{Kauzmann48}. Efforts have  since been made to describe the role of finite-$d$ fluctuations beyond the mean-field framework~\cite{effort1,effort2,effort3,effort5},
relating in particular the vanishing of $s_{\rm conf}$ to a diverging point-to-set correlation length, the key quantity characterizing nonperturbative fluctuations in glass-formers~\cite{BB04}. These nonperturbative fluctuations, however, make it difficult to examine finite-dimensional glass formers analytically. Kauzmann's intuition has since been repeatedly validated by experiments~\cite{richert1998dynamics,tatsumi2012thermodynamic}, but the conceptual and technical limits of these results have not been lifted. Current experiments access essentially the same restricted temperature range as in Kauzmann's work. Theory and experiments thus currently fail to assess the status of the Kauzmann transition in finite $d$, or whether new mechanisms qualitatively change the underlying physics~\cite{tarjus2005frustration,chandler2010dynamics}.

In this context, computer simulations are especially valuable. They allow direct measurements of both the configurational entropy and the point-to-set correlation length for realistic models of finite-dimensional glass formers~\cite{rmp11}. The recent development of the swap Monte Carlo algorithm (SWAP) further allows the exploration of a temperature regime that experiments cannot easily access~\cite{Ninarello2017}. This has consolidated and extended Kauzmann's experimental findings for three-dimensional glass formers~\cite{BCCNOY17}. Here, we find that SWAP efficiency is so strong in $d=2$ that it provides access to a temperature regime equivalent to experimental timescales $10^{18}$ larger than the age of the universe. This remarkable advance reveals the existence of a thermodynamic glass transition occurring at $T_\mathrm{K}=0$ for $d=2$, accompanied by an entropy crisis and the divergence of the point-to-set correlation length. Our results thus illuminate the dimensionality dependence of the glass transition and shed light on recent investigations about the nature of glassy dynamics in $d=2$~\cite{FS15,VKCW17,IFKKMK17}.

\begin{figure}
\centering{
\includegraphics[width=0.52\linewidth]{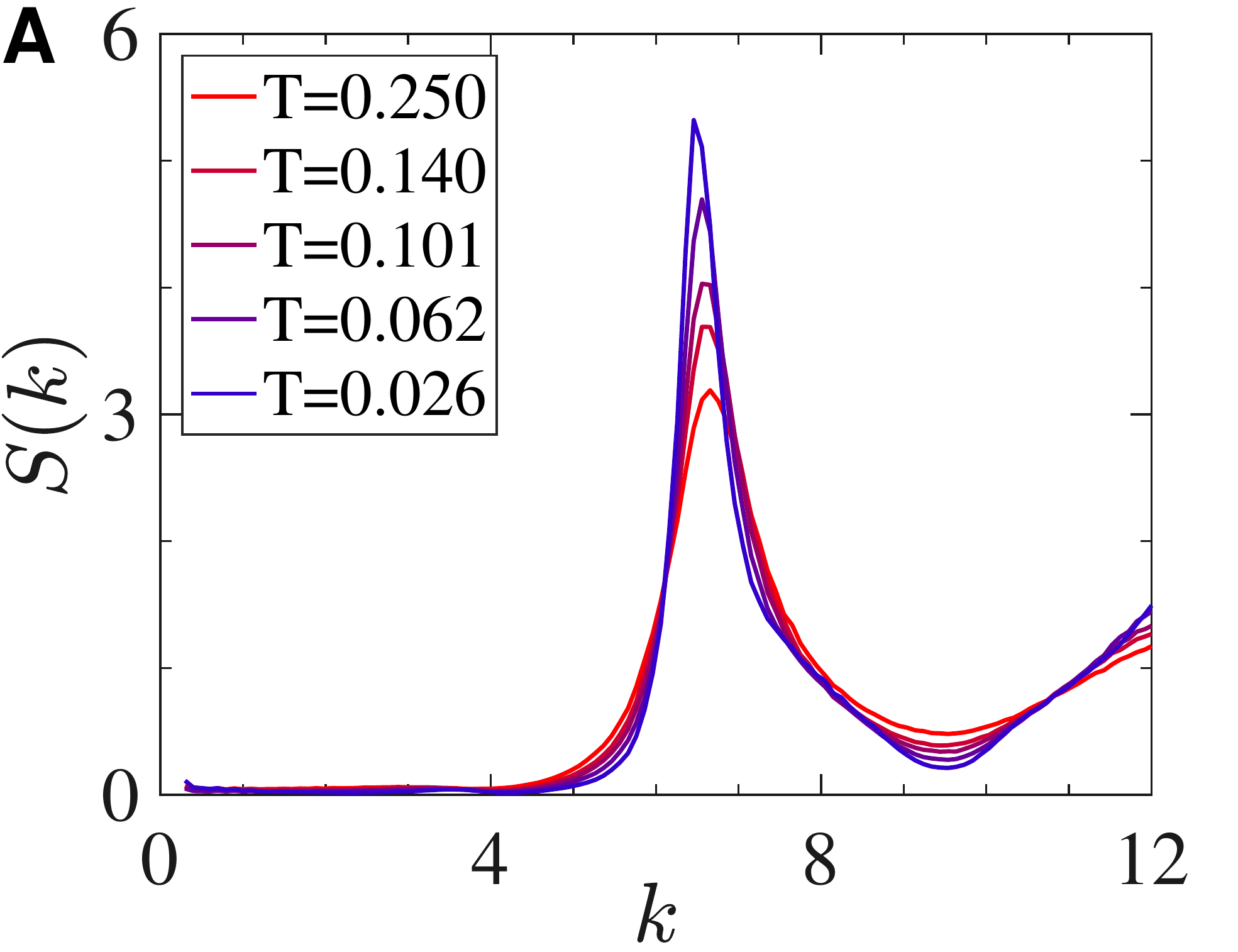}
\includegraphics[width=0.4\linewidth]{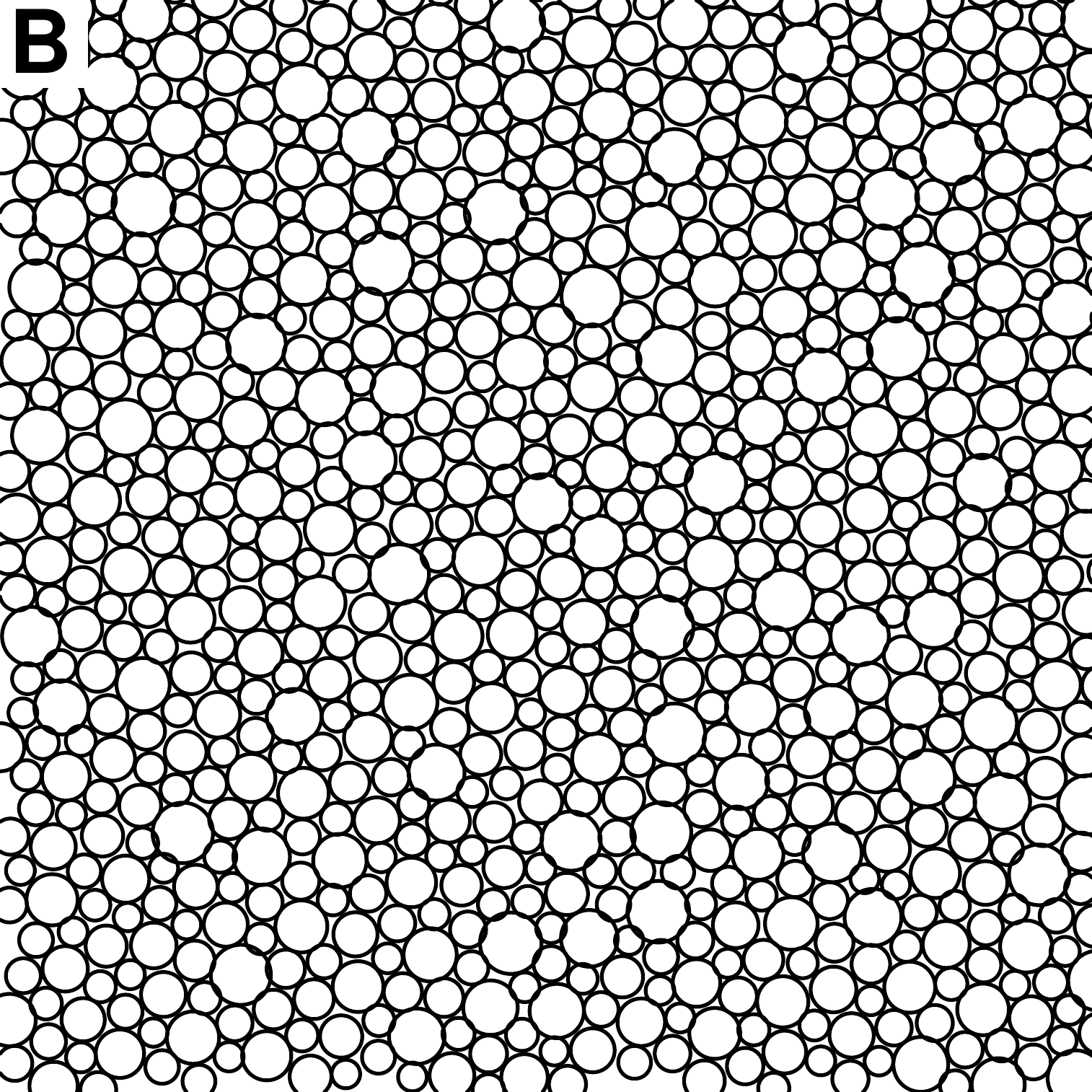}
\includegraphics[width=1.\linewidth]{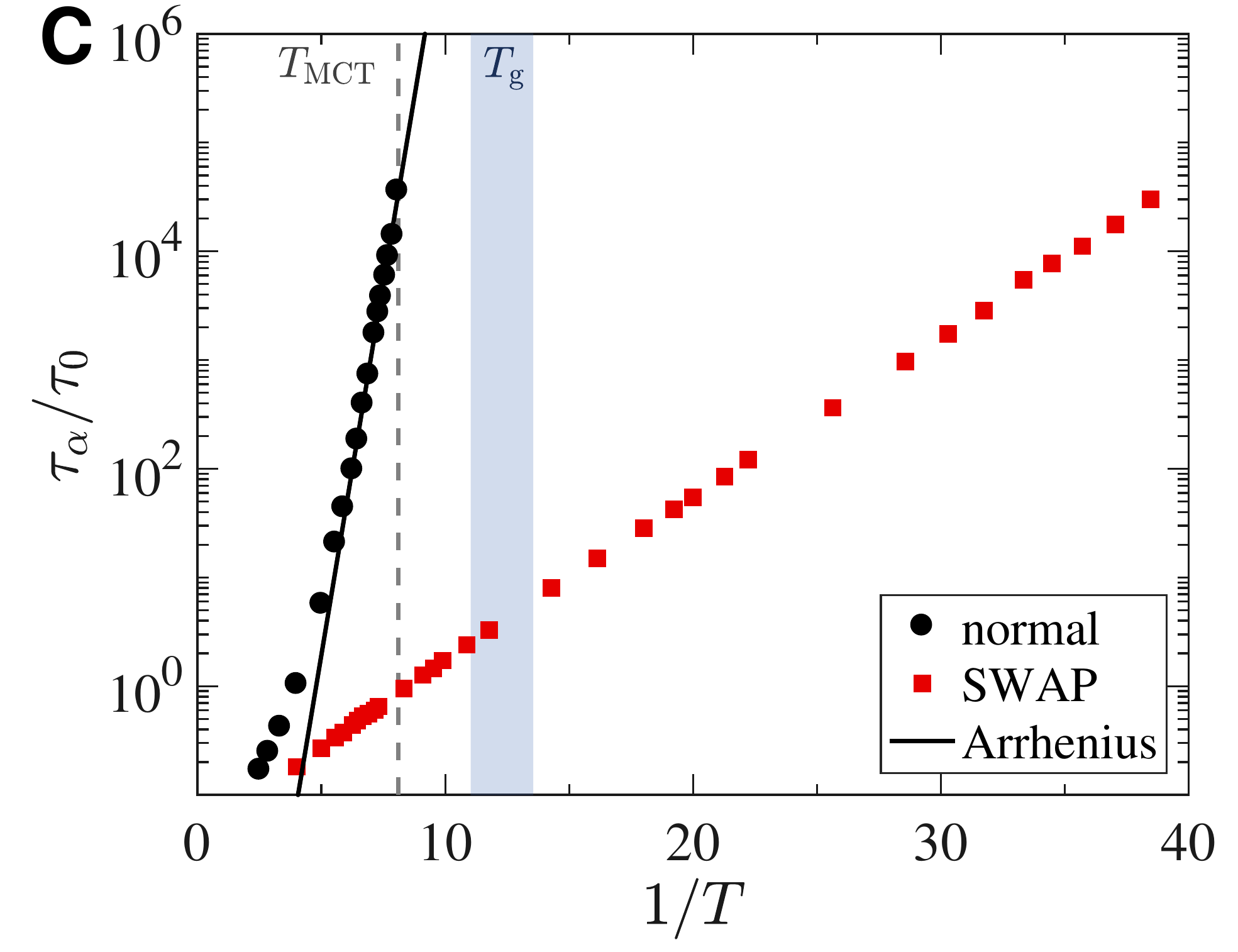}}
\caption{{\bf Statics and dynamics of the $d=2$ glass former.}
(A) The smooth evolution of the static structure factor from $T_{\rm onset}$ down to the lowest studied temperature $T=0.026$ indicates that the system remains fully amorphous at all $T$.
(B) Snapshot of an equilibrium configuration at $T=0.026$.
(C) Arrhenius representation of the structural relaxation time $\tau_\alpha$ using SWAP and normal Monte Carlo dynamics, rescaled by the relaxation time at the onset temperature. The mode-coupling temperature, $T_{\rm MCT}$ (gray dashed line), and the estimated range of experimental glass temperature, $T_\mathrm{g}$ (navy strip), are indicated. The Arrhenius fit to the low-$T$ data provides a lower bound for the growth of $\tau_\alpha$. SWAP can equilibrate systems down to $T\approx 0.3 T_\mathrm{g}$, where the Arrhenius fit gives $\tau^{\rm normal}_{\alpha}/\tau_0 \sim10^{46}$.}
\label{ceiling}
\end{figure}

More specifically, we study a two-dimensional mixture of soft particles interacting with a $1/r^{12}$ purely repulsive power-law potential and a size polydispersity chosen to minimize demixing, fractionation, and crystallisation (see Supplementary Materials for details of models, methodologies, and additional corroborating results including ones for $d=2$ hard disks). The average particle diameter is used as unit length, and the strength of the interaction potential as unit temperature.  SWAP is implemented following the methodology recently developed for $d=3$~\cite{Ninarello2017}. Systems ranging from $N=300$ to $N=20000$ particles within a periodic box are used to carefully track finite-size effects in both dynamics and thermodynamics. 
We mainly present results of $N=1000$.
Figure~\ref{ceiling}A shows that the static structure factor $S(k)$ evolves smoothly over a broad temperature range, from the onset temperature $T_{\rm onset}=0.250$ down to $T=0.026$, which is the lowest temperature for which our strict equilibrium criteria are met. The typical low-temperature configuration depicted in Fig.~\ref{ceiling}B shows that particles of different sizes are well mixed, and that local ordering is extremely weak. In fact, no crystallisation event was ever observed in our simulations, and the correlation lengths extracted from the pair correlation function for translational and bond-orientational orders evolve modestly with $T$ (see SM). In other words, the model is an excellent glass former.

The bulk dynamics and equilibration are captured by
the bond-orientational order time correlation, $C_\psi(t)$.  The $1/e$ decay of $C_\psi(t)$ robustly defines bulk relaxation timescales $\tau_\alpha$ both for SWAP and normal Monte Carlo dynamics (Fig.~\ref{ceiling}C). 
We normalize these timescales by $\tau_0\equiv\tau^{\rm normal}_\alpha(T_{\rm onset})$.
In agreement with earlier works~\cite{FS15}, we find that translational correlation functions suffer large finite-size effects, but that subtracting long-range Mermin-Wagner translational fluctuations results in system-size independent measurements~\cite{VKCW17,IFKKMK17} consistent with bond-orientational dynamics.
The normal dynamics exhibits a well-known super-Arrhenius growth of $\tau_\alpha$. Fitting its temperature evolution to a power-law divergence situates the mode-coupling crossover at $T_{\rm MCT} = 0.123$, which is roughly the lowest temperature accessible with this dynamics. Following Ref.~\cite{Ninarello2017}, we estimate the narrow range within which the experimental glass temperature takes place as $T_\mathrm{g} \in [ 0.0738, 0.0907]$. (Henceforth we set $T_\mathrm{g}=0.082$.)
The lower end of this interval stems from an Arrhenius fit which provides a lower bound to the true $\tau_\alpha$.
By all estimates, SWAP dynamics is clearly much faster than the normal one. The speedup is about 5 orders of magnitude at $T_{\rm MCT}$, 10 at $T_\mathrm{g}$, and the Arrhenius lower bound suggests a 42 order-of-magnitude speedup at $T=0.026$. Using an atomistic $\tau_0 = 10^{-10} {\rm s}$ converts this estimate to $\tau_\alpha = 10^{36} {\rm s}$, which is approximately $10^{18}$ times the age of the universe. Such a `cosmological' speedup leaves no doubt that SWAP dynamics fully bypasses the slowdown associated with the glass transition in $d=2$. 

\begin{figure}
\centering{
\includegraphics[width=0.5\textwidth]{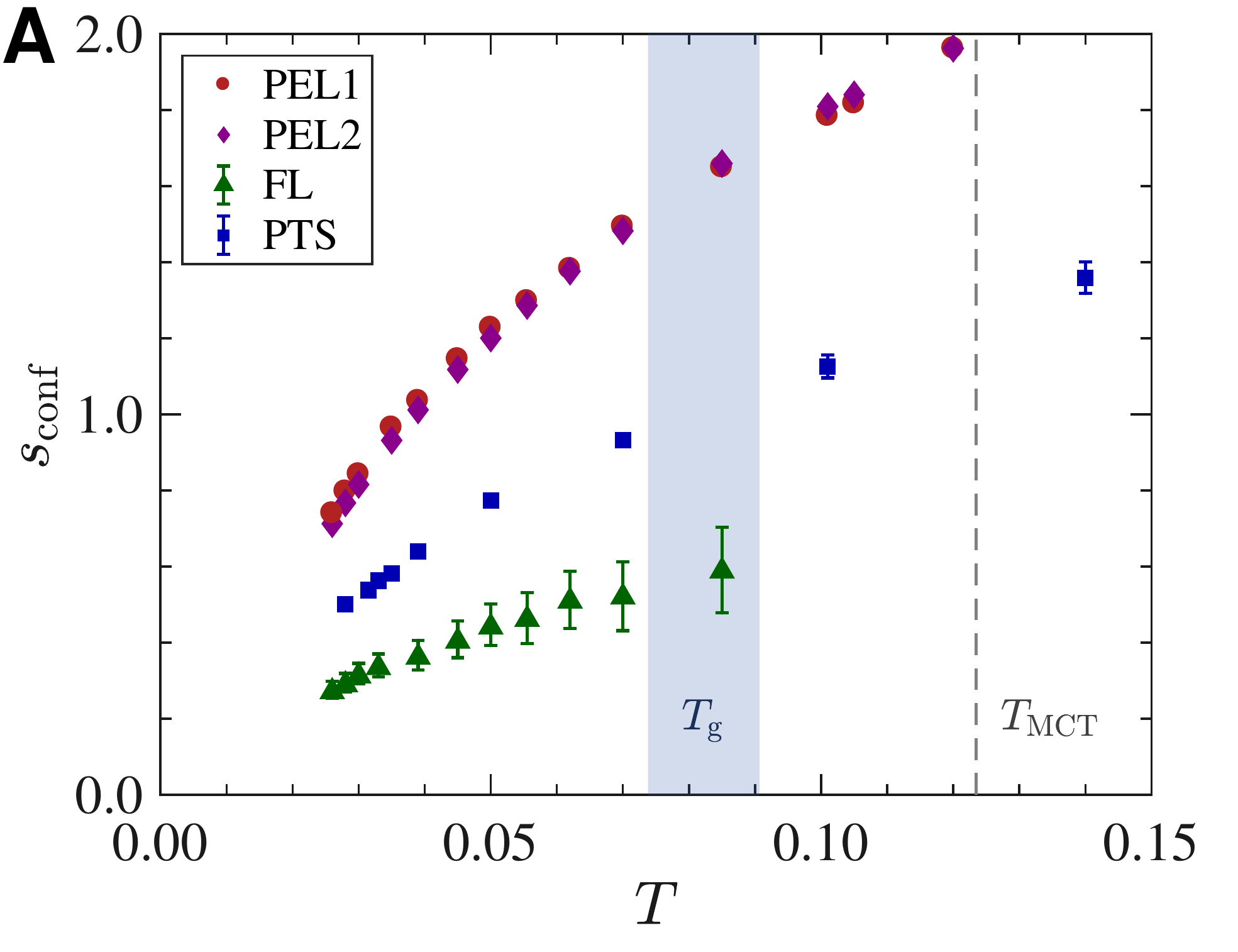}
\includegraphics[width=4.2cm]{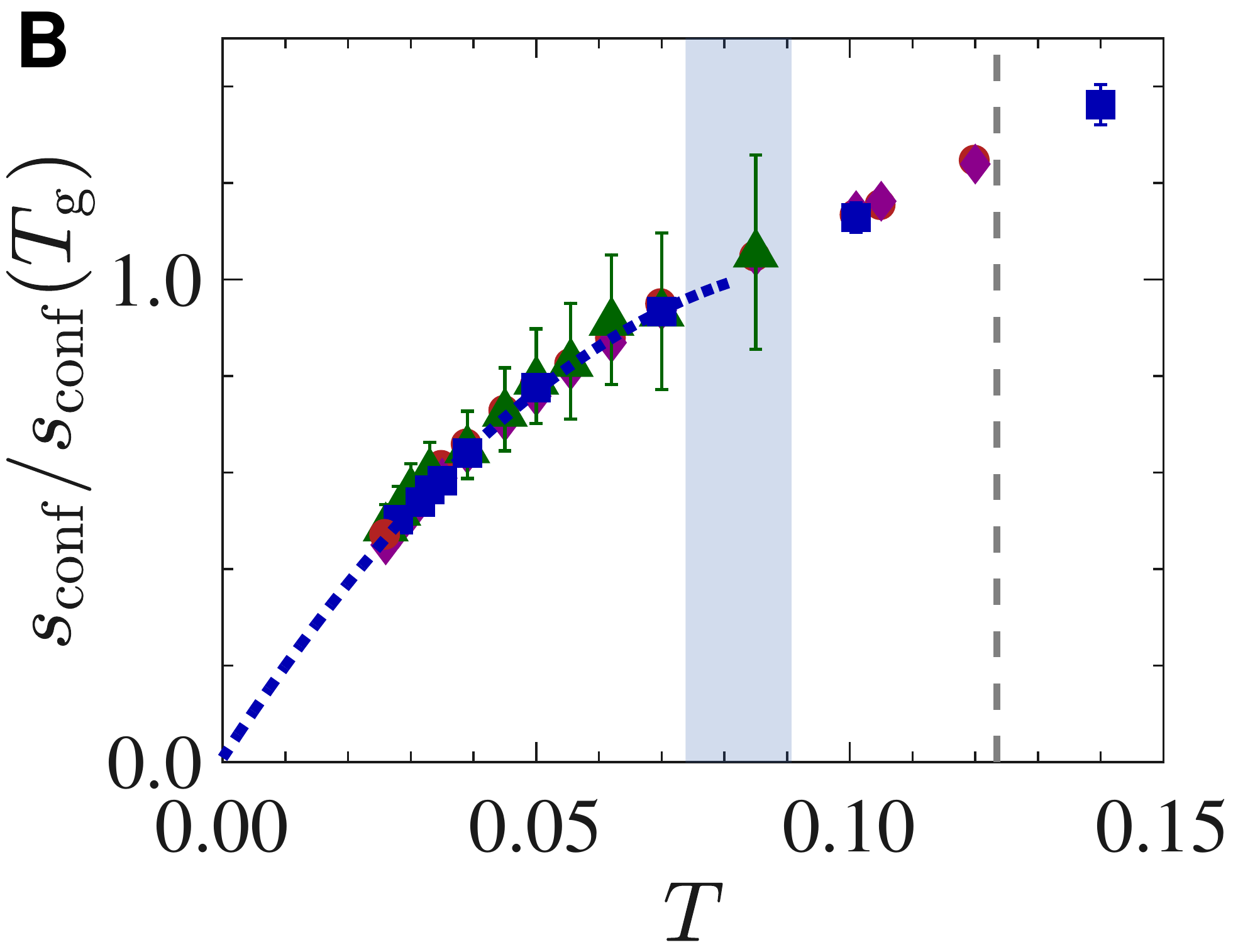}
\includegraphics[width=4.2cm]{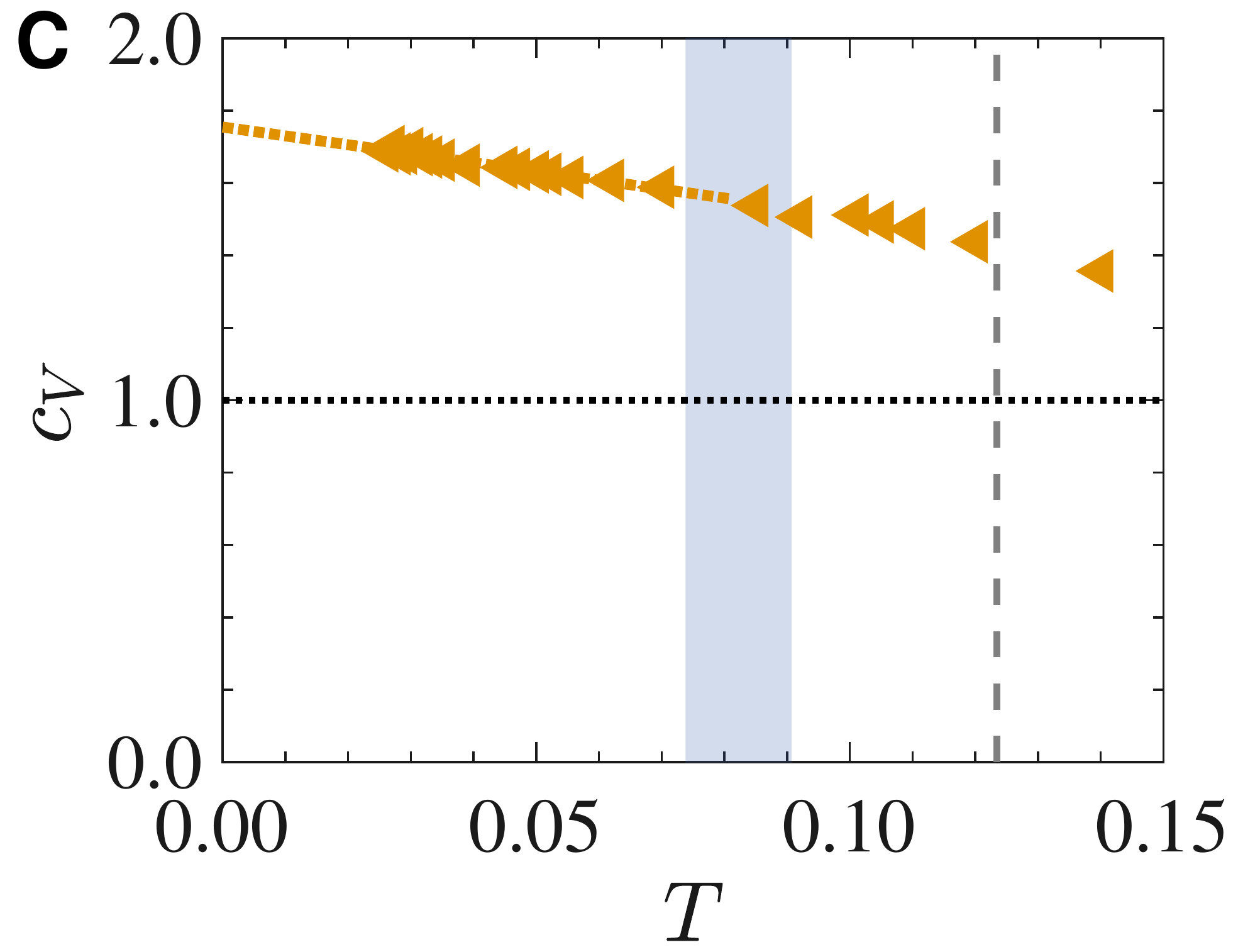}
}
\caption{{\bf Zero-temperature Kauzmann transition.}
(A) Decrease of the configurational entropy with temperature using the potential energy landscape (PEL), Frenkel-Ladd (FL), and point-to-set (PTS) length estimates. 
(B) Once rescaled by their value at $T_\mathrm{g}$, all estimates evolve nearly identically, well fitted by a quadratic function of $T$ for $T<T_\mathrm{g}$ (dashed blue line indicates the quadratic fit for the point-to-set estimate). The results are consistent with a linearly vanishing $s_\mathrm{conf}$ at $T_\mathrm{K}=0$. 
(C) The specific heat increases monotonically above the Dulong-Petit law for $d=2$ (dashed horizontal line), which is also consistent with a thermodynamic transition at $T_\mathrm{K}=0$.}
\label{absolutezero}
\end{figure}

This computational advance permits the study of the $d=2$ configurational entropy and its relationship to the putative entropy crisis. Following and extending earlier work on $d=3$ systems~\cite{BCCNOY17}, we obtain independent estimates of $s_{\rm conf}$ using state-of-the-art methodologies, see Fig.~\ref{absolutezero}A. The first estimate stems from subtracting the vibrational contribution, measured by minimizing the potential energy of the system to an inherent structure and obtaining its vibrational spectrum, from the total liquid entropy~\cite{Kob}. This potential energy landscape (PEL) approach needs to be complemented, for polydisperse systems, with an independent measure of the mixing entropy~\cite{misaki2017}. Because minor but systematic additional adjustments are then required, two sets of PEL estimates are reported in Fig.~\ref{absolutezero}A. The two are quantitatively close and similarly decrease with $T$, which confirms that methodological details do not affect our results in any essential way. This approach considerably extends $s_\mathrm{conf}$ measurements from $1.5T_\mathrm{g}$ in earlier $d=2$ simulations~\cite{2dsastry} down to a temperature 5 times smaller, $0.3T_\mathrm{g}$. 

Our second estimate directly measures the glass entropy by performing a thermodynamic integration from the well-controlled harmonic solid limit. This approach, which is inspired by the Frenkel-Ladd method for crystals~\cite{Frenkel}, was recently adapted to polydisperse amorphous solids~\cite{misaki2018}. Because it does not count the number of inherent structures but measures instead the entropy of constrained glassy states, it is also very close in spirit (although not equivalent~\cite{misaki2018}) to the free-energy measurement~\cite{cosloentropy} that makes use of the Franz-Parisi potential~\cite{FP}. The Frenkel-Ladd estimate is smaller than the PEL ones, as expected, and exhibits a similar temperature dependence. 

\begin{figure}
\begin{center}
\includegraphics[width=8cm]{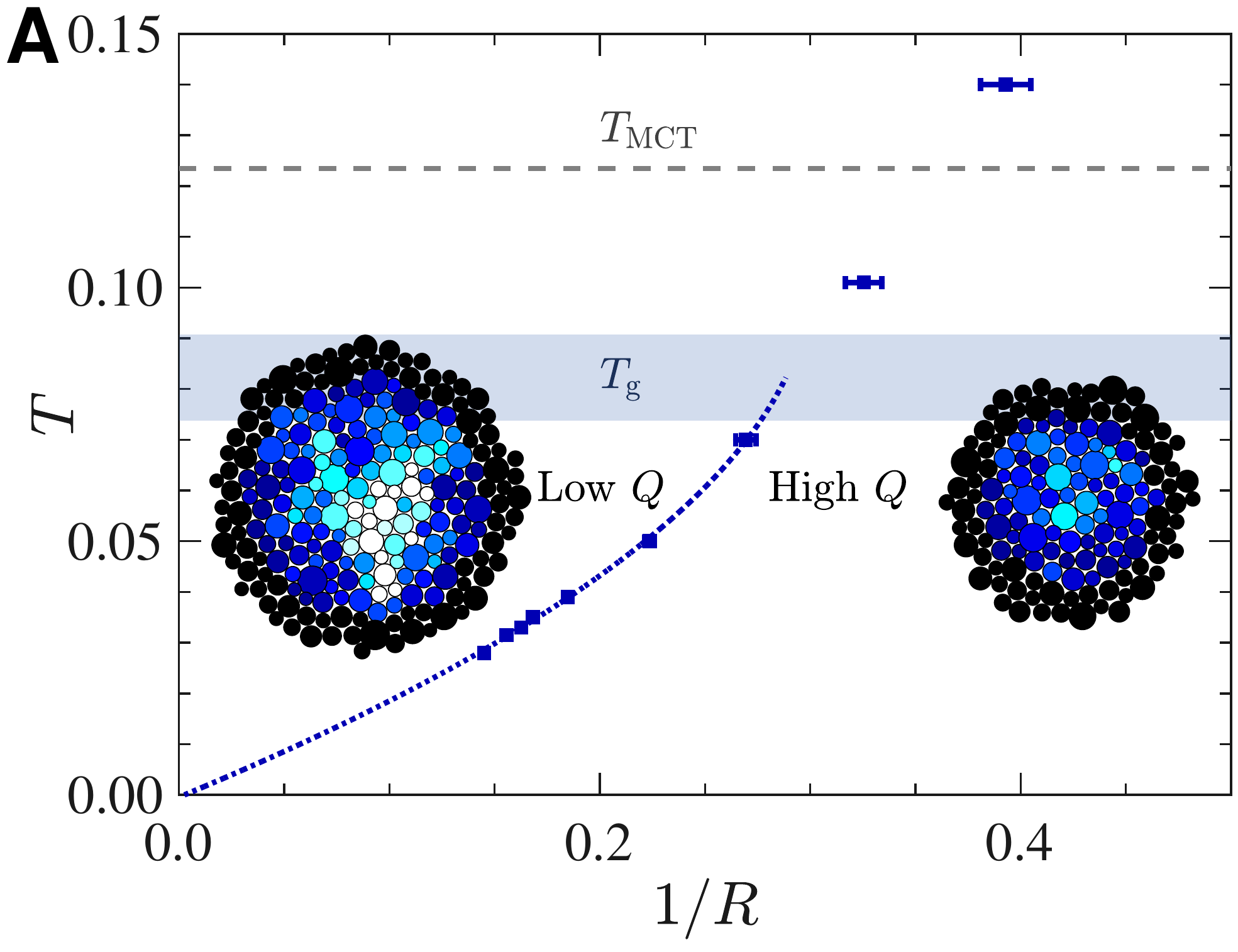}
\hspace*{-0.1cm}
\includegraphics[width=8.3cm]{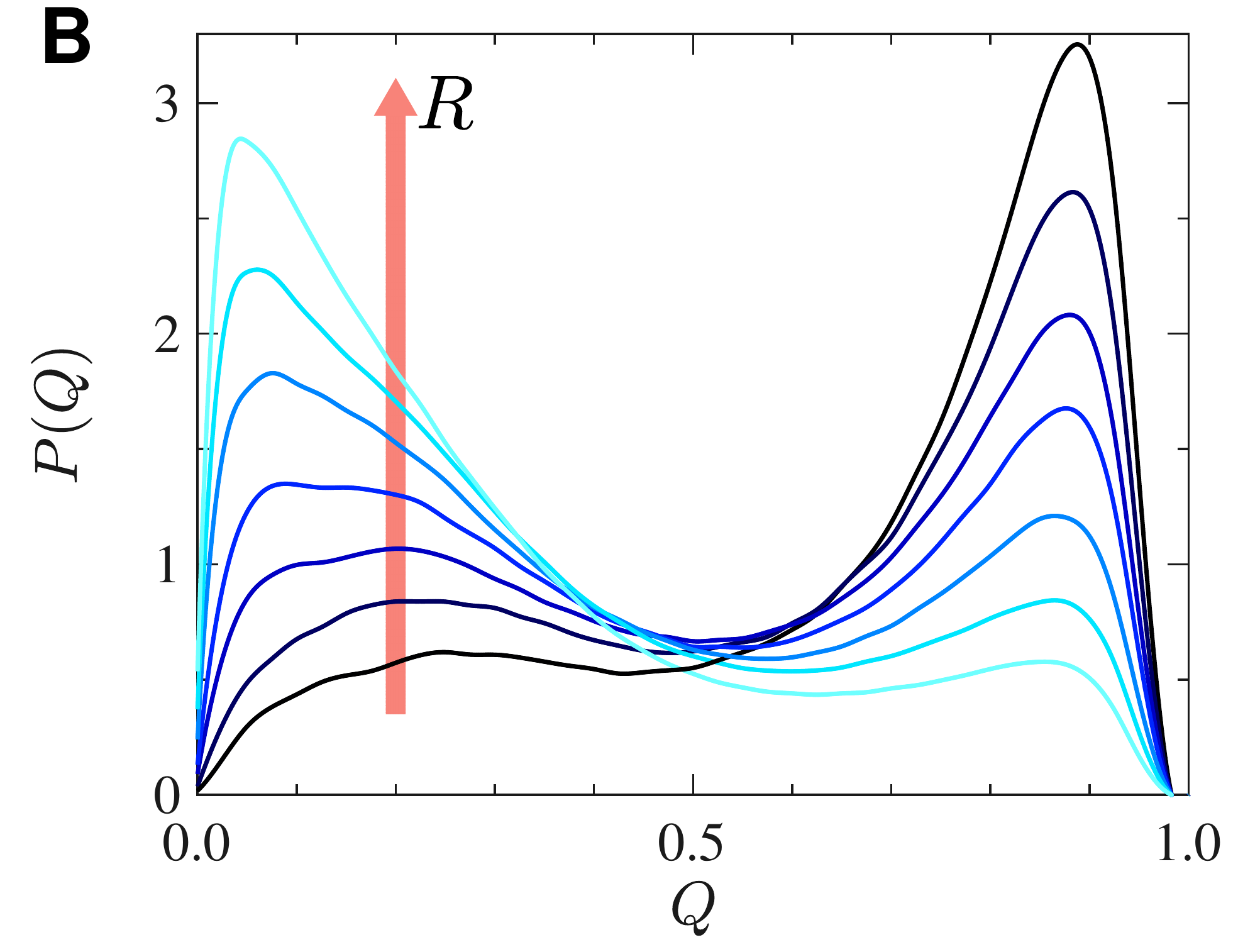}
\end{center}
\caption{{\bf Approaching the random first order transition.} 
(A) Phase diagram showing the low-$Q$ region for large cavities and high-$Q$ region for small cavities, separated by a crossover line determined by the point-to-set correlation length.  (Inset) Representative configurations with overlap field for $T=0.035$ at $R=6.6$ (low $Q$, white) and $4.8$ (high $Q$, dark).
(B) Evolution of the probability distribution function of overlap $P(Q)$ at $T=0.035$ from $R=4.8$ to $R=6.6$. Bimodality signals a first-order-like phase coexistence.}  
\label{PTS}
\end{figure}

From the data in Fig.~\ref{absolutezero}A, $s_{\mathrm{conf}}$ seemingly vanishes close to $T_\mathrm{K}=0$. This behavior sharply contrasts with that of three-dimensional glass formers, for which evidence suggests that $T_\mathrm{K}>0$~\cite{Kauzmann48,richert1998dynamics,tatsumi2012thermodynamic,BCCNOY17}. The impending entropy crisis is 
expected to give rise to large-scale fluctuations with a growing point-to-set correlation length~\cite{BB04}. We use the computational tools developed in \cite{BBCGV08,BCY16,BCCNOY17} to analyze the thermodynamic properties of liquids confined within spherical cavities of radius $R$ drawn from a reference equilibrium configuration. The distribution $P(Q)$ of the core cavity overlap $Q$ among the confined equilibrium fluid configurations is then analyzed. The point-to-set correlation length, $\xi_{\rm PTS}$, is determined from the decay with $R$ of the average overlap. This length is then transformed into a third estimate, $s_{\mathrm{conf}} \propto \xi_{\mathrm{PTS}}^{-(d-\theta)}$ with $\theta=1$. In $d=2$ this choice of $\theta$ is natural because it both saturates the bound $\theta \leq d-1$~\cite{BB04} and satisfies the wetting relation $\theta=d/2$~\cite{KTW89}. The resulting $s_{\rm conf}(T) =  \xi_{\rm PTS}(T_\mathrm{g}) / \xi_{\rm PTS}(T)$ in Fig.~\ref{absolutezero}A has a similar temperature evolution as the other estimates. 

Figure~\ref{absolutezero}B shows that rescaling all configurational entropies by their value at $T_\mathrm{g}$ collapses the entire set of measurements.
A quadratic fit $s_{\rm conf}(T) = aT + bT^2+c$ to the low temperature behavior, $T<T_\mathrm{g}$, yields  $|T_\mathrm{K}|\leq 0.003$ for all data sets.
These $T_\mathrm{K}$ estimates are 10 times smaller than our lowest temperature $T=0.026$ and 30 times smaller than $T_\mathrm{g}$.
All known alternatives to an entropy crisis invoke a change in the concavity of $s_{\rm conf}$ and should be accompanied by a maximum in the specific heat $c_V$~\cite{stillinger1988supercooled,tarjus2005frustration,debenedetti2003model}; we observe neither the concavity (Fig.~\ref{absolutezero}A) nor the maximum (Fig.~\ref{absolutezero}C). As $T \to T_\mathrm{K}$, the specific heat instead monotonically increase towards a finite value that is larger than the Dulong-Petit law. All these observations are therefore consistent with the occurrence of a non-trivial entropy crisis at $T_\mathrm{K}=0$.

The thermodynamic glass transition at $T_\emph{K}=0$ corresponds both to an entropy crisis and to a divergence of the point-to-set correlation length. We illustrate the physical meaning of this length scale in Fig.~\ref{PTS}A in the form of a $(T,1/R)$ diagram reminiscent of both the Franz-Parisi thermodynamic construction~\cite{FP} and of the random pinning approach~\cite{pinning}. Upon decreasing the cavity size at a given temperature, the system crosses over from a low-$Q$ regime at large $R$ to a high-$Q$ regime at small $R$, as illustrated by the snapshots in Fig.~\ref{PTS}A. For any $T>0$, this crossover occurs when $R = \xi_{\rm PTS}$. It represents a finite-size version of the random first-order glass transition, and corresponds to a rarefaction of the number of locally available states as $R$ decreases. This crossover is reflected by the evolution of $P(Q)$ in Fig.~\ref{PTS}B, which exhibits features reminiscent of phase coexistence near an incipient first-order transition. The observed crossover becomes sharper as $T$ decreases because it occurs over a growing correlation length and transforms into a genuine thermodynamic phase transition as $T \to T_\mathrm{K} = 0$. In absolute values, $\xi_{\rm PTS} \approx 7$ at $T=0.028$, which represents a very large static correlation length for glassy models~\cite{BCY16,BBCGV08,BCCNOY17}. It implies that large clusters comprising about 140 particles are statically correlated, and should thus move collectively to restructure the fluid. These results are consistent with the sharp decay of the configurational entropy in Fig.~\ref{absolutezero} and the dramatic increase of the relaxation time in Fig.~\ref{ceiling}.

In summary, our dynamic and thermodynamic measurements all indicate that our two-dimensional glass formers exhibit a zero-temperature equilibrium glass transition at $T_\mathrm{K}=0$. 
Our results identify the thermodynamic properties underlying the nature of glassy dynamics in $d=2$~\cite{FS15,VKCW17,IFKKMK17}.
More importantly, they show that a thermodynamic transition can occur in finite-dimensional systems, and that the lower critical dimension for the long-range amorphous order is $d_{\rm L}=2$. This finding lends indirect support to previous observations in $d=3$~\cite{BCCNOY17} and will surely guide future analytical work.

\acknowledgments 

We thank G.~Tarjus for stimulating discussions. This research was supported by a grant from the Simons Foundation (\#454933, Ludovic Berthier, \#454937, Patrick Charbonneau). Part of the computations was carried out through the Duke Compute Cluster.


\newpage

\onecolumngrid

\newpage

\onecolumngrid

\appendix

\section*{Supplementary Information}

\section{Model}
\label{SM_model}

The glass-forming model we consider in the main text consists of particles with purely repulsive soft-sphere interactions, and a continuous size polydispersity.
Particle diameters, $\sigma_i$, are randomly drawn from a distribution of the form:
$f(\sigma) = A\sigma^{-3}$, for $\sigma \in [ \sigma_{\rm min}, \sigma_{\rm max} ]$, where $A$ is a normalization constant.
The size polydispersity is quantified by $\delta=\sqrt{\overline{\sigma^2} - \overline{\sigma}^2}/\overline{\sigma}$, where $\overline{\cdots}\equiv\int \mathrm{d} \sigma f(\sigma) (\cdots)$,
and is here set to $\delta = 0.23$ by imposing $\sigma_{\rm min} / \sigma_{\rm max} = 0.45$. The average diameter, $\overline{\sigma}$, sets the unit of length. The soft-sphere interactions are pairwise and described by an inverse power-law potential
\begin{eqnarray}
v_{ij}(r) &=& v_0 \left( \frac{\sigma_{ij}}{r} \right)^{12} + c_0 + c_1 \left( \frac{r}{\sigma_{ij}} \right)^2 + c_2 \left( \frac{r}{\sigma_{ij}} \right)^4, \label{eq:soft_v} \\
\sigma_{ij} &=& \frac{(\sigma_i + \sigma_j)}{2} (1-\epsilon |\sigma_i - \sigma_j|), \label{eq:non_additive}
\end{eqnarray}
where $v_0$ sets the unit of energy (and temperature with Boltzmann constant $k_\mathrm{B}=1$), and $\epsilon=0.2$  
quantifies the degree of non-additivity of particle 
diameters.
We introduce $\epsilon>0$ to the model in order to
suppress fractionation and thus enhance its glass-forming ability~\cite{BCNO16,Ninarello2017}.
The constants, $c_0$, $c_1$ and $c_2$, enforce a vanishing potential and the continuity of its first- and second-order derivatives of the potential at the cut-off distance
$r_{\rm cut}=1.25 \sigma_{ij}$ .
We simulate a system with $N$ particles within a square cell of area $V$ under periodic boundary conditions, at number density $\rho=N/V=1.01$. Most simulations have $N=1000$, but systems with $N=300$, $3000$, $8000$ and $20000$ are also studied.

For the point-to-set length measurement, we also study a two-dimensional hard-disk model, for which the 
pair interaction is zero for non-overlapping particles and infinite otherwise. 
The system has the same size distribution $f(\sigma)$ and size polydispersity $\delta$ as the soft-disks described above.
Given these parameters, the system is then uniquely characterized by its area fraction 
$\phi = \pi N \overline{\sigma^2} /(4V)$, and we 
frequently report the data using the
reduced pressure $Z=P/(\rho k_{\rm B}T)$, where $\rho$, 
$k_{\rm B}$, and $T$ are 
the number density, Boltzmann constant and temperature, respectively. Without loss of generality, we set $k_B$ and  $T$ to unity for the hard-disks. 
The pressure $P$ is calculated from the contact value of the pair correlation function properly scaled for a polydisperse system~\cite{santos2002contact}. 
We use $N=1000$ for this model.

\section{Observables}
\label{SM_observables}
We monitor the system structure with two common liquid state quantities: the pair-distribution function $g(r)$, and the structure factor $S(k)=\langle \rho_{\bf -k} \rho_{\bf k} \rangle/N$, where $\rho_{\bf k}=\sum_{i}e^{i\mathbf{k}\cdot\mathbf{r}_i}$ is the Fourier-space density.
Orientational correlations are also considered, and are quantified using the six-fold bond-orientational order parameter~\cite{Royall2015,RT15}
\begin{equation}
  \label{eq:psi6}
\psi_6 = \frac{1}{N} \sum_{j=1}^{N}  \psi^j_6  \ \  {\rm where} \ \ \ \psi^j_6 = \frac{1}{n_j} \sum^{n_j}_{k=1}\exp(i6\theta_{jk}),
\end{equation}
where the sum is performed over the $n_j$ first neighbors of the $j$-particle. These neighbors are defined as particles with $r_{ij}/\sigma_{ij}<1.33$, which is location of the
distance of the first minimum in the rescaled radial distribution function $g(r/\sigma_{ij})$.
The angle $\theta_{jk}$ then measures the orientation of the axis between the two particles with respect to the $x$-axis. Because these correlations are orientationally invariant the choice of $x$-axis is made without loss of generality.
Orientational correlations are then monitored through the two-point bond-orientational correlation function
\begin{equation}
  \label{eq:g6}
  g_6(r)=\langle \psi_6(r)\psi_6^*(0)\rangle,
  \end{equation}
where $\psi_6(r)= \sum_{i=1}^N\delta(|\mathbf{r}-\mathbf{r}_i|)\psi_6^i$.
The radial decay of the hexatic order correlation function,
$g_6(r)/g(r)$~\cite{RT15}, provides an hexatic correlation length $\xi_6$, as discussed in Sec.~\ref{sec:structure}.

Translational dynamics is characterized by first measuring the intermediate scattering function
\begin{equation}
  \label{Fs}
  F_\mathrm{s}(k,t)= \frac{1}{N}\left\langle \sum_{j=1}^{N} \exp\left[ i{\bf k}\cdot ({\bf r}_j(t) - {\bf r}_j(0))  \right]\right\rangle
\end{equation}
at the wave number $k$ corresponding to the first peak of $S(k)$. The relaxation time of the density fluctuations, $\tau_\alpha^{\rm TR}$, is then
extracted from the exponential decay of the scattering function, i.e., $F_\mathrm{s}(k, \tau_\alpha^{\rm TR})=e^{-1}$.
Orientational dynamics is characterized similarly, replacing the Fourier-space density by the bond-orientational correlation function in Eq.~(\ref{eq:psi6}) defined by
\begin{equation}
   \label{C6} 
  C_{\psi_6}(t)= \frac{1}{N}\left\langle \sum^{N}_{i=0} \psi^i_6(t) \left [ \psi^i_6(0)  \right]^* \right\rangle.
\end{equation}
In order to extract the bond-orientational relaxation time $\tau_{\alpha}$, we use $C_{\psi_6}(\tau_{\alpha})=e^{-1}$.

\section{Equilibration and the glass-ceiling}
Normal Monte-Carlo (MC) simulations allow only local particle displacements, drawing a random displacement vectors on the $(x,y)$ axis in the interval $[-\Delta r_\mathrm{max},\Delta r_\mathrm{max}]$ with $\Delta r_\mathrm{max}=0.6$ and moving a randomly chosen particle following a Metropolis acceptance criterion.
Compounding $N$ such displacement attempts defines a MC step, which is used as unit of time in this work.
To ensure equilibration, we monitor both static and dynamical observables.
Starting from a high-temperature liquid configuration, we quench the system at the final temperature and wait for the potential energy of the system to stop aging on a time window of $\sim 10^6$ MC steps. We first estimate $\tau_{\alpha}$ on simulations long enough to allow few decorrelations of $C_{\psi_6}(t)$, and then perform simulations for $220\tau_{\alpha}$.
The system is left to equilibrate during the first $20\tau_{\alpha}$; static and dynamical observables are computed over the following $200\tau_{\alpha}$.
Swap MC simulations include attempts at exchanging random pairs of particle diameters, which replace particle displacements with probability $p_{\rm swap}=0.2$. This algorithm defines the SWAP dynamics.
The same equilibration and measuring protocol as for normal MC is then followed. 
Static observables monitor ordering and phase separations in the system, as discussed in Sec.~\ref{sec:structure},
whereas dynamical observables quantify the relaxation and equilibration timescales.

\begin{figure}
  \centering
  \includegraphics[width=0.5\textwidth]{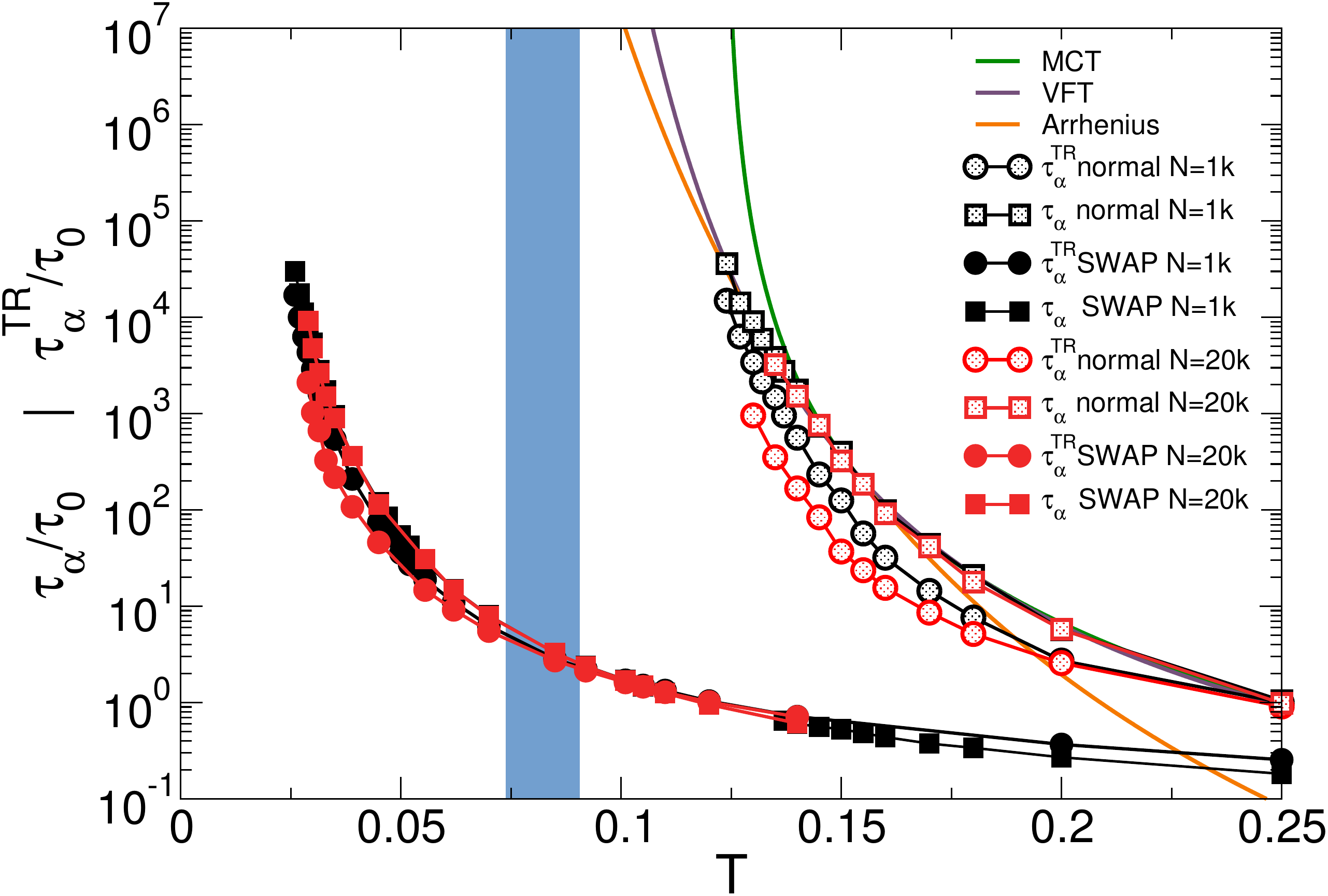}
  \caption{Relaxation times as a function of the temperature for both normal and SWAP dynamics. The $y$-axis is rescaled by
    the relaxation time of normal dynamics at the onset temperature $\tau_{\alpha}(T_{\rm onset}=0.25)=\tau_0=2592~{\rm MC steps}$. Empty (full) symbols indicate normal (SWAP) dynamics.
    Circles (squares) denote results for translational (orientational) relaxation times $\tau_{\alpha}^{\rm TR}$ ($\tau_{\alpha}$) for $N=1000$ and $N=20000$ systems. The MCT, VFT and Arrhenius fits (see text) are given as green, purple and orange solid lines, respectively. These fits help estimate the glass ceiling region, {\it i.e.} the
    lower bound for the region accessible in equilibrium experiments, which is denoted as a blue box.}
      \label{fig:ceilingSM}
\end{figure}

In Fig.~\ref{fig:ceilingSM}, we report orientational $\tau_{\alpha}$ and translational $\tau_{\alpha}^{\rm TR}$ relaxation times for both normal and SWAP dynamics. Because the relaxation of local orientational degrees of freedom is slower, the associated timescale is used as reference.
We perform three different fits to the $\tau_{\alpha}$ results for the physical dynamics, in order to extract the temperatures relevant to the dynamical slowing down. First, we fit $\tau_{\alpha}$ to a power-law function, as is predicted in the context of the mode-coupling theory~\cite{Goetze2008},
\begin{equation}
  \tau_{\alpha}\propto (T-T_{\rm MCT})^{-\gamma},
    \label{eq:MCT}
\end{equation}
over the interval $\tau_{\alpha}\in(\tau_0,10^3\tau_0)$. The resulting $T_{\rm MCT}=0.123$ roughly corresponds to the lowest temperature at which normal dynamics can reach equilibrium in simulations of reasonable duration~\cite{Ninarello2017}.

Next, we estimate the laboratory glass transition temperature, $T_\mathrm{g}$, at which experiments with atomic and molecular glass formers cannot be equilibrated anymore. At $T_\mathrm{g}$, relaxation times have increased
by 12 orders of magnitude with respect to their value at the onset of the supercooled dynamics~\cite{ANGELL96}.
We thus fit the relaxation times both to a Vogel-Fulcher-Tallman (VFT) law
\begin{equation}
  \tau_{\alpha}\propto \exp\left(\frac{A}{T-T^\mathrm{VFT}}\right),
  \label{eq:VFT}
\end{equation}
and to an Arrhenius law
\begin{equation}
  \tau_{\alpha}\propto \exp\left(\frac{B}{T}\right),
    \label{eq:ARR}
\end{equation}
where $A$ and $B$ are fitting constants.
These two expressions respectively overestimate and underestimate the increase of relaxation times in experimental glass-formers~\cite{Elmatad2009, Hecksher2008}.
We fit Eq.~(\ref{eq:VFT}) using the whole temperature range $T<T_{\rm onset}$, whereas we fit Eq.~(\ref{eq:ARR}) only to $T<0.16$ to ensure that the result serves as a proper lower bound on the relaxation time.
Extrapolating up to the temperature at which $\log_{10}(\tau_{\alpha}/\tau_0) \simeq 12$ gives $T_\mathrm{g}^\mathrm{VFT}=0.0907$ and $T_\mathrm{g}^\mathrm{Arr}=0.0738$.
These two temperature are, by construction, upper and lower bounds for $T_\mathrm{g}$, and thus define an experimental
glass-ceiling region~\cite{BCCNOY17} in Fig.~\ref{fig:ceilingSM}. In all cases, SWAP dynamics equilibrates well beyond this experimentally limited regime, reaching $T=0.026$.
Figure \ref{fig:ceilingSM} also shows the fitting curves. The mode-coupling power-law prediction describes the growth of the relaxation times only within the
first three orders of magnitude of the glassy regime, but at lower temperatures it overestimates the results by many orders of magnitude.
Whereas Eq.~(\ref{eq:VFT}) adequately describes these same results over more than four orders of magnitude, an Arrhenius law captures barely two orders of magnitude. 

\section{Structural correlations}
\label{sec:structure}

In Section~\ref{sec:PTS}, we show that $\xi_\mathrm{PTS}$ increases as temperature decreases. Ref.~\cite{RT15}, however, showed that for some computational models made of polydisperse particles, correlation lengths related to the degree of order present increase faster than $\xi_\mathrm{PTS}$. In particular, Ref.~\cite{RT15} analyzed
the two-points positional and bond-orientational correlations, paying particular attention to the radial decay of the functions $g(r)-1$ and $g_6(r)/g(r)$, respectively 

\begin{figure}
  \includegraphics[width=0.48\textwidth]{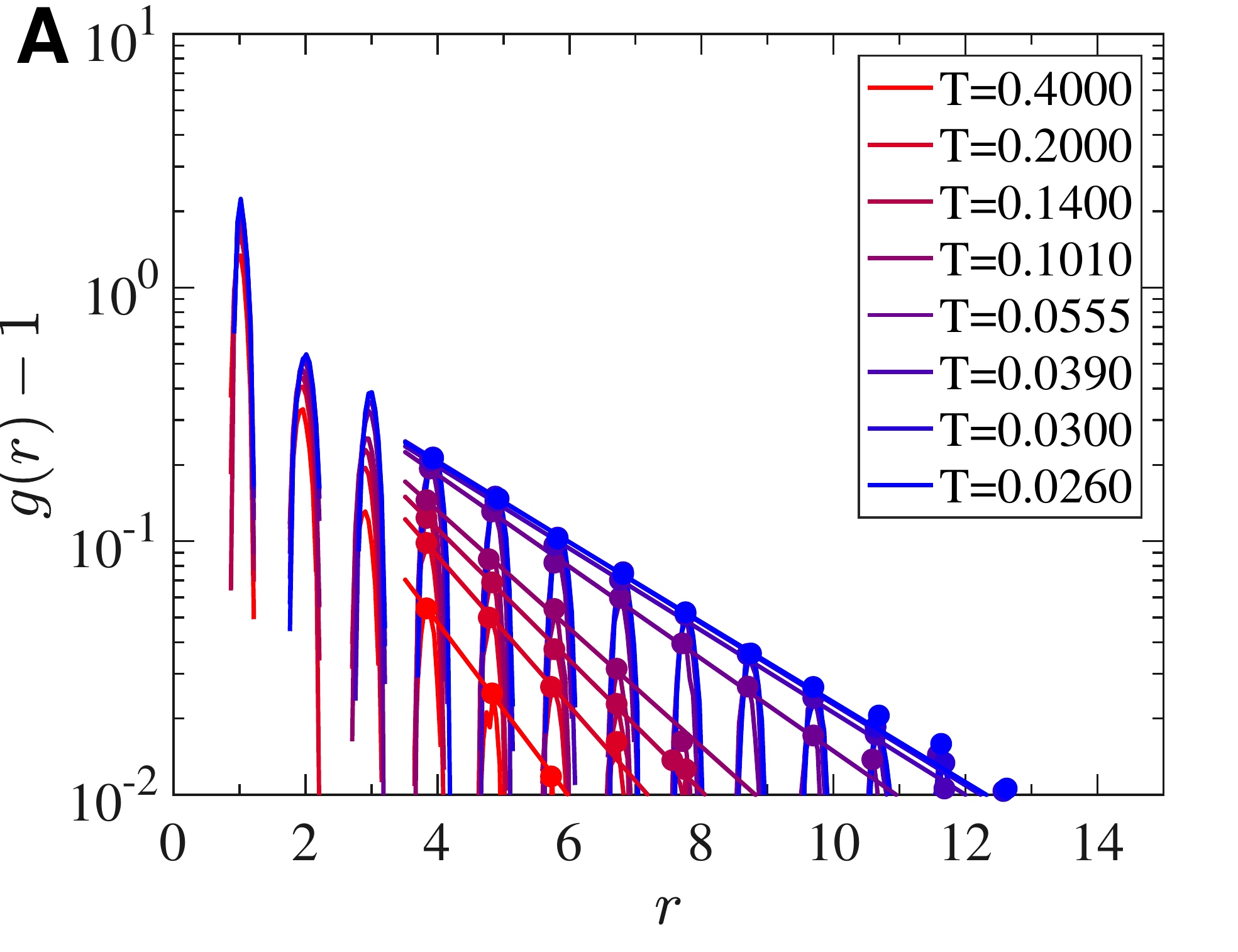}
  \includegraphics[width=0.48\textwidth]{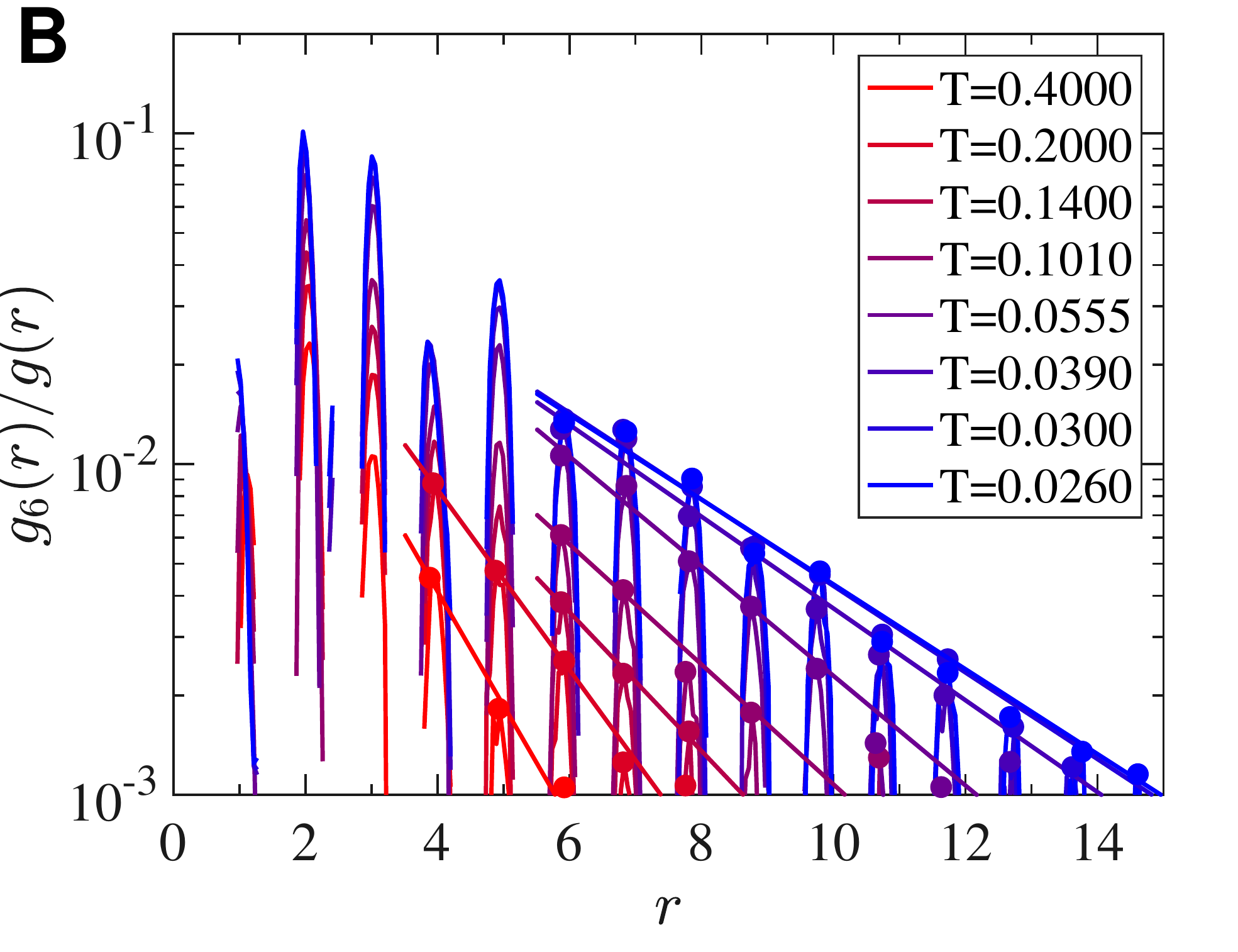}
  \centering
  \caption{Two-point (A) positional and (B) bond-orientational correlation functions. Colors denote different temperatures from red (high) to blue (low). Peak maxima are fitted with an exponential form, $C_{\mathrm{s},6}\exp(-r/\xi_{\mathrm{s},6})$, in order to extract positional and bond-orientational static correlation lengths $\xi_\mathrm{s}$ and $\xi_6$, respectively.}
    \label{hexatic_correlation}
\end{figure}

\begin{figure}
  \centering
    \includegraphics[width=0.5\textwidth]{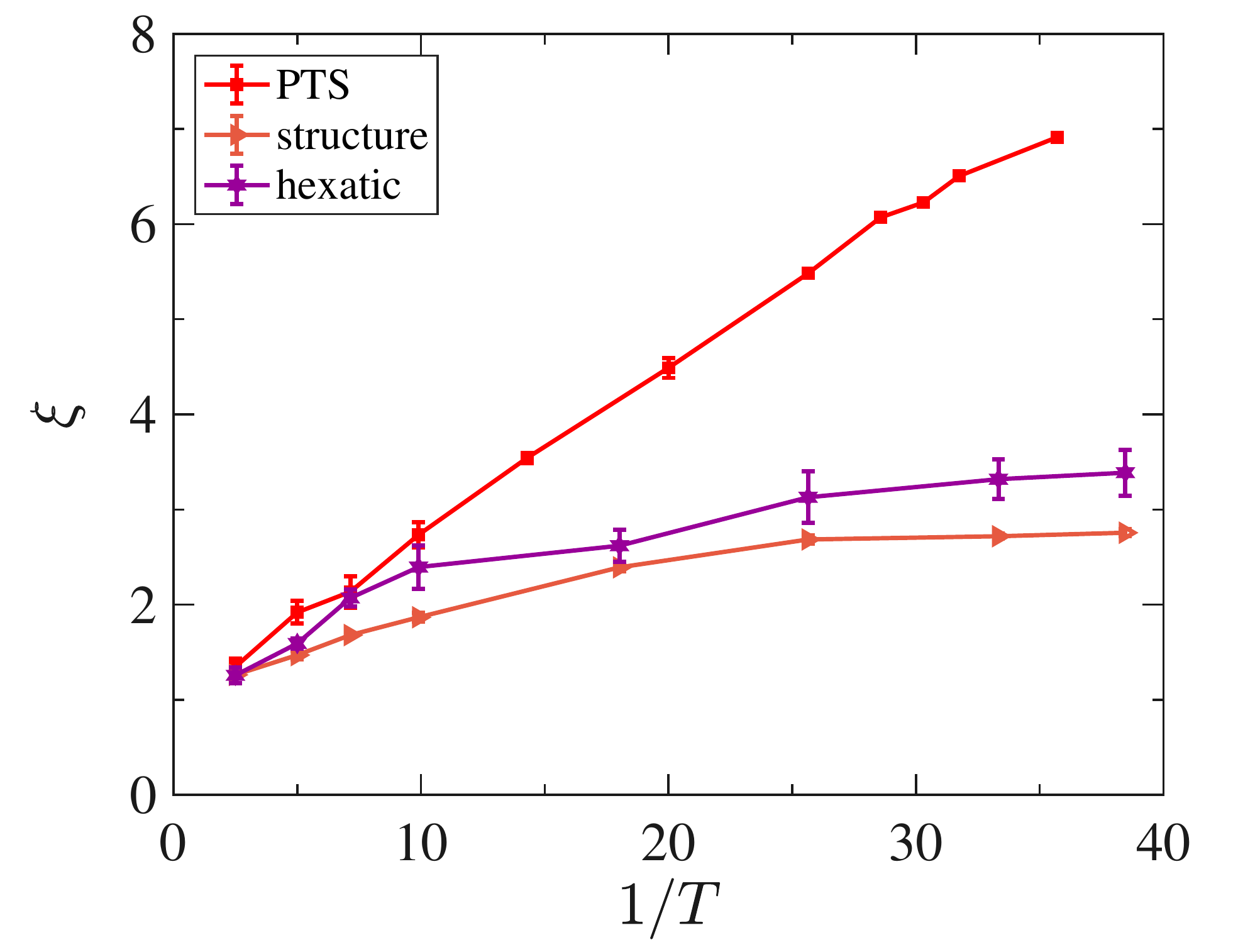}
  \caption{Growth of different static correlation lengths with temperature: positional order length, $\xi_\mathrm{s}$, bond-orientational order length, $\xi_{6}$, and  point-to-set length, $\xi_\mathrm{PTS}$ (see Sec.~\ref{sec:PTS}). The increase of the first two lengths is mild compared to that of $\xi_\mathrm{PTS}$.}
    \label{hexatic_comp}
\end{figure}

Results for these two quantities are reported in Fig.~\ref{hexatic_correlation}. Here, following Ref.~\cite{RT15}, Delaunay neighbors are obtained from a radical Voronoi tessellation. Both functions exhibit clear peaks at distances corresponding to the correlation shells,
but their temperature evolution is relatively mild. We fit the peak points with an exponential function of the form $C_{\mathrm{s},6} \exp(-r/\xi_{\mathrm{s},6})$ in order to extract a correlation function both for positional
$\xi_\mathrm{s}$ and bond-orientational $\xi_6$ correlations. The temperature evolution of the resulting static correlation lengths is shown in Fig.~\ref{hexatic_comp} together with that of $\xi_\mathrm{PTS}$.
Over the whole temperature range, we observe an increase by a factor $\approx 2.2$ and $\approx 2.7$ for $\xi_\mathrm{s}$ and $\xi_6$ with saturation at low temperature, which is considerably smaller than the factor of $\approx 5.1$ increase observed for $\xi_\mathrm{PTS}$. Coupled with the additional verifications for potential crystallization and fractionation, this result rules out the presence of significant structural order in our system, even at extremely low temperatures. Our observations are also remarkably different from those of Ref.~\cite{RT15}; they show that good glass formers are not affected by increases in positional and bond-orientational order.

\section{Configurational entropy}

The configurational entropy, $s_{\rm conf}$, is defined as
\begin{equation}
s_{\rm conf}=s_{\rm tot} - s_{\rm glass},
\end{equation}
where $s_{\rm tot}$ and $s_{\rm glass}$ are the total entropy and the entropy of a typical glass state, respectively.
We separately measure $s_{\rm tot}$ and $s_{\rm glass}$ by thermodynamic integration based on the scheme developed in Ref. ~\cite{misaki2018}.


\subsection{Setting}

Consider a $M$-component polydisperse system. (A system with $M = N$ is said to have a continuous polydispersity.) If $N_m$ is the number of particles of the $m$-th species, then the fraction of the $m$-th species is $X_m=N_m/N$, and hence $\sum_{m=1}^M N_m=N$ and $\sum_{m=1}^M X_m=1$.  
For simplicity, we set all particles masses to unity. We denote particle positions as ${\bf r}^N=({\bf r}_1, {\bf r}_2, \cdots, {\bf r}_N)$, and the set of their diameter as $\Sigma^N=\{ \sigma_1, \sigma_2, \cdots, \sigma_N  \}$.  
In order to consider permutations of particle diameters as additional degrees of freedom, 
we introduce a permutation $\pi$ to the set $\Sigma^N$. A specific sequence of particle diameters is denoted $\Sigma_{\pi}^N$, e.g., $\Sigma_{\pi}^N=(\sigma_{3}, \sigma_8, \sigma_5,\cdots )$. A total of $N!$ possible such permutations exists, and for a system with continuous polydispersity, all such permutations are distinguishable. 


The system potential energy, $U$, depends both on particle positions ${\bf r}^N$ and on the permutation $\pi$, and is thus formally denoted $U (\Sigma_{\pi}^N, {\bf r}^N)$.
For notational simplicity, we write $U ({\bf r}^N)=U (\Sigma_{\pi^*}^N, {\bf r}^N)$ for the reference system with $\Sigma_{\pi^*}^N$.
The resulting canonical partition function at inverse temperature $\beta=1/T$ is
\begin{equation}
\mathcal{Z} =  \frac{1}{N!} \sum_{\pi} \frac{1}{\Pi_{m=1}^M N_m! \Lambda^{Nd}} \int_V \mathrm{d} {\bf r}^N  e^{-\beta U (\Sigma_{\pi}^N, {\bf r}^N)},
\label{eq:partition_function_new}
\end{equation}
where $\Lambda=\sqrt{2\pi \beta \hbar^2}$
is the thermal de Broglie wavelength with the unit mass. Without loss of generality, we set the Planck constant $\hbar=1$.
Note that Eq.~(\ref{eq:partition_function_new}) should be distinguished from the conventional partition function, $Z$, in which only particle positions ${\bf r}^N$ are degrees of freedom,
\begin{equation}
Z =  \frac{1}{\Pi_{m=1}^M N_m! \Lambda^{Nd}} \int_V \mathrm{d} {\bf r}^N  e^{-\beta U ({\bf r}^N)}.
\label{eq:partition_function_conventional}
\end{equation}
The following subsections describe how Eq.~(\ref{eq:partition_function_new}) can be used to compute both the total and the glass entropies.

\subsection{Total entropy}


The partition function $\mathcal{Z}$ in Eq.~(\ref{eq:partition_function_new}) for the target system $\beta U(\Sigma_{\pi}^N, {\bf r}^N)$  reduces to the conventional partition function $Z$ without permutations in Eq.~(\ref{eq:partition_function_conventional}), because diameter permutations are always compensated by position permutations in absence of constraint, i.e.,
\begin{equation}
\mathcal{Z} =  \frac{1}{N!} \sum_{\pi} \frac{1}{\Pi_{m=1}^M N_m! \Lambda^{Nd}} \int_V \mathrm{d} {\bf r}^N  e^{-\beta U (\Sigma_{\pi}^N, {\bf r}^N)}
= \frac{1}{\Pi_{m=1}^M N_m! \Lambda^{Nd}} \int_V \mathrm{d} {\bf r}^N e^{-\beta U ({\bf r}^N)} = Z\, .
\label{eq:partition_function_total}
\end{equation}
The total entropy computation is therefore equivalent to what has been observed in previous studies~\cite{coluzzi2000lennard,ozawa2015equilibrium}.

\begin{figure}[htbp]
\includegraphics[width=0.48\columnwidth]{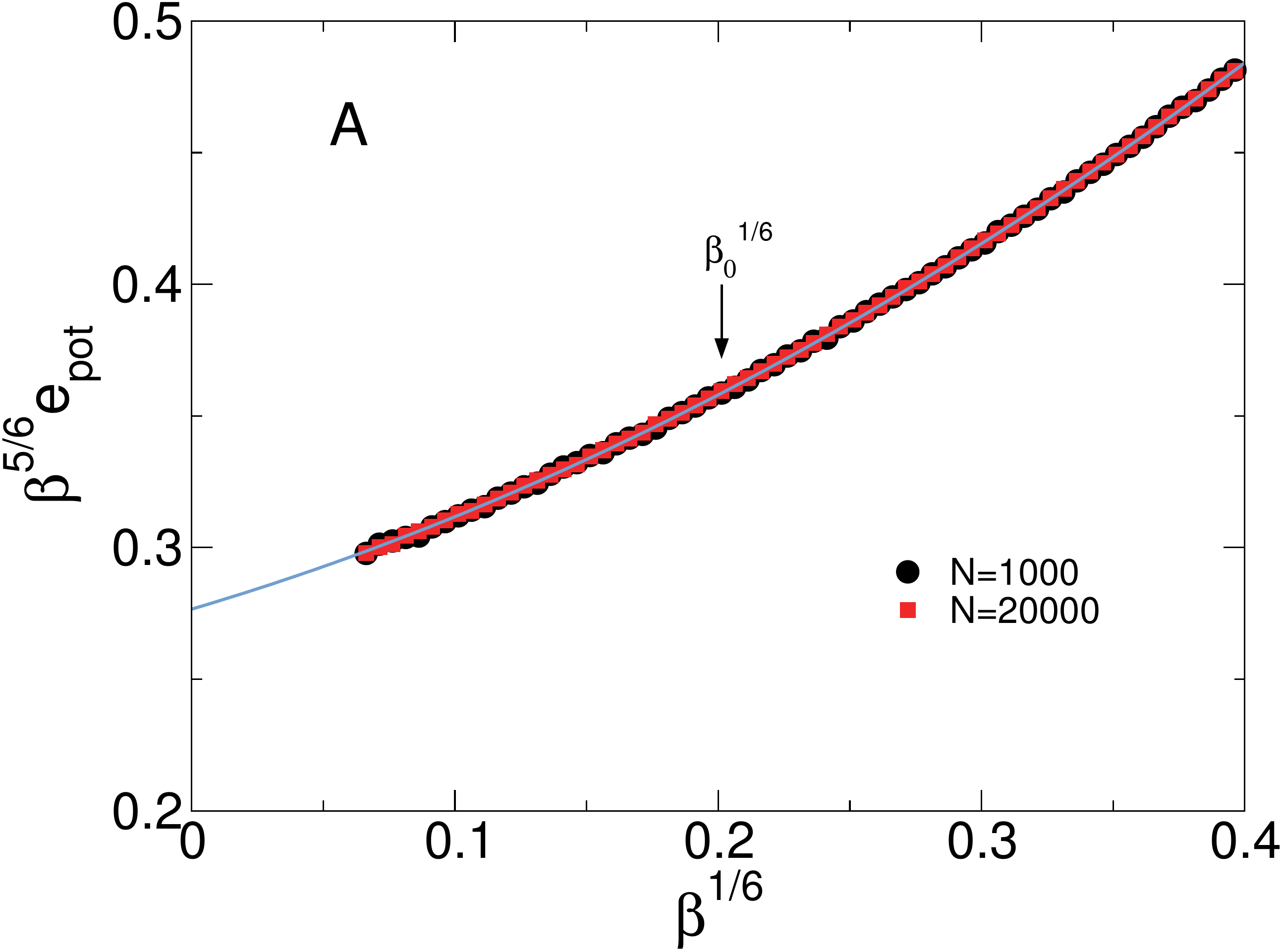}
\includegraphics[width=0.48\columnwidth]{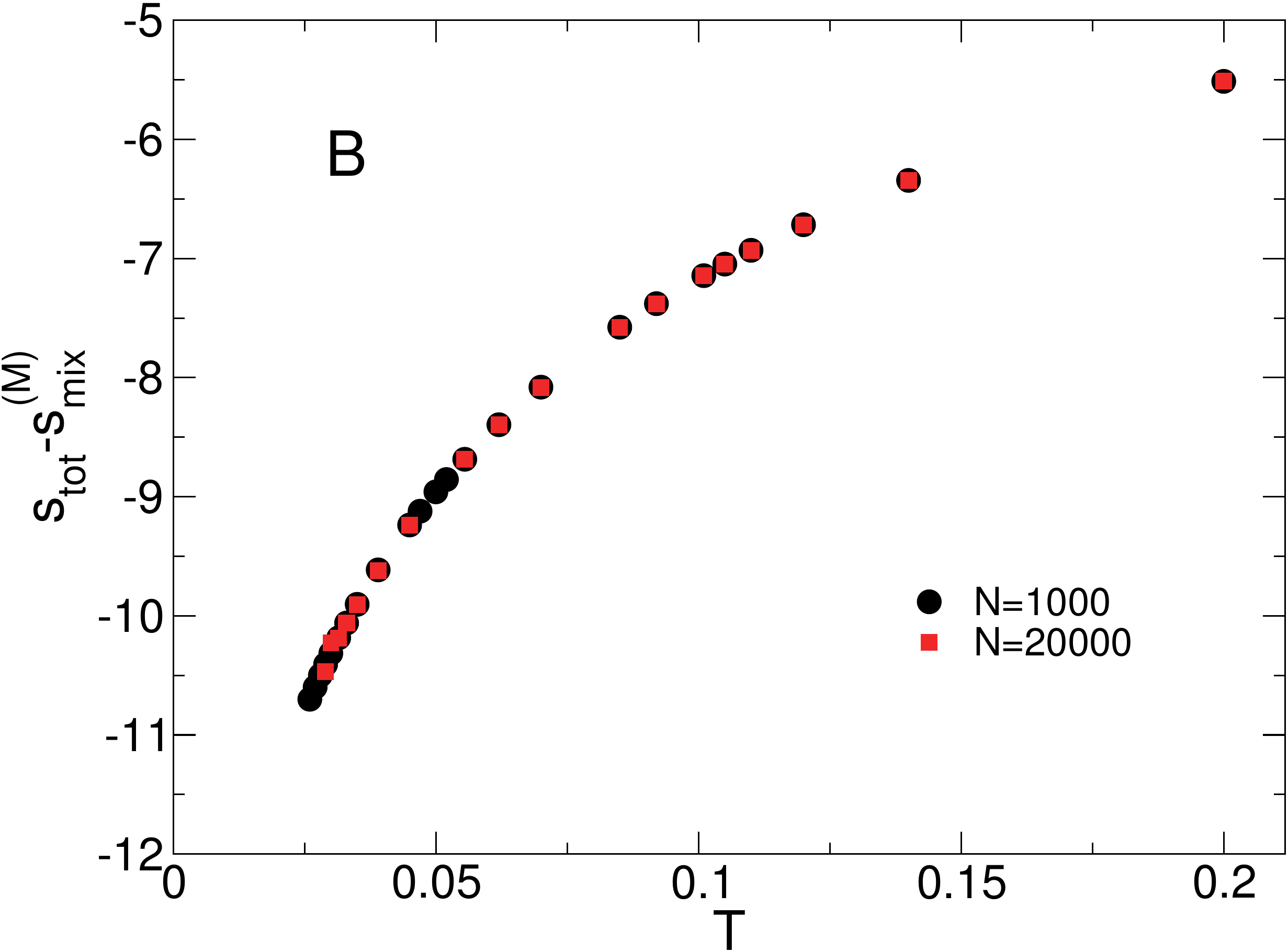}
\caption{
(A) High-temperature average potential energy results, $\beta^{5/6} e_{\rm pot}(\beta)$ for N=1000 (black circles) and N=20000 (red squares) systems, along with the resulting fitting form (blue line) with parameters $A$, $B$, and $C$ as in Eq.~(\ref{eq:I_F}). Including $D$ in the fit has no noticeable numerical impact.
The vertical arrow denotes our choice of $\beta_0^{1/6}=0.2013$. The total entropy results are unchanged for any reasonable choice of $\beta_0$.
(B) The resulting temperature dependence of $s_{\rm tot}-s_{\rm mix}^{(M)}$.
}
\label{fig:s_tot}
\end{figure}

Using a high-temperature $\beta\to 0$ ideal gas as an exactly solvable reference system, we perform a thermodynamic integration over (inverse) temperature up to the target temperature $\beta$,
\begin{equation}
s_{\rm tot} = \frac{(d+2)}{2} - \ln \rho - \ln \Lambda^d + \beta e_{\rm pot}(\beta) -  \int_{0}^{\beta} d\beta' e_{\rm pot}(\beta') + s_{\rm mix}^{(M)},
\label{eq:s_tot}
\end{equation} 
where $s_{\rm mix}^{(M)} = - \sum_{m=1}^M X_m \ln X_m$ is the ideal gas mixing entropy per particle and $e_{\rm pot}(\beta)$ is the average potential energy per particle.
The integration in Eq.~(\ref{eq:s_tot}) requires special care, because $e_{\rm pot}(\beta)$ diverges in the high-temperature limit~\cite{coluzzi2000lennard,ozawa2015equilibrium}.
We sidestep the difficulty by introducing an intermediate temperature $\beta_0$  that separates the very high temperature regime, $\beta' \in [0, \beta_0]$, from the rest, $\beta' \in (\beta_0, \beta]$. We thus write 
\begin{equation}
I \equiv \int_{0}^{\beta} d\beta' e_{\rm pot}(\beta') = \int_{0}^{\beta_0} d\beta' e_{\rm pot}(\beta') + \int_{\beta_0}^{\beta} d\beta' e_{\rm pot}(\beta') 
\equiv I_{\rm F} + I_{\rm N},
\label{eq:def_integral}
\end{equation}
where $I_{\rm N}$ is obtained by usual thermodynamic integration, and $I_{\rm F}$ is obtained by fitting the $e_{\rm pot}(\beta)$ to a polynomial, and then analytically
integrating the resulting function~\cite{coluzzi2000lennard,ozawa2015equilibrium}. 
The specific polynomial form we use for the high-temperature expansion of a system of soft spheres with interaction potential $v(r) \propto r^{-n}$ (in $d$ dimensions) is
\begin{equation}
e_{\rm pot}(\beta) = A \beta^{(d/n)-1} + B \beta^{(2d/n)-1} + C \beta^{(3d/n)-1} + D \beta^{(4d/n)-1} + \cdots,
\label{eq:highT_expansion}
\end{equation}
where the constants $A$, $B$, $C$, and $D$ are determined by fitting, as in Fig.~\ref{fig:s_tot}A.
Using Eqs.~(\ref{eq:def_integral}) and (\ref{eq:highT_expansion}), we then get
\begin{equation}
I_{\rm F} = \int_{0}^{\beta_0} d\beta' e_{\rm pot}(\beta') = \frac{n}{d} A \beta_0^{d/n} + \frac{n}{2d} B \beta_0^{2d/n} + \frac{n}{3d} C \beta_0^{3d/n} + \frac{n}{4d} D \beta_0^{4d/n} + \cdots.
\label{eq:I_F}
\end{equation}
which only depends on the fit parameters, $A$, $B$, $C$, and $D$.
Figure \ref{fig:s_tot}B presents the results for $s_{\rm tot}-s_{\rm mix}^{(M)}$ obtained by this procedure. Comparing results for systems with $N=1000$ and $N=20000$ confirms the absence of size dependence.

\subsection{Glass entropy}
We evaluate the entropy of glass states by Frenkel-Ladd (FL) thermodynamic integration~\cite{Frenkel,coluzzi1999thermodynamics,sastry2000evaluation,angelani2007configurational}, which requires imposing a harmonic potential with spring constant $\alpha$ on particle positions. The process then entails integrating the long-time limit of the mean-squared displacement starting from a strong $\alpha_{\rm max}$, at which the system behaves as an Einstein solid, and reaching a weak  $\alpha_{\rm min}$, at which the system is self caged. More specifically, we set
\begin{equation}
\beta U_{\alpha}(\Sigma_{\pi}^N, {\bf r}^N, {\bf r}_0^N) = \beta U(\Sigma_{\pi}^N, {\bf r}^N) + \alpha \sum_{i=1}^{N} | {\bf r}_i - {\bf r}_{0 i} |^2,
\label{eq:hamiltonian}
\end{equation}
where ${\bf r}_0^N$ is the template configuration from the equilibrium configuration of the target system.

As for the total entropy, we start from the partition function in Eq.~(\ref{eq:partition_function_new}) for the glass state,
\begin{equation}
\mathcal{Z}_{\alpha} =  \frac{1}{N!} \sum_{\pi} \frac{N!}{\Pi_{m=1}^M N_m! \Lambda^{Nd}} \int_V \mathrm{d} {\bf r}^N e^{-\beta U_{\alpha}(\Sigma_{\pi}^N, {\bf r}^N, {\bf r}_0^N)}.
\label{eq:Z_alpha}
\end{equation}
Note that the numerator of Eq.~(\ref{eq:Z_alpha}) is now multiplied by $N!$, because a given template configuration, ${\bf r}_0^N$, selects a single glass basin from the position phase space, while there exists $N!$ identical such choices, generated by permuting ${\bf r}_0^N$. 
Note also that the presence of the template configuration ${\bf r}_0^N$ prevents diameter permutations from being compensated by position permutation. The identity in Eq.~(\ref{eq:partition_function_total}) therefore does not hold in the glass state. 
The integration limit, $\lim_{\alpha_{\rm min} \to 0}$, also requires special conceptual and practical considerations.
Although for FL integration of a crystal $\alpha_{\rm min}$ is chosen to be infinitesimally small, here an additional constraint is that the system should remain within a glass basin and should thus not melt.
The practical implementation of this constraint is detailed in Sec.~\ref{sec:deltaintegration}. 

We compute the entropy $s_{\alpha}=\beta e_{{\rm tot}, \alpha} - \beta f_{\alpha}$, where $e_{{\rm tot}, \alpha}$ is the total energy and $f_{\alpha}=- (\beta N)^{-1} \ln \mathcal{Z}_{\alpha}$ is the free energy.
The glass entropy of the target system is then
\begin{equation}
s_{\rm glass} = \lim_{\alpha_{\rm min} \to 0} \overline{s_{\alpha_{\rm min}}},
\label{eq:S_glass_def}
\end{equation}
where $\overline{\cdots}$ here denotes averaging over template configurations ${\bf r}_0^N$. 

For convenience, we also define the following statistical averages,
\begin{eqnarray}
\left\langle (\cdots) \right\rangle_{\alpha}^{\rm T, S} &=& \frac{\frac{1}{N!} \sum_{\pi} \int_V \mathrm{d} {\bf r}^N  (\cdots) e^{ -\beta U_{\alpha}(\Sigma_{\pi}^N, {\bf r}^N, {\bf r}_0^N)}  }{\frac{1}{N!} \sum_{\pi} \int_V \mathrm{d} {\bf r}^N  e^{-\beta U_{\alpha}(\Sigma_{\pi}^N, {\bf r}^N, {\bf r}_0^N)}}, \label{eq:T_S} \\
\left\langle (\cdots) \right\rangle_{\alpha}^{\rm T} &=& \frac{\int_V \mathrm{d} {\bf r}^N  (\cdots) e^{ - \beta U_{\alpha}({\bf r}^N, {\bf r}_0^N)}   }{\int_V \mathrm{d} {\bf r}^N  e^{ - \beta U_{\alpha}({\bf r}^N, {\bf r}_0^N)}  }, \label{eq:T} \\
\left\langle (\cdots) \right\rangle_{\beta}^{\rm S} &=& \frac{ \frac{1}{N!} \sum_{\pi} (\cdots) e^{ -\beta U(\Sigma_{\pi}^N, {\bf r}_0^N) }  }{ \frac{1}{N!} \sum_{\pi}  e^{ -\beta U(\Sigma_{\pi}^N, {\bf r}_0^N) }},
\end{eqnarray}
where the superscripts denote statistical averages over positions (T) and permutations (S), evaluated by Monte Carlo (MC) simulations with standard translations and diameter swaps, respectively.
Note that any diameter permutation can be expressed as the product of the swaps of two diameters, hence permutation-phase space is properly sampled by swap MC simulations.

Following the conventional Frenkel-Ladd prescription~\cite{Frenkel} for Eq.~(\ref{eq:Z_alpha}),  we obtain 
\begin{equation}
s_{\rm glass} =  \frac{d}{2} - \ln \Lambda^d - \frac{d}{2} \ln \left(\frac{\alpha_{\rm max}}{\pi}\right) +  \lim_{\alpha_{\rm min} \to 0} \int_{\alpha_{\rm min}}^{\alpha_{\rm max}} \mathrm{d} \alpha \Delta_{\alpha}^{\rm T,S} + s_{\rm mix}^{(M)} -  \overline{ {\bf s}_{\rm mix}({\bf r}_0^N, \beta)}, 
\label{eq:S_glass_final}
\end{equation}
where $\Delta_{\alpha}^{\rm T, S}$ are constrained mean-squared displacements
\begin{equation}
\Delta_{\alpha}^{\rm T, S} = \frac{1}{N} \overline{\left\langle \sum_{i=1}^{N} | {\bf r}_i - {\bf r}_{0 i} |^2 \right\rangle_{\alpha}^{\rm T,S}},
\end{equation}
and ${\bf s}_{\rm mix}({\bf r}_0^N, \beta)$ is a mixing entropy contribution defined by
\begin{equation}
{\bf s}_{\rm mix}({\bf r}_0^N, \beta) = - \frac{1}{N} \ln \left( \frac{1}{N!} \sum_{\pi} e^{-\beta \left[ U(\Sigma_{\pi}^N, {\bf r}_0^N) - U({\bf r}_0^N) \right] } \right).
\label{eq:s_mix_def_new}
\end{equation}
This generalization of the standard FL integration method to systems with continuous polydispersity includes two novel physical features.
First, the mean-squared displacement $\Delta_{\alpha}^{\rm T,S}$ has to be evaluated by MC simulations of both translational and swap displacements, and is thus generally distinct from the standard mean-squared displacement, $\Delta_{\alpha}^{\rm T}$. 
Because $\Delta_{\alpha}^{\rm T,S}$ accounts for the non-vibrational contributions due to diameter permutations as well as for the purely vibrational contribution, $\Delta_{\alpha}^{\rm T,S} \geq \Delta_{\alpha}^{\rm T}$. 
Including the non-vibrational contribution also markedly improves the estimation of the glass entropy~\cite{ozawa2018ideal,misaki2018}, as we will see below. 
Second, the expression contains terms related to the mixing entropy, $s_{\rm mix}^{(M)} - \overline{ {\bf s}_{\rm mix}}$.
The diverging term, $s_{\rm mix}^{(M=N)} = \ln N \to \infty$, in Eq.~(\ref{eq:S_glass_final}) then exactly cancels the corresponding term in $s_{\rm tot}$ in Eq.~(\ref{eq:s_tot}). The remaining mixing entropy contribution, $\overline{{\bf s}_{\rm mix}}$ in $s_{\rm conf}$, is finite even for systems with continuous polydispersity.
Therefore, with this scheme continuous polydispersity does not present any conceptual or technical difficulty~\cite{misaki2018}. 

The key remaining tasks in order to compute $s_{\rm glass}$ involve measuring the mixing entropy contribution $\overline{{\bf s}_{\rm mix}}$ and integrating $\Delta_{\alpha}^{\rm T,S}$. Both are detailed below.

\subsubsection{Mixing entropy $\overline{{\bf s}_{\rm mix}}$}

\begin{figure}[htbp]
\includegraphics[width=0.48\columnwidth]{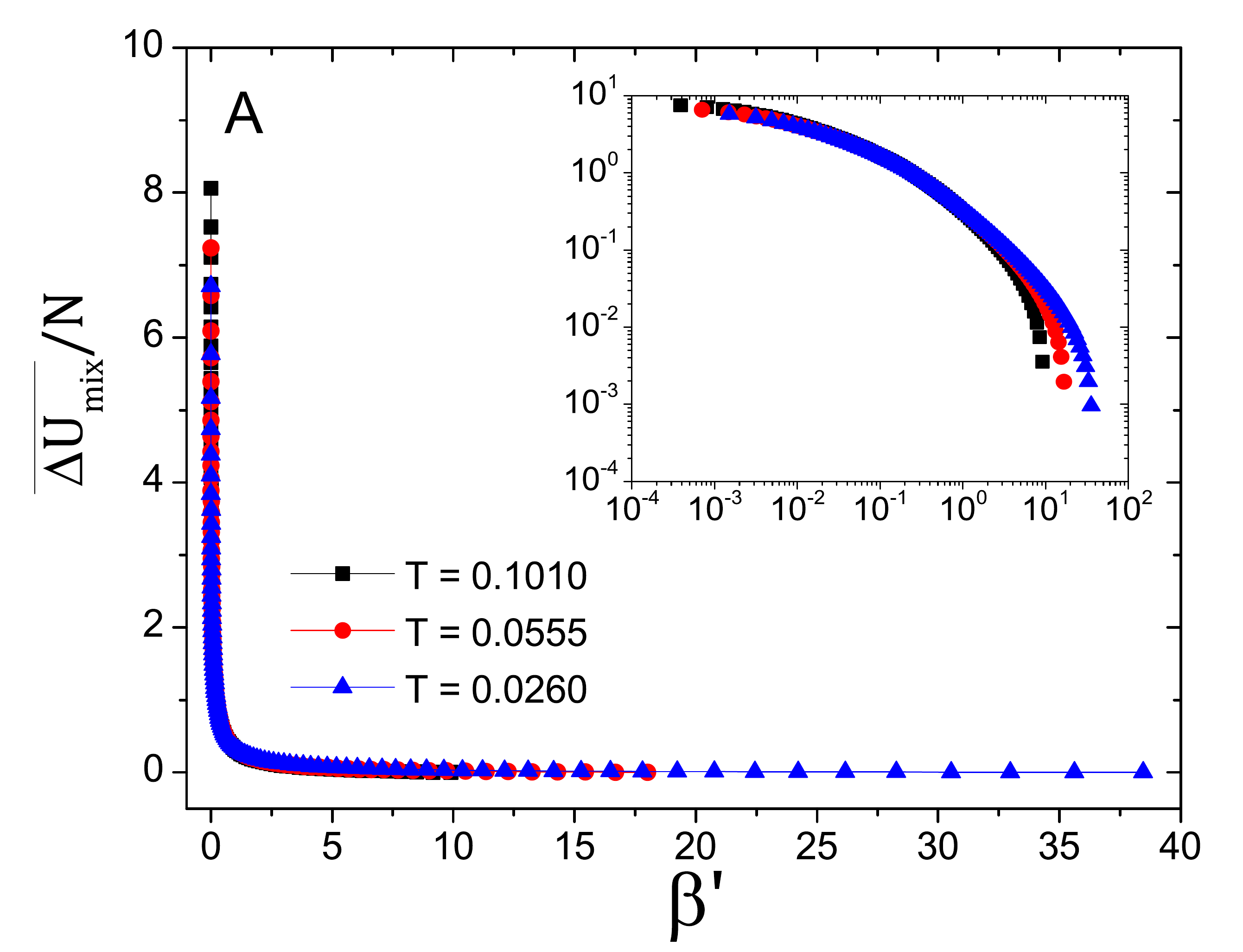}
\includegraphics[width=0.48\columnwidth]{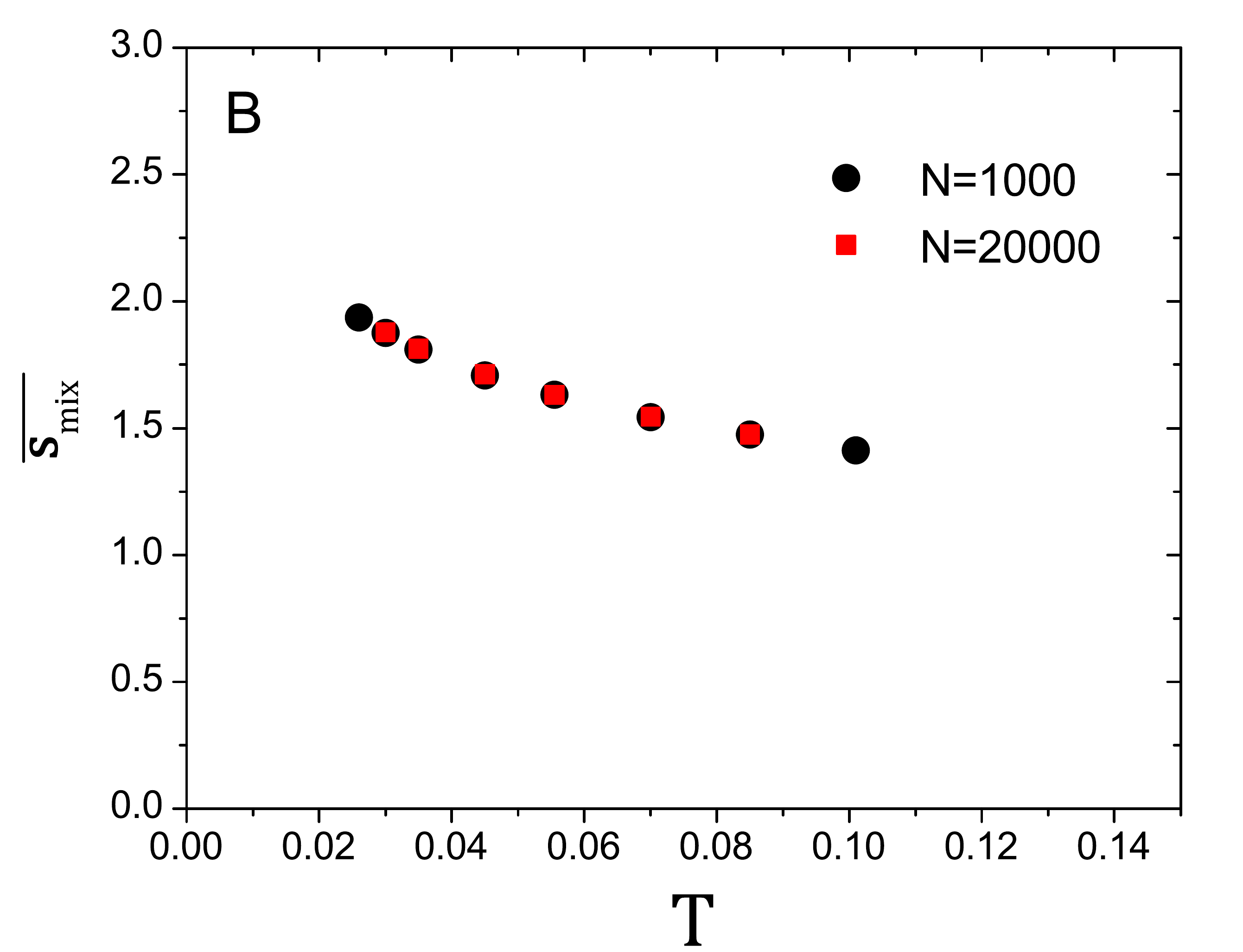}
\caption{
(A) Temperature evolution of $\overline{\Delta U_{\rm mix}({\bf r}_0^N, \beta)}$ for several equilibrium template configurations. (Inset) Same data on a logarithmic scale. 
(B) $\overline{{\bf s}_{\rm mix}}$ obtained by Eq.~(\ref{eq:s_mix_integral}) as a function of the temperature for N=1000 (black circles) and N=20000 (red squares) systems.
}
\label{fig:s_mix}
\end{figure}

The mixing entropy contribution, $\overline{{\bf s}_{\rm mix}}$, is determined by thermodynamic integration,  
\begin{equation}
\overline{{\bf s}_{\rm mix}({\bf r}_0^N, \beta)} =  \frac{1}{N} \int_0^{\beta} \mathrm{d} \beta' \overline{\Delta U_{\rm mix}({\bf r}_0^N, \beta')},
\label{eq:s_mix_integral}
\end{equation}
where $\Delta U_{\rm mix}$ is a potential energy difference defined by
\begin{equation}
\Delta U_{\rm mix}({\bf r}_0^N, \beta') =  \left\langle U(\Sigma_{\pi}^N, {\bf r}_0^N) \right\rangle_{\beta'}^{\rm S} - U({\bf r}_0^N).
\end{equation}
In practice, to get $\Delta U_{\rm mix}({\bf r}_0^N, \beta')$ the system is gradually heated from the target temperature $\beta$ to an infinite temperature $\beta \to 0$ using MC simulations with a fraction $p_{\rm swap}=1$ of the diameter swaps. Particles are thus kept at the same position as in the template configuration ${\bf r}_0^N$. 
As shown in Fig.~\ref{fig:s_mix}A, $\overline{\Delta U_{\rm mix}}/N$ takes very small values at large $\beta$, but sharply increases upon approaching $\beta\to 0$.
Note that $\overline{\Delta U_{\rm mix}}/N$ remains finite at $\beta \to 0$, hence so does $\overline{{\bf s}_{\rm mix}}$.
The resulting $\overline{{\bf s}_{\rm mix}}$ then increases slightly as temperature decreases, as seen in Fig.~\ref{fig:s_mix}B.
We confirm the absence of size dependence by comparing results for systems with $N=1000$ and $N=20000$.

\subsubsection{Integration of $\Delta_{\alpha}^{\rm T,S}$}
\label{sec:deltaintegration}
\begin{figure}[htbp]
\includegraphics[width=0.48\columnwidth]{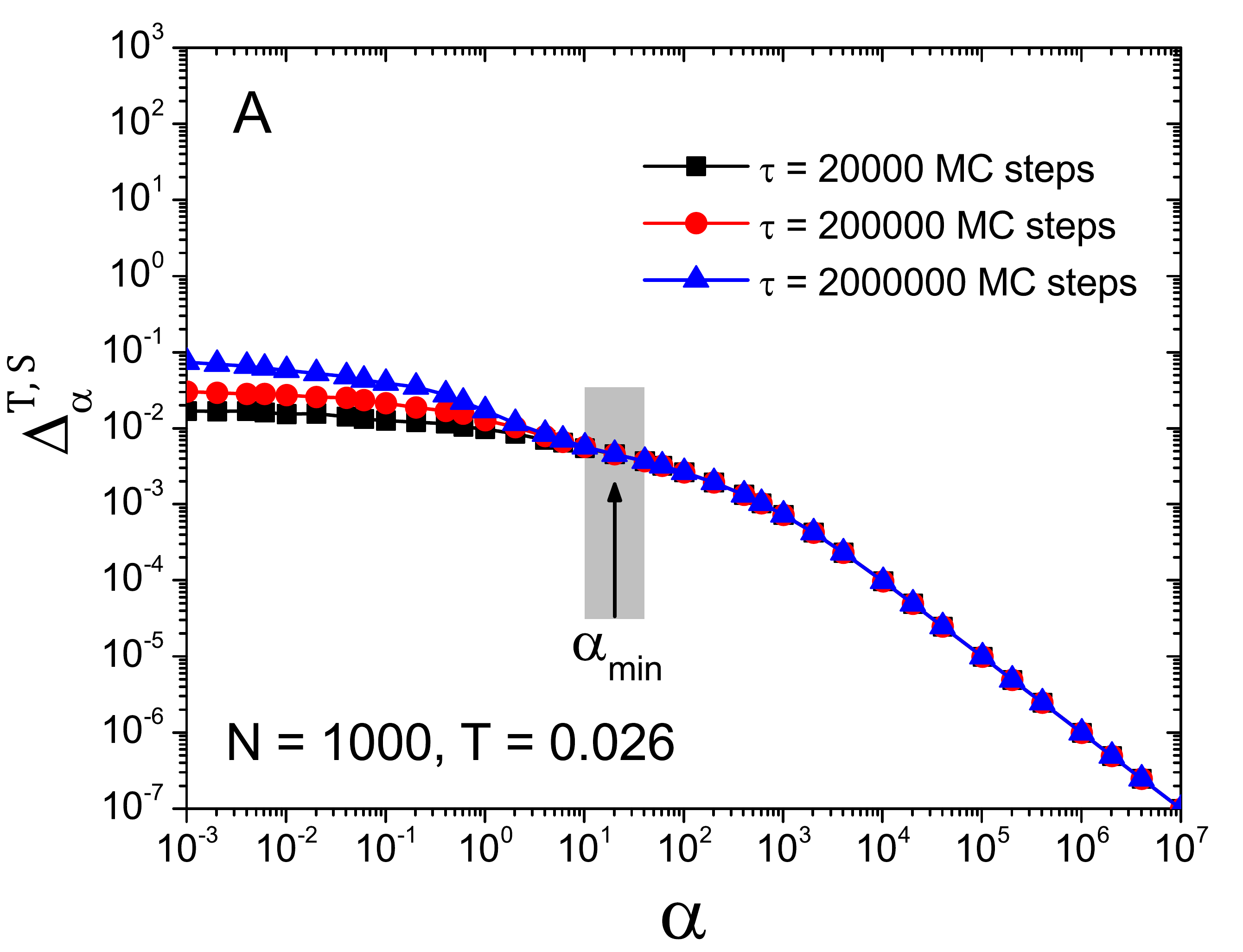}
\includegraphics[width=0.48\columnwidth]{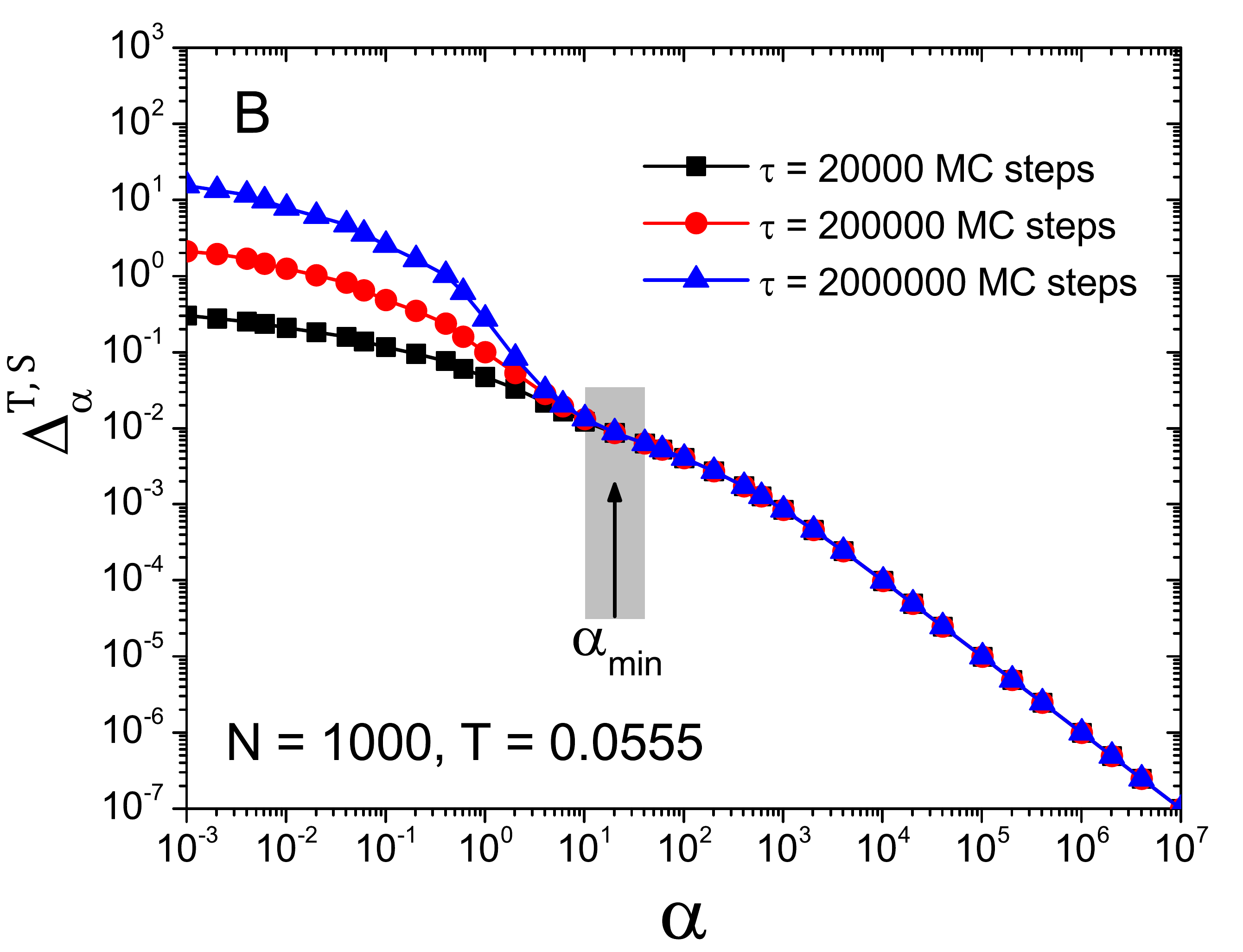}
\caption{
Timescale dependence of $\Delta_{\alpha}^{\rm T, S}$ for $N=1000$ at (A) $T=0.026$ and (B) $T=0.0555$.
For both temperatures, no $\tau$ dependence is observed up to $\alpha_{\rm min} \in [10.1, 40.5]$ (shaded region). 
}
\label{fig:timescale_MSD}
\end{figure}

Starting from $\alpha=\alpha_{\rm max}$, we perform MC simulations with decreasing $\alpha$ in steps of $\delta (\log_{10} \alpha) \simeq 0.18-0.4$.
For each data point, we perform $\tau=2 \times 10^5$ MC steps, measuring $\Delta_{\alpha}^{\rm T, S}$ only in the second half of the simulation. 
Figure~\ref{fig:Delta}A shows the evolution of $\Delta_{\alpha}^{\rm T, S}$ with $\alpha$. 
At large $\alpha$, the system is an Einstein solid with $\Delta_{\alpha}^{\rm T, S}=1/\alpha$, but upon decreasing $\alpha$, $\Delta_{\alpha}^{\rm T, S}$ plateaus. The system is then self caged.
Further decreasing $\alpha$, however, makes the harmonic constraint too weak to prevent the glass state from melting, thus implicitly defining $\alpha_{\rm min}$. The ensuing particle diffusion explains the upturn of $\Delta_{\alpha}^{\rm T, S}$.
In order to perform the integration in Eq.~(\ref{eq:S_glass_final}), a practical manipulation of the limit must be used for $\alpha<\alpha_{\rm min}$. We consider 
\begin{eqnarray}
\lim_{\alpha_{\rm min} \to 0} \int_{\alpha_{\rm min}}^{\alpha_{\rm max}} \mathrm{d} \alpha \Delta_{\alpha}^{\rm T,S}  &\simeq& \alpha_{\rm min}  \Delta_{\alpha_{\rm min}}^{\rm T,S}  + \int_{\alpha_{\rm min}}^{\alpha_{\rm max}} \mathrm{d} \alpha \Delta_{\alpha}^{\rm T,S} \nonumber \\ 
&=& \alpha_{\rm min}  \Delta_{\alpha_{\rm min}}^{\rm T,S}  + (\ln 10) \int_{\log_{10} \alpha_{\rm min}}^{\log_{10} \alpha_{\rm max}} \mathrm{d} (\log_{10} \alpha)\alpha \Delta_{\alpha}^{\rm T,S}.
\label{eq:integration_Delta}
\end{eqnarray}
While $\alpha_{\rm max}$ should straightforwardly be chosen in the Einstein solid regime, e.g., we use $\alpha_{\rm max} \simeq 1 \times 10^7$,
the choice of $\alpha_{\rm min}$ is not unambiguous. Based on the above discussion, 
we understand that $\alpha_{\rm min}$ should be within the plateau regime of $\Delta_\alpha^{\rm T,S}$, 
where $\Delta_{\alpha}^{\rm T, S}$ does not depend on $\tau$.
As seen in Fig.~\ref{fig:Delta}A, if $\alpha$ is too small, $\Delta_{\alpha}^{\rm T, S}$ increases at large $\tau$. 
In order to identify the regime of proper equilibration in the plateau region, the $\tau$-dependence of $\Delta_{\alpha}^{\rm T, S}$ is presented in Fig.~\ref{fig:timescale_MSD} A, B.
The shaded region denotes the regime in which the time needed to obtain well averaged observables has no detectable $\tau$ dependence. This corresponds to the regime within which $\alpha_{\mathrm{min}}$ can be safely chosen.
The choice of $\alpha_{\rm min}$ nonetheless affects $s_{\rm glass}$, especially at high temperatures, where a plateau never fully forms.
The systematic uncertainty associated with this choice is captured by the errorbars for $s_{\rm glass}$ in the shaded region, $\alpha_{\rm min} \in [10.1, 40.5]$,  
of Fig.~\ref{fig:Delta}A. 
The edges of the errorbar in Fig.~\ref{fig:Delta}C correspond to $s_{\rm glass}$ extracted from the two extremes of the shaded region, $\alpha_{\rm min} =10.1$ and $40.5$.
As expected, these error bars become smaller as temperature decreases, thus validating our choice of $\alpha_{\rm min}$.
Since $s_{\rm conf}$ in the main text depends on the chosen $\alpha_{\rm min}$ in the determination of $s_{\rm glass}$, we display the errorbars corresponding to $s_{\rm conf}$ from $\alpha_{\rm min}$-values chosen inside the plateau region, in the
same way as in Fig.~\ref{fig:Delta}C.

\subsubsection{Effetc of Mermin-Wagner (MW) fluctuations}

\begin{figure}[htbp]
\includegraphics[width=0.32\columnwidth]{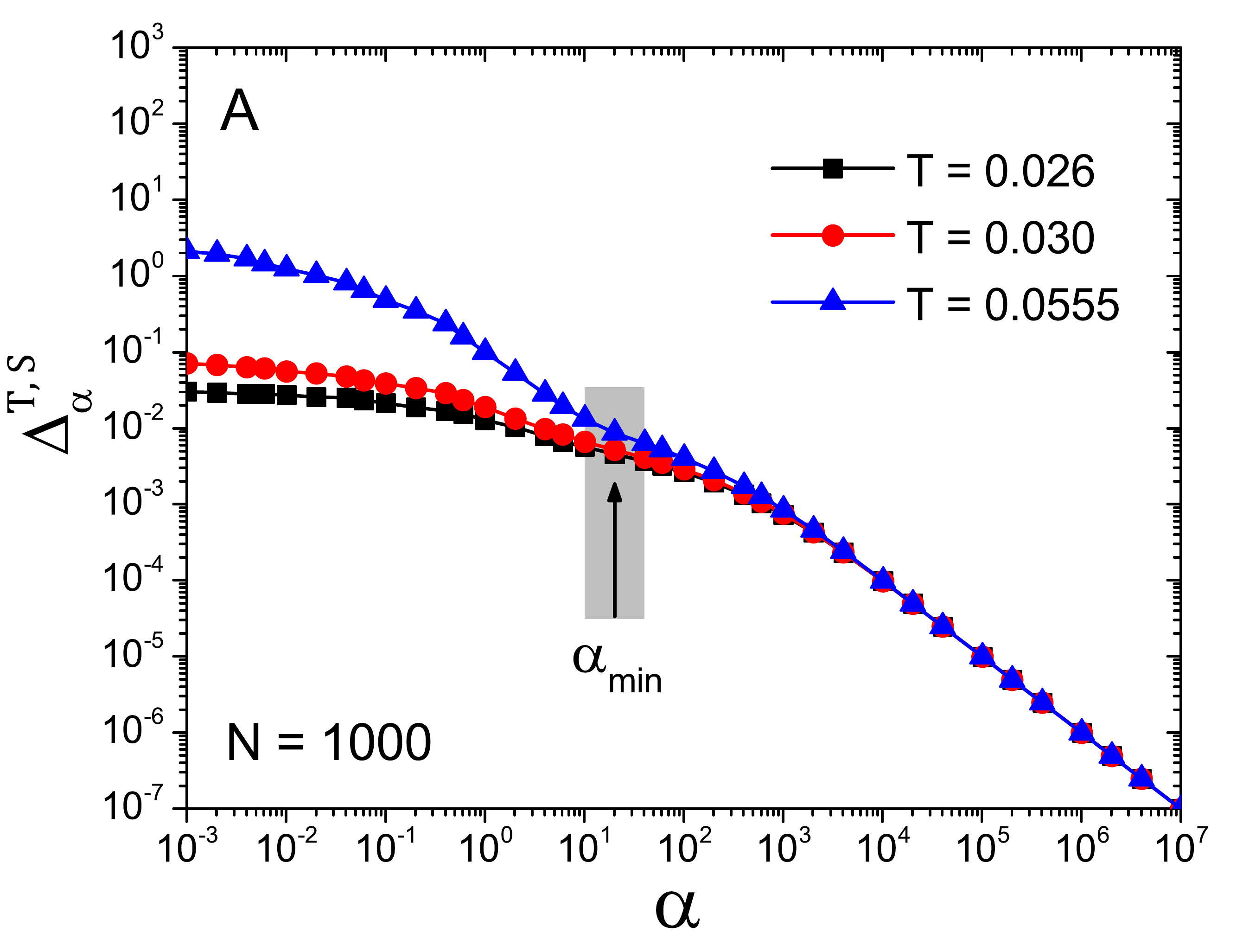}
\includegraphics[width=0.32\columnwidth]{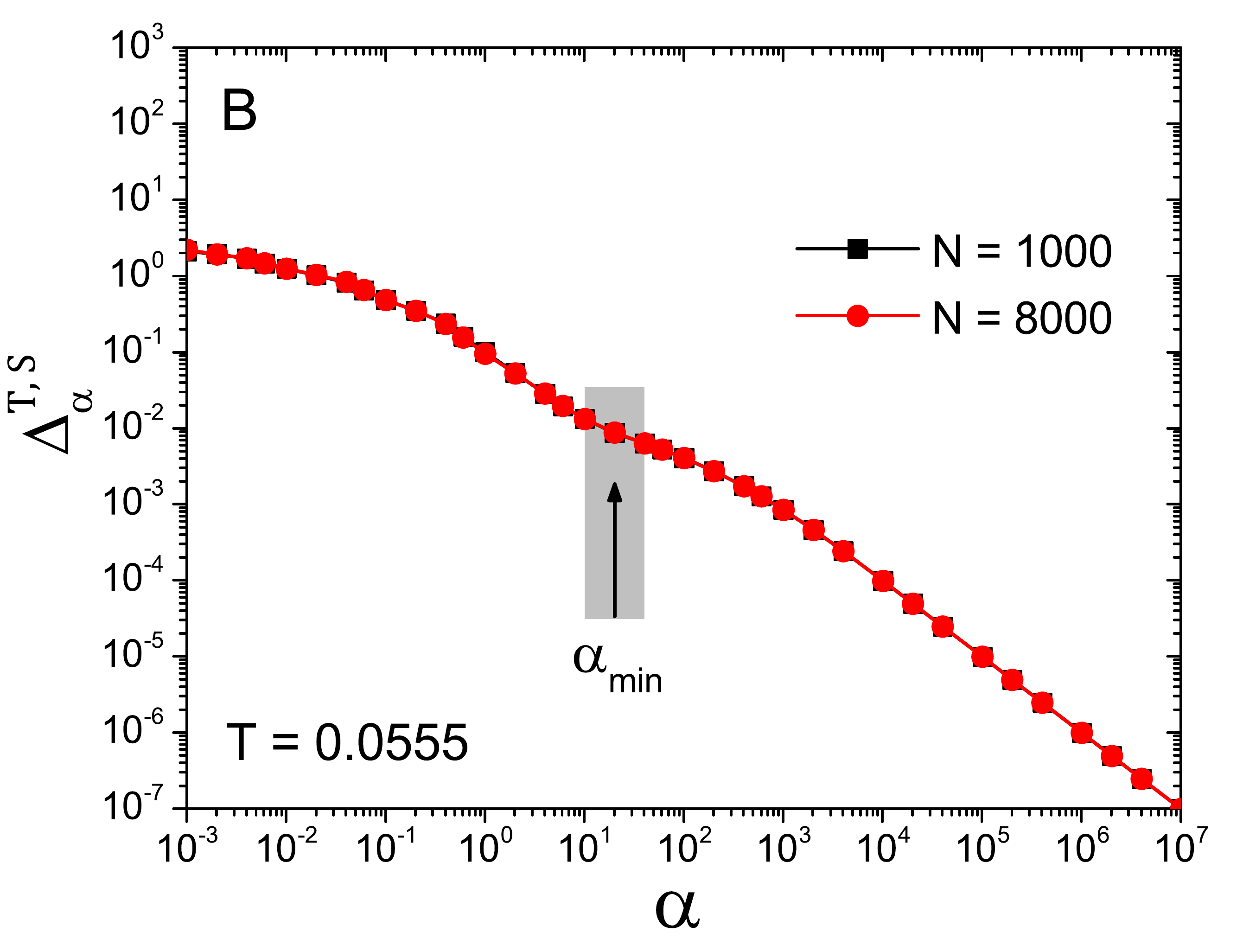}
\includegraphics[width=0.32\columnwidth]{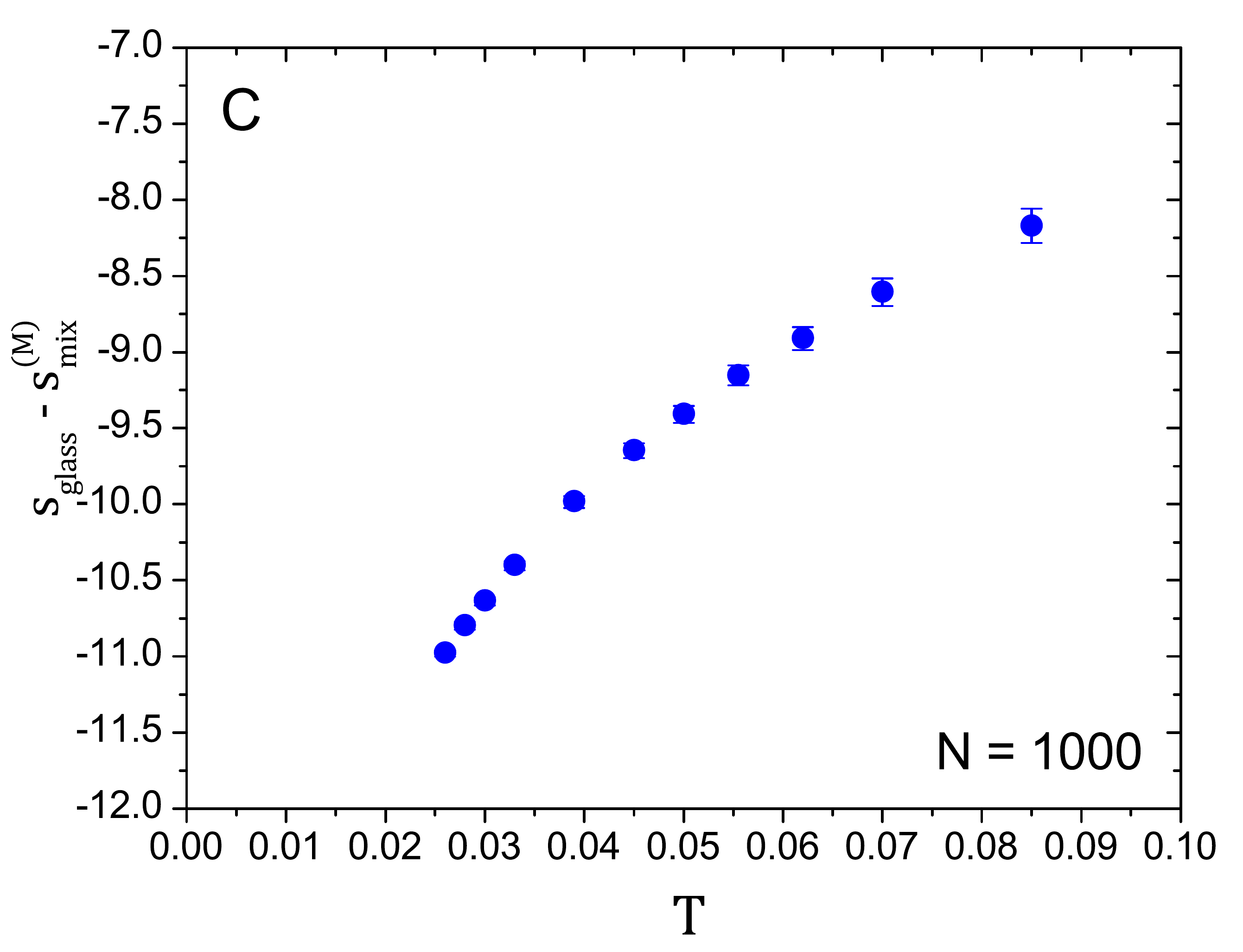}
\caption{
(A) Mean-squared displacement $\Delta_{\alpha}^{\rm T, S}$ in the Frenkel-Ladd construction at several temperatures for $N=1000$.
The shaded region denotes the potential range for $\alpha_{\rm min}$, and the arrow the specific choice of $\alpha_{\rm min}=20.3$. 
(B) System size dependence of $\Delta_{\alpha}^{\rm T, S}$ for $T=0.0555$.
(C) $s_{\rm glass}$ obtained by Eq.~(\ref{eq:S_glass_final}) using $\alpha_{\rm min}=20.3$ for $N=1000$. The diverging mixing entropy term, $s_{\rm mix}^{(M)}$, is subtracted.
The span of the errorbars corresponds to $s_{\rm glass}$ for $\alpha_{\rm min} \in [10.1, 40.5]$.
}
\label{fig:Delta}
\end{figure}

Note that because $\Delta_{\alpha}^{\rm T, S}$ essentially coincides with the plateau height of the dynamically measured mean-squared displacement, one may also expect MW fluctuations to contribute significantly~\cite{shiba2018isolating}. 
To assess the relevance of MW fluctuations, we consider the system size dependence of $\Delta_{\alpha}^{\rm T, S}$. Figure~\ref{fig:Delta}B presents no notable finite-size effect down to very small $\alpha$. This suggests that imposing a very weak harmonic constraint suppresses MW fluctuations without disturbing the overall thermodynamics of the system. This process is thus akin to the effect of pinning a few percent of the particles as was reported in Ref.~\cite{IFKKMK17}.

\subsection{Potential energy landscape approach}

\begin{figure}[htbp]
\begin{center}
  \includegraphics[width=0.432\linewidth]{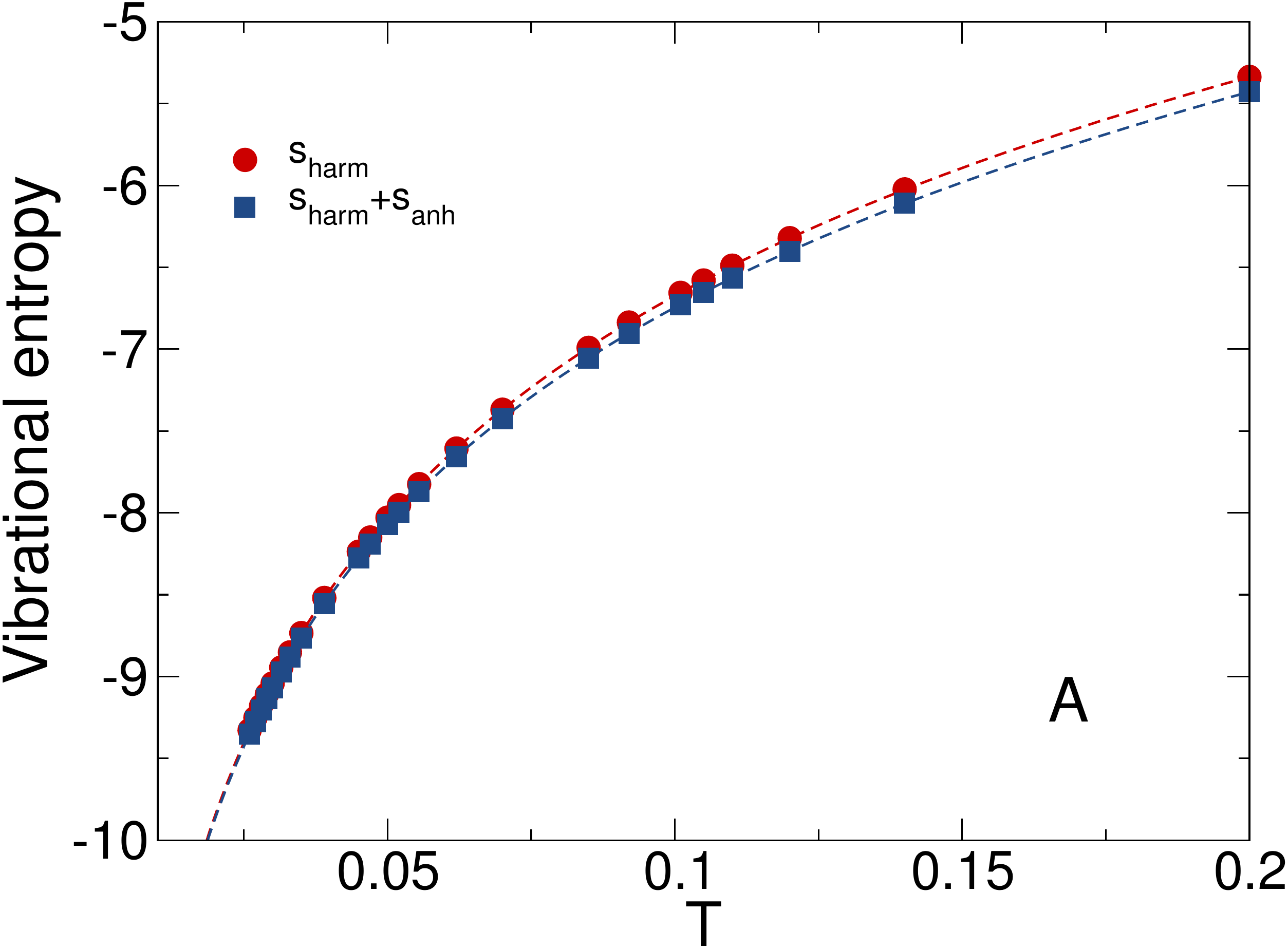}\hspace{0.03\linewidth}
  \includegraphics[width=0.45\linewidth]{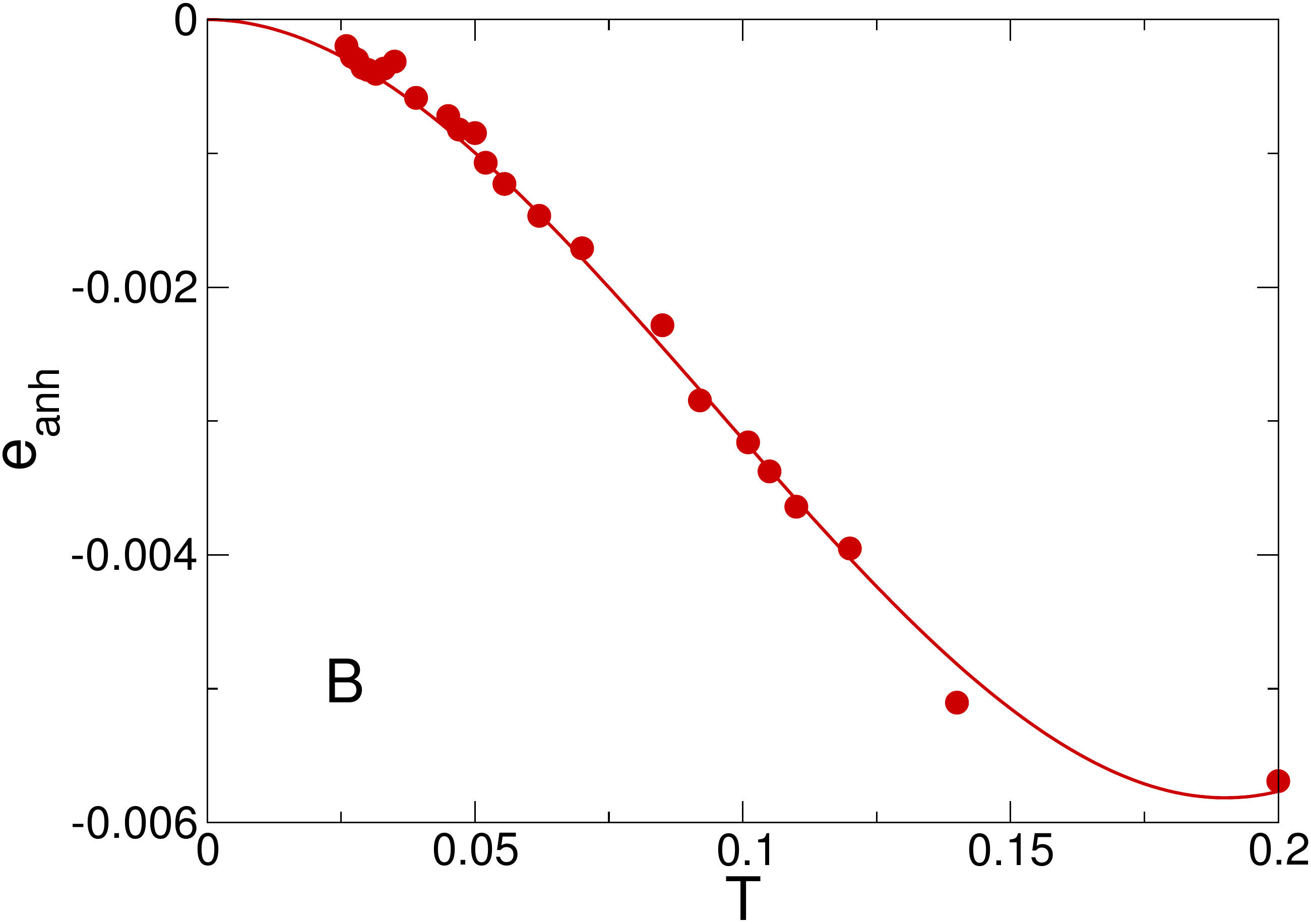}
\caption{
(A) Vibrational entropy results for $d=2$
including the harmonic contribution $s_{\rm harm}$ (dots) and also its anharmonic correction $s_{\rm anh}$ (squares).
(B) The anharmonic contribution of the potential energy $e_{\rm anh}$.
}
\label{fig:s_vib}
\end{center}
\end{figure}

We also consider an alternate approach for estimating $s_{\rm conf}$ based on the potential energy landscape (PEL)~\cite{sciortino2005potential}.
In this approach, the glass entropy is obtained from information about the inherent structures (IS) of the glass state.
In order to evaluate the impact of polydispersity on $s_{\rm conf}$, we employ an effective $M^*$-component approximation as in Ref.~\cite{misaki2017}. This approach provides an effective mixing entropy $s_{\rm mix}^*=s_{\rm mix}^{(M^*)}$.
(The numerical determination of $M^*$ is explained below.)
We then compute the glass entropy $s_{\rm glass}$ by 
$s_{\rm glass}=s_{\rm harm} + s_{\rm anh}$,
where $s_{\rm harm}$ and $s_{\rm anh}$ are the harmonic vibrational entropy and its anharmonic correction, respectively~\cite{sciortino2005potential}.
The harmonic term is computed as
\begin{equation}
s_{\rm harm} = \frac{1}{N} \left\langle \sum_{a=1}^{d(N-1)} \left\{1 - \ln (\beta \hbar \omega_a) \right\} \right\rangle_{\rm IS},
\label{eq:s_vib_IS}
\end{equation}
where $\langle \cdots \rangle_{\rm IS}$ is an average over IS configurations obtained by the conjugate gradient method and $\omega_a=\sqrt{\lambda_a/m}$ is the square root of  eigenvalue $\lambda_a$ of the Hessian of this IS.
Figure~\ref{fig:s_vib}A shows $s_{\rm harm}$ as a function of $T$ for $d=2$.

The anharmonic contribution to the potential energy is
$e_{\rm anh}(T) = e_{\rm pot}(T) - e_{\rm IS}(T) - \frac{d}{2} T$,
where $e_{\rm IS}$ is the inherent structure energy, and the last term is the harmonic contribution to the energy.
From $e_{\rm anh}(T)$, we also have
\begin{equation}
s_{\rm anh}(T) = \int_0^T dT' \frac{1}{T'} \frac{\partial e_{\rm anh}(T')}{\partial T'},
\label{eq:S_anh}
\end{equation}
where we used the fact that the system is perfectly harmonic at low $T$, i.e., $s_{\rm anh}(T=0)=0$. A low-temperature expansion,
$e_{\rm anh}(T)=\sum_{k=2} c_k  T^k$,
has $T$-independent coefficients, $c_k$.
Substituting this expansion into Eq.~(\ref{eq:S_anh}) gives
\begin{equation}
s_{\rm anh}(T) = \sum_{k=2} \frac{k}{k-1} c_k T^{k-1}.
\label{eq:S_anh2}
\end{equation}

The fit of $e_{\rm anh}$ with parameters $c_2$ and $c_3$ is shown in Fig.~\ref{fig:s_vib}B, and the resulting $s_{\rm harm} + s_{\rm anh}$ is shown in Fig.~\ref{fig:s_vib}A.
The resulting anharmonic contribution is $|s_{\rm anh}| < 0.1$ in the temperature range of interest.

\subsubsection{MW fluctuations effects}

The glass entropy measured using the PEL approach also is not affected by MW fluctuations.
Consider first the mean-squared displacement of standard solids, $\langle |{\bf u}|^2 \rangle$. 
For a monodisperse crystalline solid, one can write 
\begin{equation}
\langle |{\bf u}|^2 \rangle=\frac{d k_{\rm B}T}{m} \int_{2 \pi c/L}^{\infty} \mathrm{d} \omega \frac{g(\omega)}{\omega^2},
\end{equation}
where $g(\omega)$ and $c$ are the vibrational density of states and the velocity of sound, respectively.
Because one expects a Debye scaling $g(\omega) \propto \omega^{d-1}$ at low $\omega$, in $d=2$,  $\langle |{\bf u}|^2 \rangle \sim \ln L \to \infty$ diverges in the thermodynamic limit.
Writing Equation~\ref{eq:s_vib_IS} using the density of state formalism,
\begin{equation}s_{\rm harm}=d \int_{2 \pi c/L}^{\infty} \mathrm{d} \omega g(\omega)\left\{1 - \ln (\beta \hbar \omega) \right\}\, ,
\end{equation}
by contrast, in $d=2$ gives the $L$-dependent term, $\frac{\ln L}{L^2}$, that vanishes as the system size increases.
Additionally, $s_{\rm anh}$ does not depend on system size because $e$ and $e_{\rm IS}$ (and thus $e_{\rm anh}$) display no system-size dependence at large enough $L$ (not shown).
The glass entropy, $s_{\rm harm}+s_{\rm anh}$, is therefore system-size independent and hence unaffected by MW fluctuations.

\subsubsection{Determination of $M^*$}

\begin{figure}[htbp]
\includegraphics[width=0.48\columnwidth]{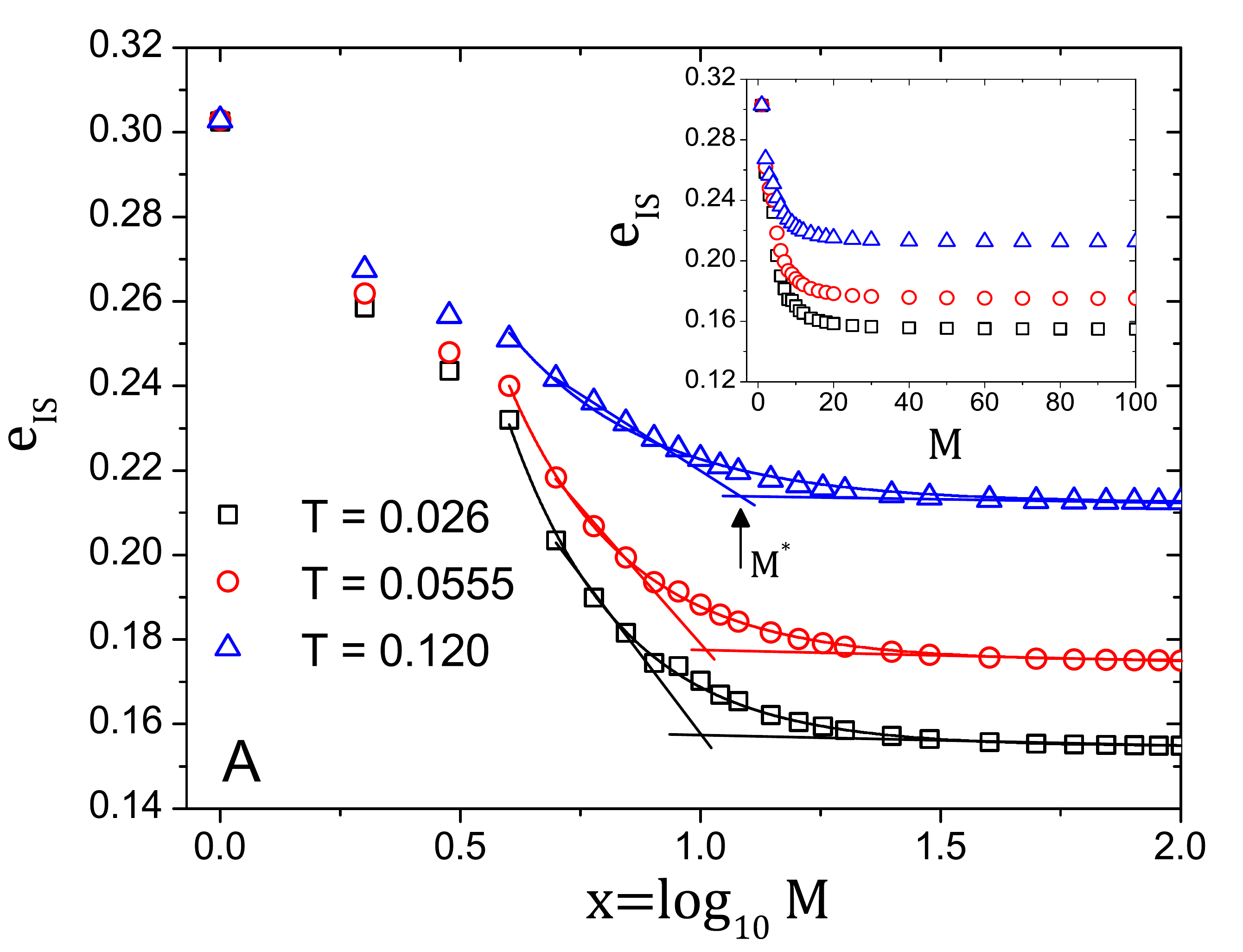}
\includegraphics[width=0.48\columnwidth]{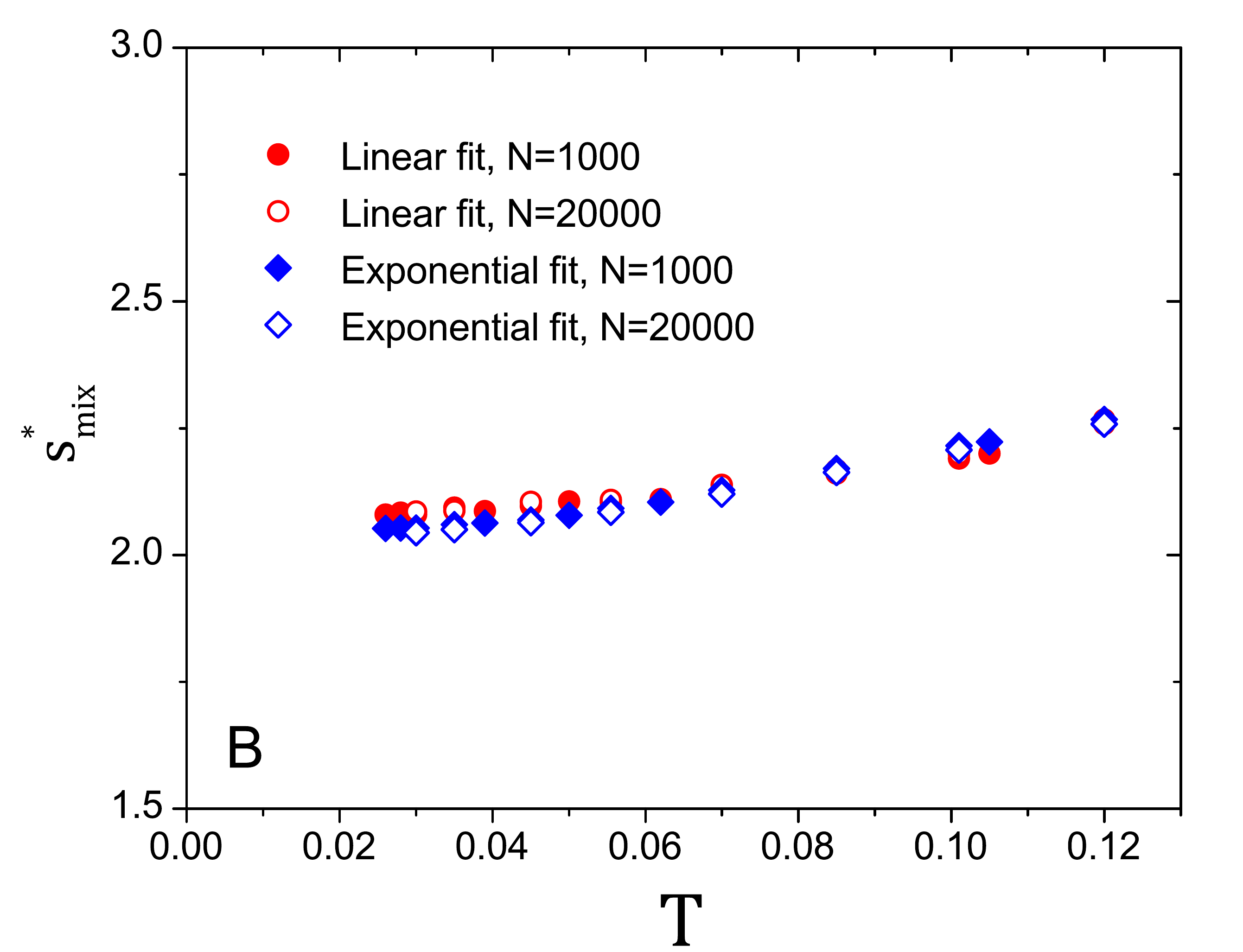}
\caption{
(A) $e_{\rm IS}$ vs. $x=\log_{10} M$ plot for the determination of $M^*$ for $N=1000$. The horizontal axis is the logarithmic plot.
The vertical arrow represents the $M^*$ value
determined by the intersection of the two straight lines.
The exponential fit is also shown.
Inset: The horizontal axis is linear.
(B) $s_{\rm mix}^*=s_{\rm mix}^{(M^*)}$ as a function of $T$ for $M^*$ determined by the linear fit and exponential fit for $N=1000$ and $20000$.
}
\label{fig:smix_star}
\end{figure}

The effective component, $M^*$, is determined based on the potential energy landscape.
As explained in Ref.~\cite{misaki2017}, 
$M^*$ should be such that 
(i) particle diameter swaps within a single effective species 
leave the potential energy basin unaffected, and 
(ii) particle diameter swaps between different species 
drive the 
system out of the original basin.
To determine $M^*$ in practice, we prepare equilibrium configurations of the original
continuously polydisperse system characterized by the distribution $f(\sigma)$. 
We then decompose $f(\sigma)$ into $M$ species (from $M=1$ to $100$), dividing $f(\sigma)$ into equal intervals $\Delta \sigma=(\sigma_{\rm max}-\sigma_{\rm min})/M$, such that each species occupies more or less the same fraction of the total volume.
For a given $M$ value, we 
systematically perform diameter swaps within each species.
We repeat such diameter swap $N$ times so that most particles experience the swap.
We then quench the obtained configuration 
to its IS, monitoring whether the system lands in a different
basin (for $M<M^*$) or not (for $M>M^*$) by measuring $e_{\rm IS}$ as a function of $M$ (or $x=\log_{10} M$) [see Fig.~\ref{fig:smix_star}A].

At large $x=\log_{10} M$, we observe nearly constant values of 
$e_{\rm IS}^{(M)} \simeq e_{\rm IS}^{(M = N)}$, which means that the swap of the diameters 
within each $M$ species marginally affects the system.
After the diameter swaps, the system thus essentially remains in the original basin.
However, with decreasing $M$, $e_{\rm IS}^{(M)}$ starts to increase significantly from $e_{\rm IS}^{(M = N)}$. 
This observation indicates that at smaller $M$, the impact of particle swaps is 
so strong that the original basin is destroyed, and the system moves to 
another basin.

From the $e_{\rm IS}$ vs.~$x=\log_{10} M$ plot, the clear crossover between large and small $M$ behaviors determines $M^*$ as the intersection of two linear fits as shown in Fig.~\ref{fig:smix_star}A.
We show the resulting $s_{\rm mix}^*=s_{\rm mix}^{(M^*)}$ as a function of the temperature in Fig.~\ref{fig:smix_star}B.
We confirm the absence of size dependence by comparing results for systems with $N=1000$ and $N=20000$.

We also employ an exponential fit as an alternative way to extract $M^*$ from the crossover.
We use the following exponentially decaying function: $e_{\rm IS}(x)=e_{\rm IS}^{(M=N)}+A \exp[-(x-x_0)/B]$, where $x_0=\log_{10}M_0$ is the starting point of the exponential fitting, and $A$ and $B$ are fitting parameters.
We set $M_0=4$ thus $x_0=0.602$.
The exponential functional form precisely captures the data points as shown in Fig.~\ref{fig:smix_star}A.
Here we define $x^*=\log_{10}M^*$ by the location where the exponential function decays sufficiently, i.e., $(e_{\rm IS}(x^*)-e_{\rm IS}^{(M=N)})/A=C$, where $C$ is an arbitrary small value.
We set $C \simeq 0.2$ so that $M^*$ by this exponential scheme corresponds to the one by the intersection of the two linear fits described above for $T=0.12$ where the linear fit scheme is good.
As shown in Fig.~\ref{fig:smix_star}B, the resulting $s_{\rm mix}^*=s_{\rm mix}^{(M^*)}$ by the exponential fit eventually follows similar temperature dependence of the linear fit, suggesting robustness of our numerical determination of $M^*$.
In the main text, $s_{\rm conf}$ using $s_{\rm mix}^*$ from the linear and exponential fit schemes are called PEL1 and PEL2, respectively.

\section{Point-to-set (PTS) correlations}
This section reports the setups and the results for point-to-set observables in soft disks; results for hard disks are reported in subsection~\ref{HD}.


\label{sec:PTS}
\subsection{PTS observables}
Similarity between two configurations within the cavity is characterized by the cavity core overlap, $\qc$, computed as in Refs.~\cite{BCY16,CDLY16,YBCT16,BCCNOY17}. (i) We assign a local overlap value to each particle through the overlap estimator function $w(z)\equiv {\rm exp}\le[-\le(\frac{z}{b}\ri)^2\ri]$ with $b=0.2$;
(ii) we perform a linear interpolation through a Delaunay tessellation to define a continuous overlap field; and (iii) we measure the cavity core overlap by taking the average of the field values within the radius $r_{\rm c}=1.0$
from the cavity center, evaluated by MC integration with $10^3$
points.

For each temperature $T$ and cavity radius $R$, the PTS correlation function
\be\label{PTScor}
Q_{\rm PTS}(R;T)=[\langle \qc\rangle]|_{T, R}\, ,
\ee
is evaluated by disorder-averaging--denoted $[\ldots]$--over $100$ cavity centers ($200$ for $0.0315\leq T\leq 0.039$ and $300$ for $T=0.028$)
and, within each cavity, thermal-averaging--denoted $\langle\ldots\rangle$--over $s_{\rm{prod}}$ pairs of equilibrated configurations (see subsection~\ref{PTSMC}).

\begin{figure*}
\centering
\hspace{-0.1in}\includegraphics[width=0.33\textwidth]{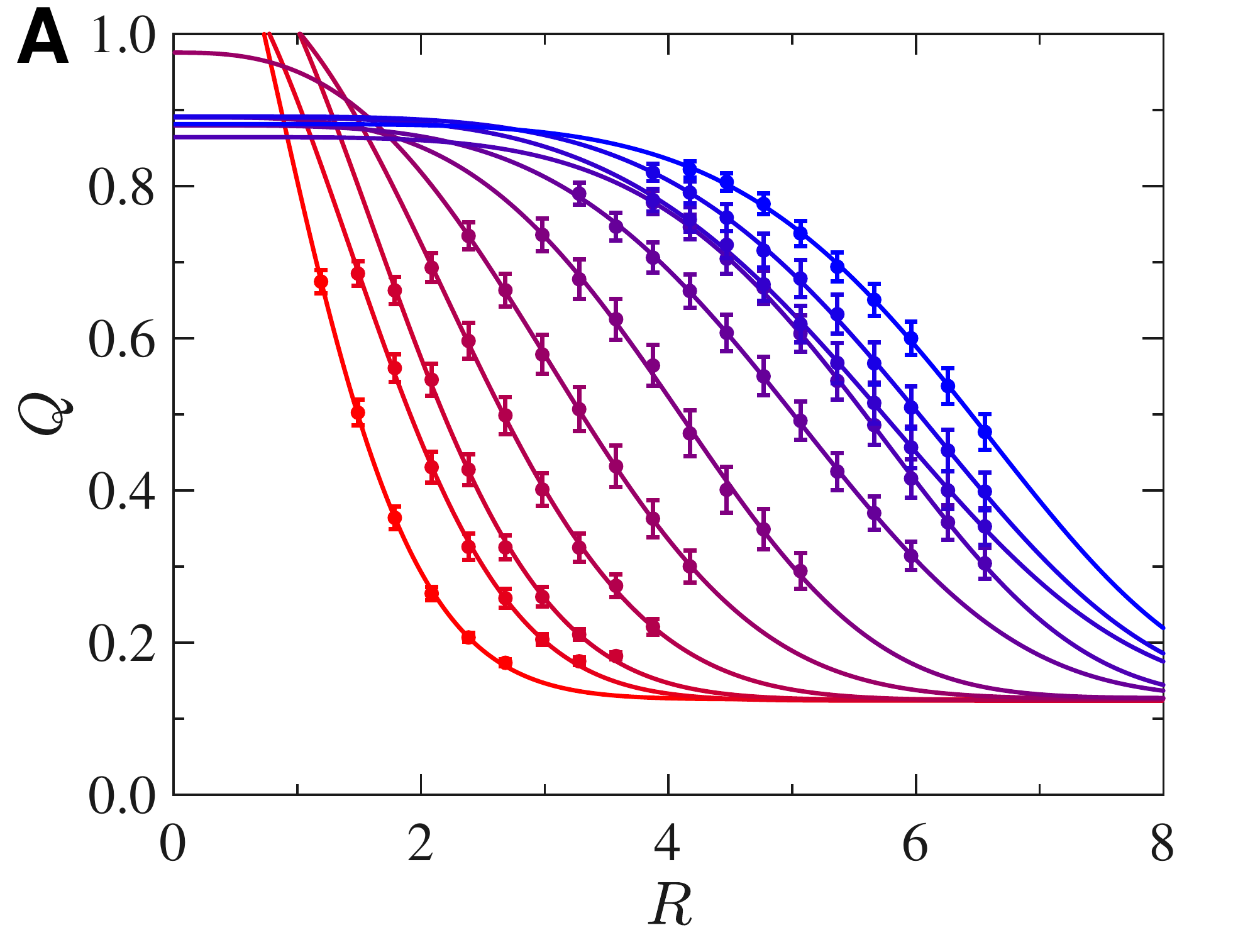}
\hspace{-0.1in}\includegraphics[width=0.33\textwidth]{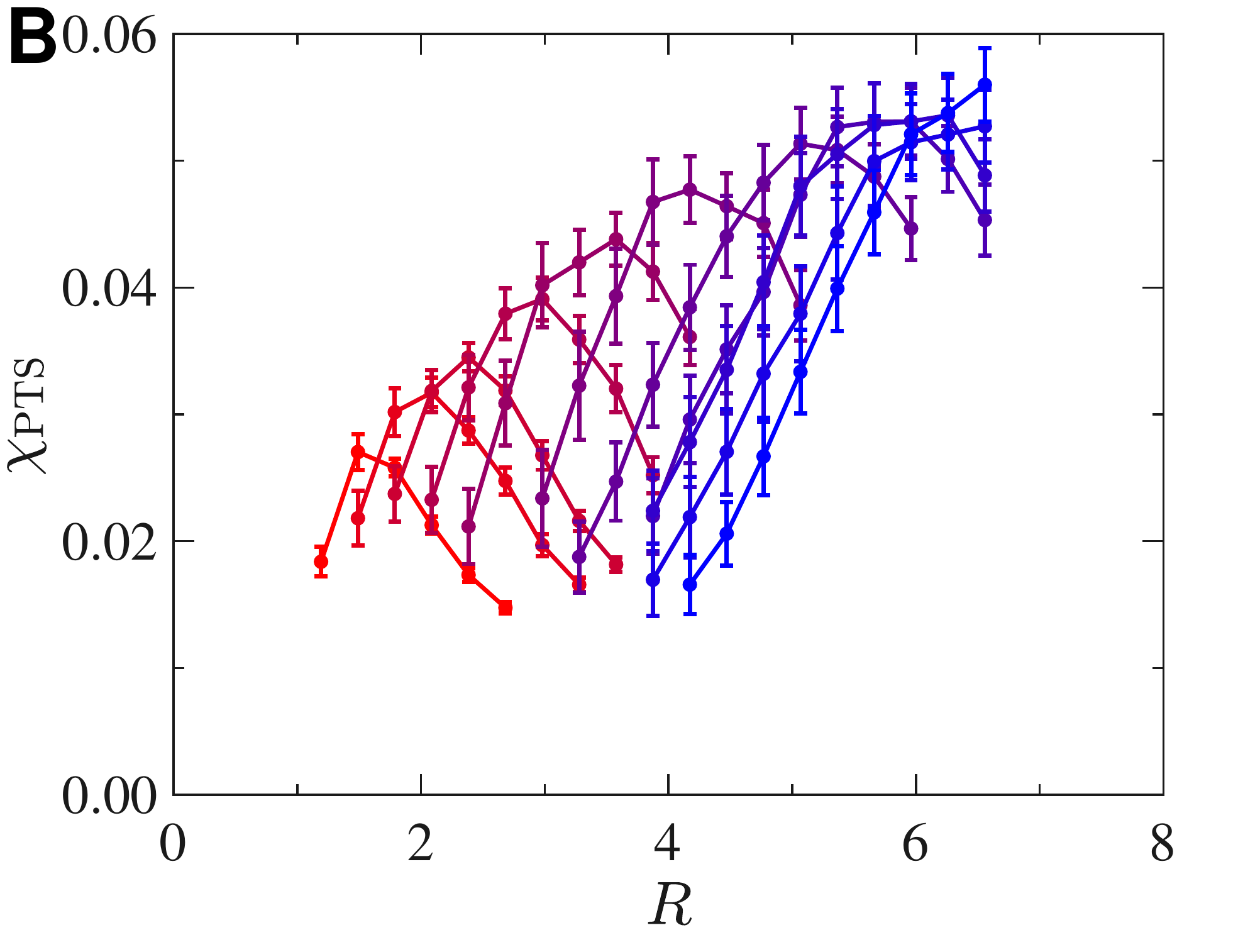}
\hspace{-0.1in}\includegraphics[width=0.33\textwidth]{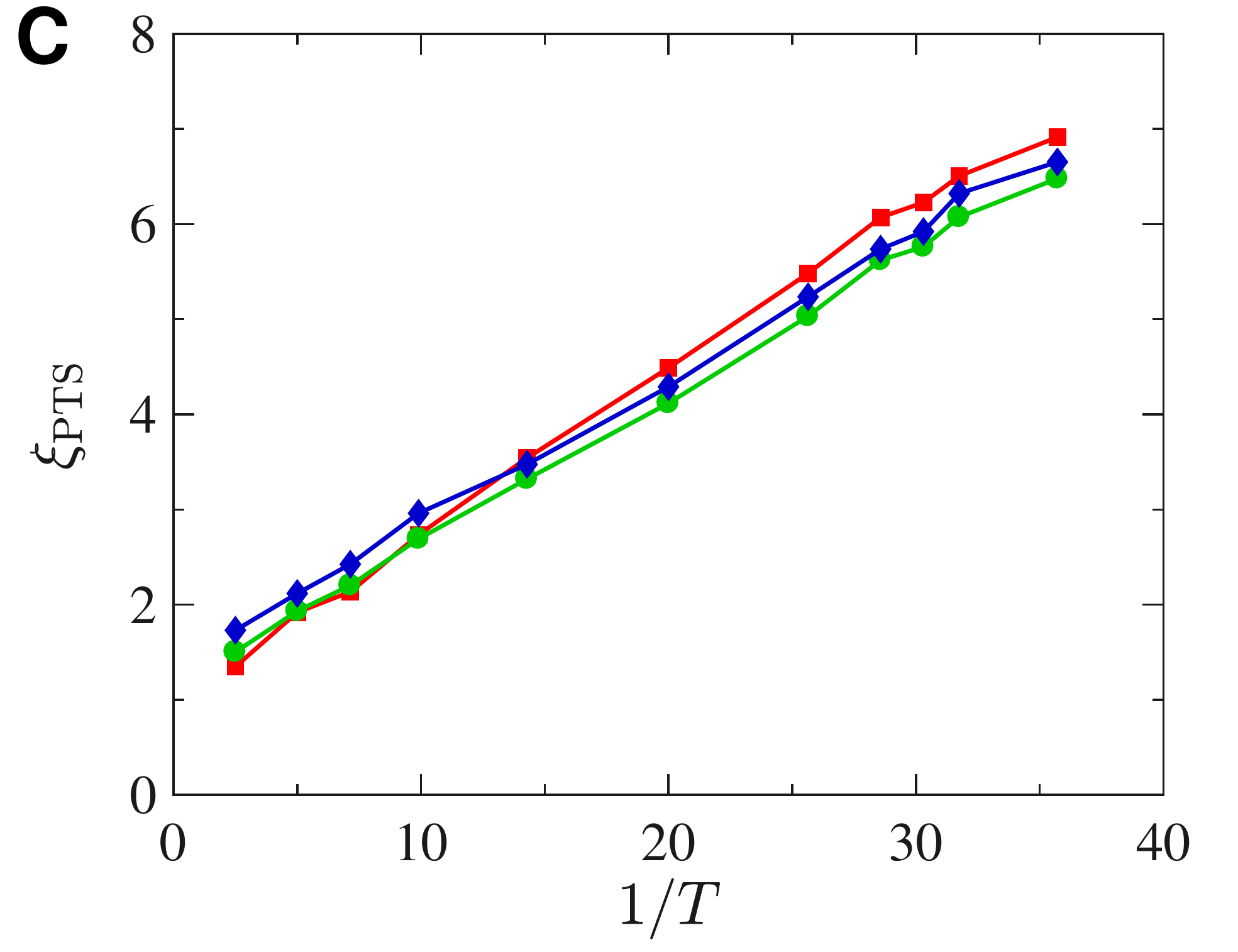}
\caption{
(A) Radial decay of the cavity PTS correlation at $T = 0.400$, $0.200$, $0.140$, $0.101$, $0.070$, $0.050$, $0.039$, $0.035$, $0.033$, $0.0315$, and $0.028$ (from red to blue) for soft disks.
Solid lines are fits to a compressed exponential.
(B) PTS susceptibilities with cavity radius $R$. Solid lines are guides for the eyes.
(C) PTS lengths $\xi^{\mathrm{fit}}_{\mathrm{PTS}}$ (red-square), $\xi^{\mathrm{th}}_{\mathrm{PTS}}$ (green-circle), and $\xi^{\mathrm{peak}}_{\mathrm{PTS}}$ (blue-diamond) as a function of the inverse temperature for soft disks. The clear linear growth suggests that $T_{\mathrm{K}}=0$ and that the RFOT exponent $\theta=1=\frac{d}{2}=d-1$ in $d=2$.
}
\label{PTSs}
\end{figure*}

One way to extract the PTS correlation length is through the compressed exponential fit, 
\begin{equation}
Q^{\mathrm{fit}}_{\rm PTS}(R;T)=A\exp[-\le\{R/\xi^{\mathrm{fit}}_{\rm PTS}(T)\ri\}^{\gamma}]+Q_{\rm PTS}^{\rm bulk}(T)\, ,
\end{equation}
with the bulk value, $Q_{\rm PTS}^{\rm bulk}$, evaluated by taking $10^5$
pairs of independent configurations in bulk samples.
Note that differently from Ref.~\cite{BCY16,CDLY16,YBCT16,BCCNOY17}, the compression exponent $\gamma$ [see Fig.~\ref{PTSs}A] is here not fixed but treated as an additional fit parameter. Its value ranges roughly from $2$ to $5$ from high to low temperatures. 
Another definition of the PTS length, $\xi^{\mathrm{th}}_{\mathrm{PTS}}$, is given by the relation $Q^{\mathrm{fit}}_{\rm PTS}(\xi^{\mathrm{th}}_{\mathrm{PTS}};T)-Q_{\rm PTS}^{\rm bulk}\equiv e^{-1}$.
A third estimate comes from the peak location of the PTS susceptibility~\cite{BCY16} [see Fig.~\ref{PTSs}B],
\begin{equation}\label{PTSsus}
\chi_{\mathrm{PTS}}(R;T)=[\langle \qc^2\rangle-\langle \qc\rangle^2]|_{T, R}\, .
\end{equation}
Specifically the peak location,  $\xi^{\mathrm{peak}}_{\mathrm{PTS}}$, is estimated through polynomial extrapolation of five maximal values. All three estimates qualitatively support the conclusion that the PTS correlation length diverges upon approaching $T=0$ in $d=2$ [see Fig.~\ref{PTSs}C and subsection~\ref{HD} for hard disks].

\subsection{PTS equilibration}
\label{PTSMC}
In order to properly and efficiently sample the cavity configurations, we employ a parallel-tempering scheme~\cite{FS01,FWT02} adapted to the cavity sampling as in Refs.~\cite{BCY16} with varying temperatures and shrinking factors $(T_{a},\lambda_{a})$ for replicas $a=1,\ldots,n$, where $a=1$ corresponds to the original ensemble. Within each replica, for a cavity containing $N_{\rm cav}$ mobile particles, one MC sweep entails $N_{\rm cav}$ MC trial moves consisting of $80\%$
local displacements--with its length uniformly sampled from $l\in[0,0.15]$--
and $20\%$
particles identity swaps. For cavity sizes $R>2.0$,
in order to accelerate runs, swap moves are attempted only for particle pairs with diameter difference $\lambda_{a}|\sigma_i-\sigma_j|<0.20$.
A replica-identity swap is then attempted every $1000$ MC sweeps on average.

As in Ref.~\cite{BCY16}, we impose the linear relation between replica temperatures and shrinking factors as $\frac{T_a-T_1}{T_{\rm dec}-T_1}=\frac{\lambda_a-\lambda_1}{\lambda_{\rm dec}-\lambda_1}$ with $T_{\rm dec}$ and $\lambda_{\rm dec}$ chosen appropriately (see Tables \ref{sample0}-\ref{sample7}).
The chosen shrinking factors, $\{\lambda_a\}_{a\geq2}$, ensure sufficient replica-swap rates. In order to achieve this sampling, replicas are added one by one, with $\lambda_1=1>\lambda_2>\ldots>\lambda_n$, each time targeting a replica-swap acceptance rate of $\sim20\%$~\cite{BCCNOY17}. 
This process is stopped upon reaching $\lambda_n<\lambda_{\rm dec}$. In Tables \ref{sample0}-\ref{sample7}, the average number of replicas used, $n_{\rm ave}=[n]$, is recorded for each given temperature and radius.

The quality of the equilibration within each cavity is assessed from monitoring the convergence of two preparation schemes~\cite{BICtest12,BCY16}: one starting from the original configuration and the other starting from a randomized configuration prepared by running $10^4$
MC sweeps with shrunk and heated cavity particles, with $(\lambda,T)=(0.6,0.5)$.
Convergence is deemed achieved when
\be\label{qon}
\langle \qc^{\rm on}\rangle\equiv\frac{1}{s_{\rm prod}}\sum_{s=s_{\rm eq}+1}^{s_{\rm eq}+s_{\rm prod}}\qc^{\rm on}\le(t_{\rm rec}s\ri)
\ee
obtained through both approaches lie within $\pm0.1$
of each other for each cavity. Here $\qc^{\rm on}(t)$ is the cavity core overlap between the original configuration and the equilibrated configuration after $t$ MC sweeps, recorded each $t_{\rm rec}=10^4$
MC sweeps. The first $s_{\rm eq}$ configurations are discarded, and thermal averages are taken over the following $s_{\rm prod}$ configurations.
With our choice of parallel-tempering parameters (see Tables \ref{sample0}-\ref{sample7}), for all temperatures and radii, at least $96\%$
of all cavities pass the convergence test. Averaging over cavities results in an even closer agreement between the two schemes, i.e., overlap estimates converge to within $\pm0.01$.

In obtaining PTS correlation functions and PTS susceptibility in Eqs.~(\ref{PTScor}) and (\ref{PTSsus}), respectively, we evaluate core cavity overlaps for $s_{\rm prod}$ pairs of configurations obtained through the two different schemes.

\subsection{Glassiness}\label{glassiness}
\begin{figure*}[htbp]
\centerline{
\hspace{-0.065in}\includegraphics[width=0.33\textwidth]{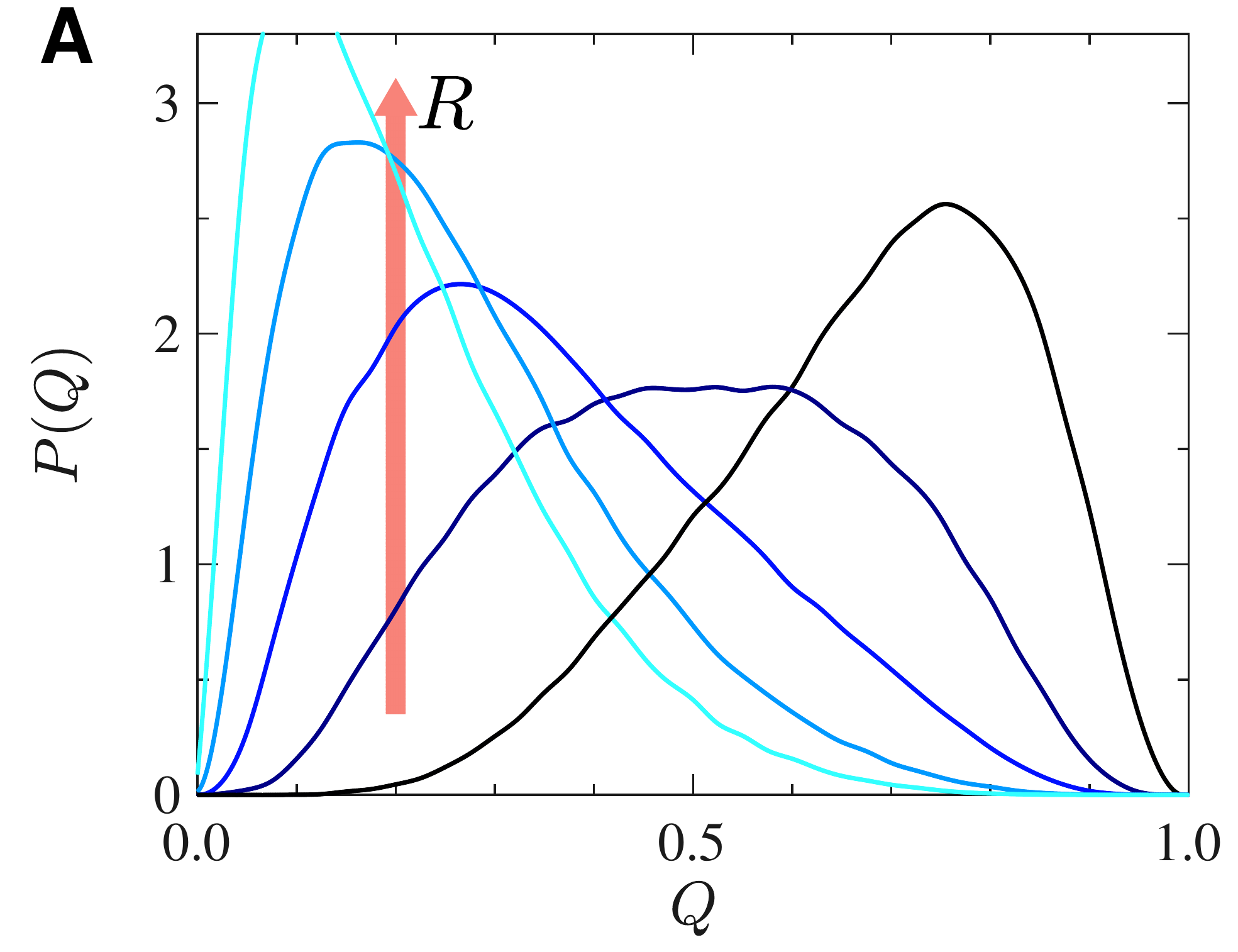}\quad%
\hspace{-0.065in}\includegraphics[width=0.33\textwidth]{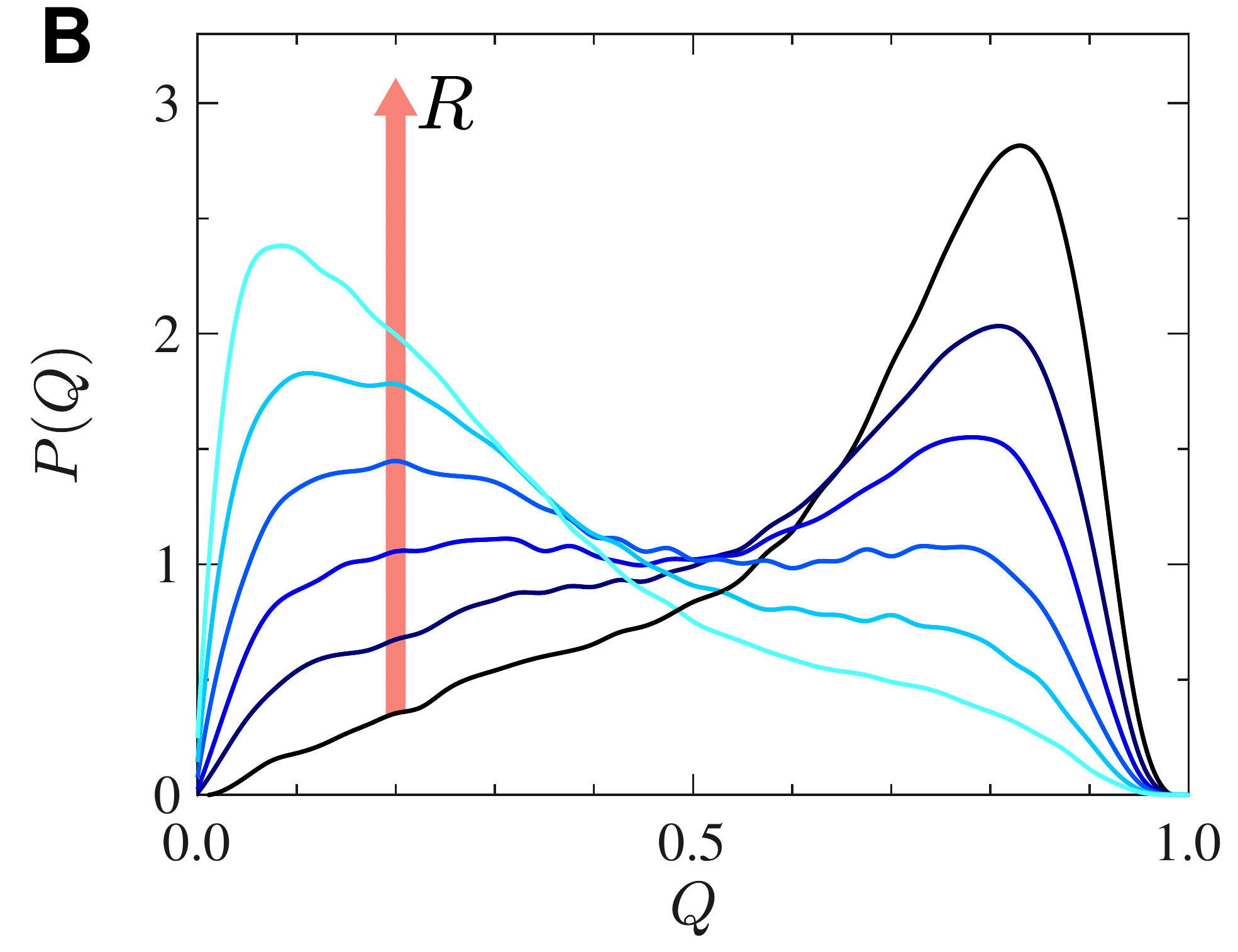}\quad%
\hspace{-0.065in}\includegraphics[width=0.33\textwidth]{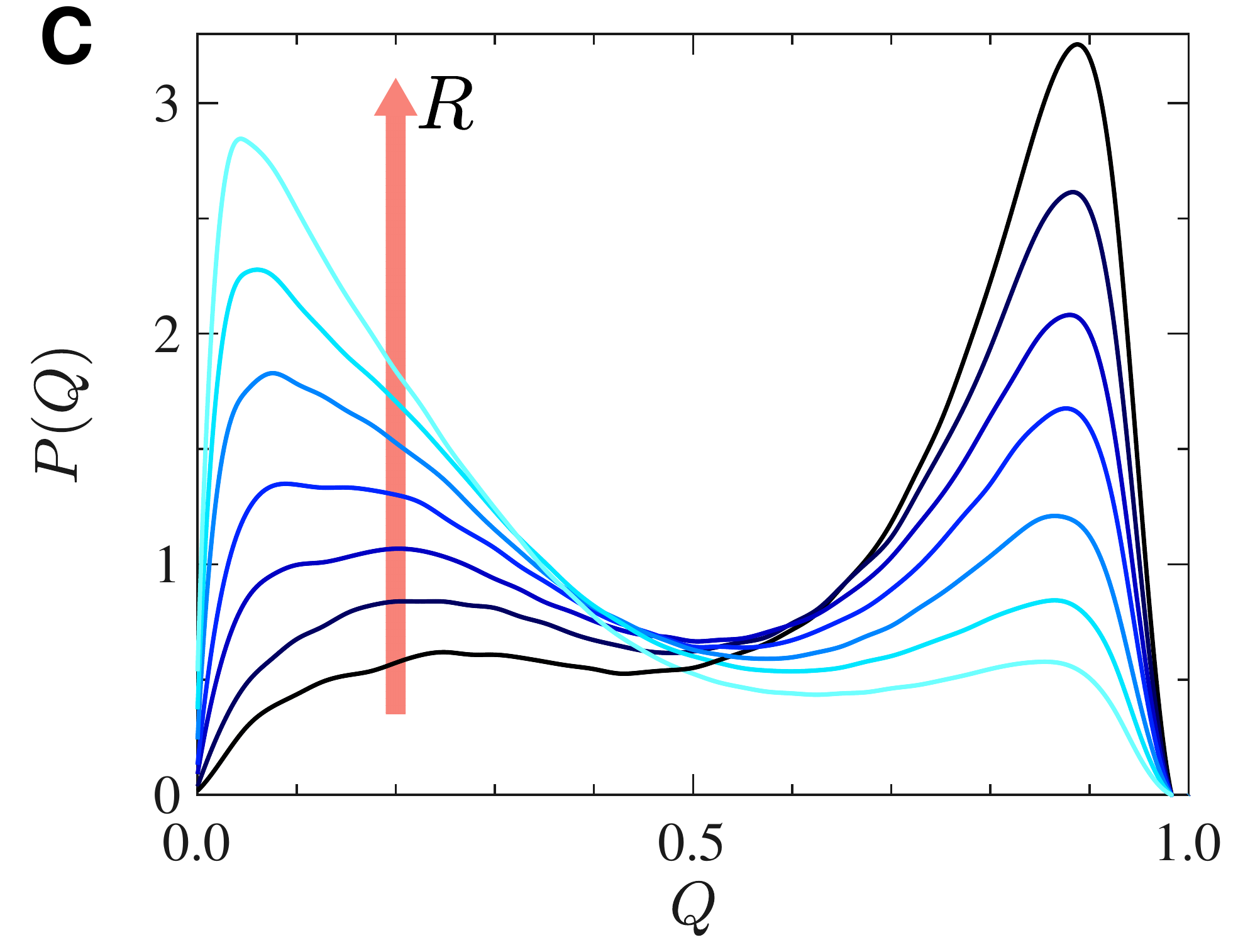}
}
\caption{Disorder-averaged probability distribution function of core overlap $P(\qc)$, at $T =0.400$ for radii $R=1.2,1.5,\ldots,2.4$ (A), $T=0.070$ for $R=2.7,3.0,\ldots,4.2$ (B), and $T=0.035$ for $R=4.8,5.1,\ldots,6.6$ (C). As temperature decreases, the bimodal structure becomes more pronounced.}
\label{bimodal}
\end{figure*}

As detailed in Ref.~\cite{YBCT16}, PTS observables and equilibration diagnose glassiness by accessing information about the underlying free-energy landscape. 
On the static side, the probability distribution function of cavity core overlaps exhibits broad fluctuations at the PTS length scale, with bimodal distribution in the deeply glassy regime (see Fig.~\ref{bimodal}).
This nontrivial signature of confinement in turn leads to a peak in the PTS susceptibility and to a nonconvex dependence of the PTS correlation, as functions of the cavity radius $R$ (see Fig.~\ref{PTSs})~\cite{BCY16}. In nonglassy systems, by contrast, these nontrivial behaviors are absent~\cite{YBCT16}. 

\begin{figure*}
\centering
\includegraphics[width=0.48\textwidth]{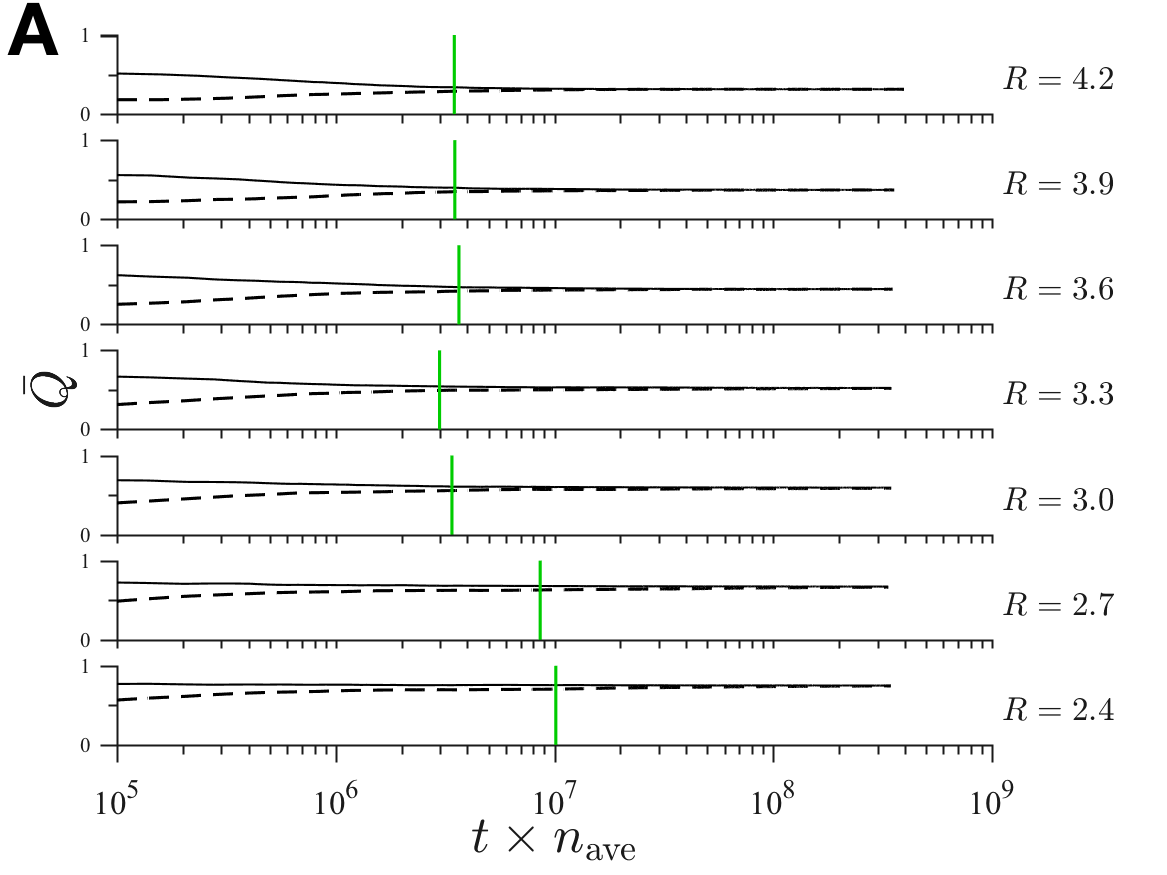}\label{PT}\quad%
\includegraphics[width=0.48\textwidth]{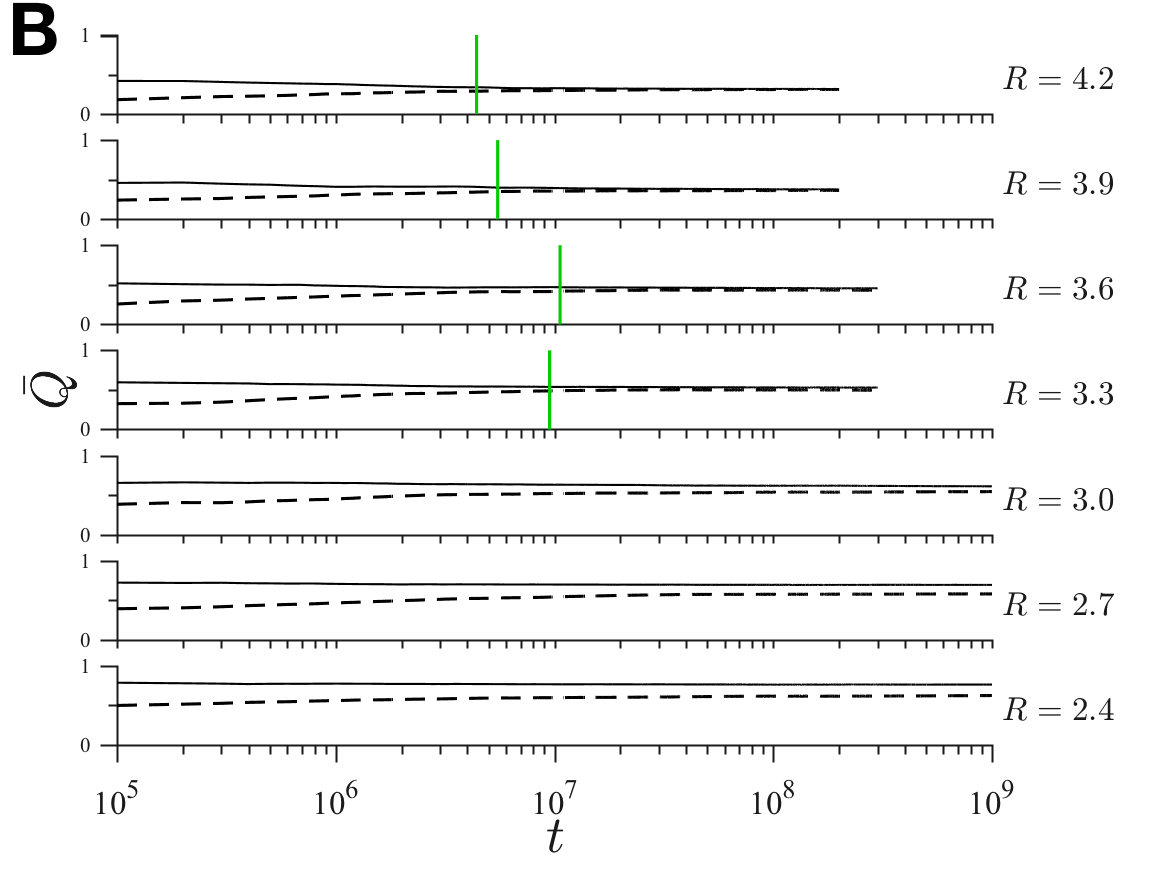}\label{NPT}%
\caption{
Running average of core overlaps~\cite{BCY16}, $\bar{Q}(t)\equiv\frac{1}{\le(t/t_{\rm rec}\ri)}\sum_{s=1}^{\le(t/t_{\rm rec}\ri)}\qc^{\rm on}\le(t_{\rm rec}s\ri)$, after $t$ MC sweeps from both the original (solid lines) and a randomized (dashed lines) configurations at $T=0.070$ for a cavity of radius $R$, averaged over $100$ such cavities. Each green vertical line denotes an estimate of an equilibration times, here defined to be the time beyond which the difference between the running averages $\bar{Q}$ from two schemes converges within $0.05$.
(A) With parallel tempering. In order to compare appropriate computational times, $x$-axis is multiplied by the average number of replicas, $n_{\mathrm{ave}}$ (see Table~\ref{sample4}).
(B) Without parallel tempering. The equilibration time rapidly grows as the cavity radius shrinks, which is interpreted as a finite-size echo of a glass transition~\cite{BCY16}. For instance, for $R\leq3.0$
equilibration is not attained even after $10^9$ MC steps.}
\label{BICs}
\end{figure*}

Proper sampling within cavity confinement grows increasingly challenging as $R$ decreases. Without parallel-tempering, the relaxation time explodes for decreasing $R$ (see Fig.~\ref{BICs}). This dynamical observation also bears out that the slowdown in our polydisperse soft-disk system is triggered by the rugged free-energy landscape characteristic of glassiness.

\subsection{Finite-size effect}\label{finite}
\begin{figure*}
\centering
\hspace{-0.1in}\includegraphics[width=0.48\textwidth]{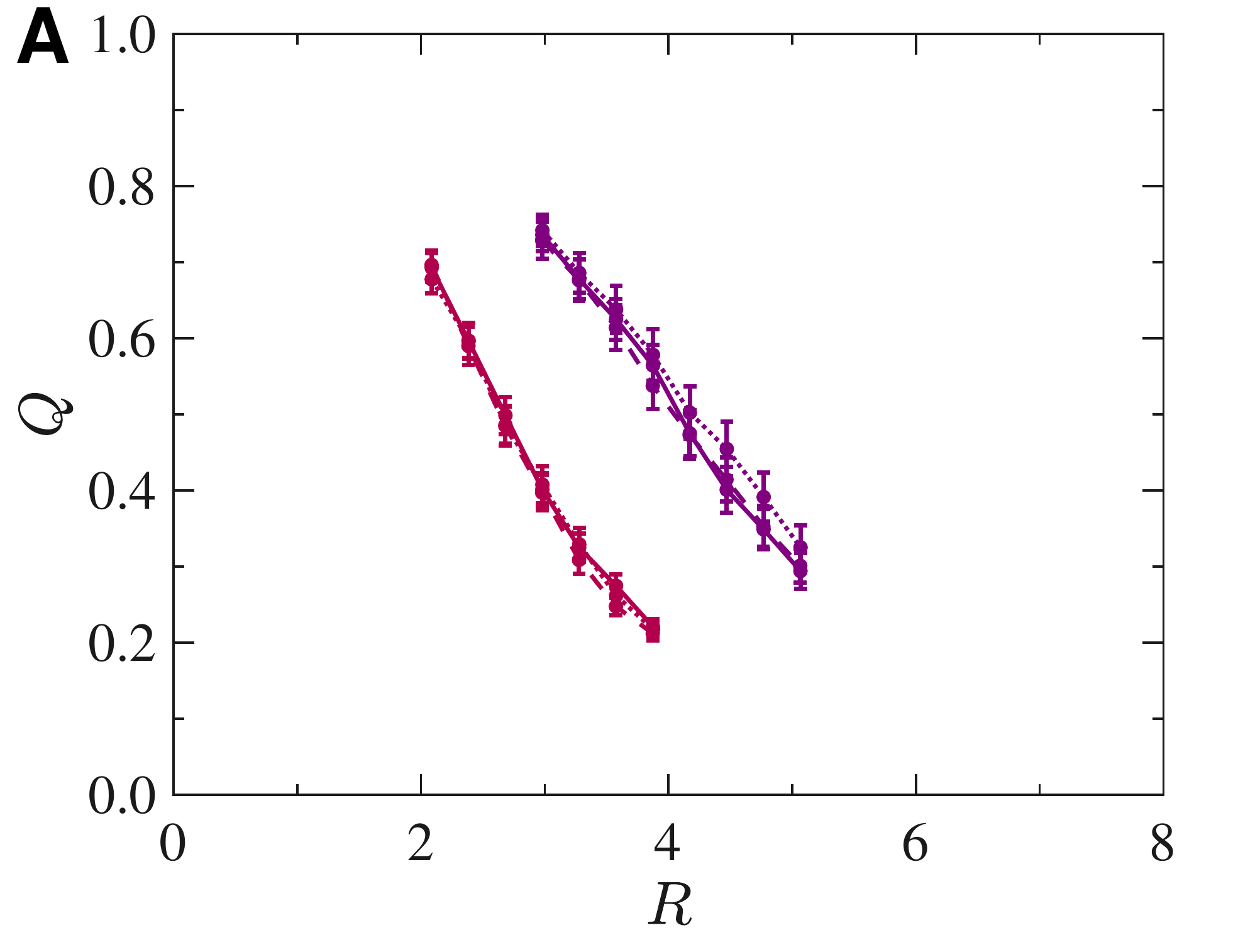}%
\hspace{-0.1in}\includegraphics[width=0.48\textwidth]{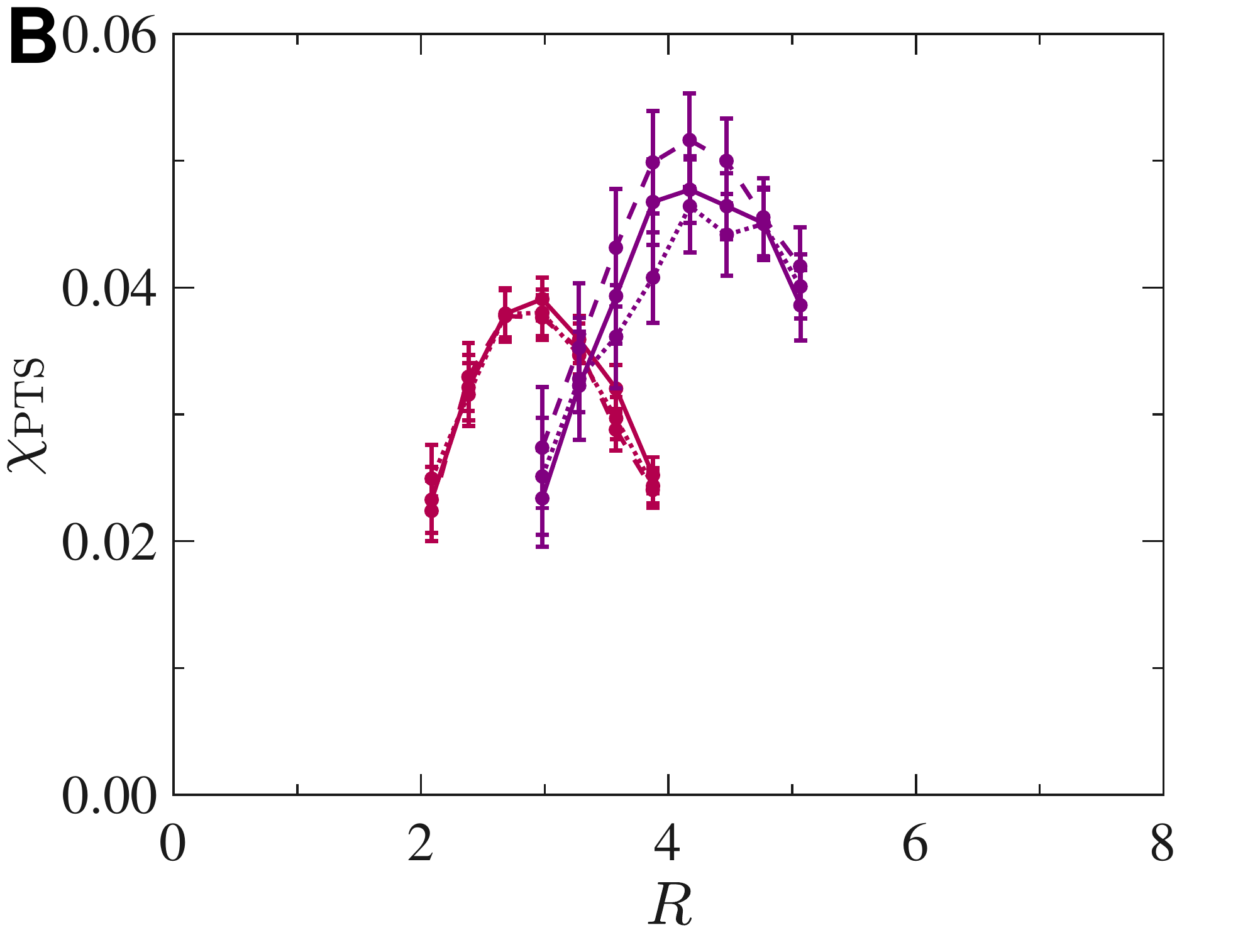}%
\caption{PTS observables at $T = 0.101$ and $0.050$ for soft disks, measured for bulk system sizes $N=300$ (dashed), $1000$ (solid), and $8000$ (dotted). The color scheme is the same as in Fig.~\ref{PTSs}.
(A) Radial decay of the cavity PTS correlation.
(B) PTS susceptibilities with cavity radius $R$.}
\label{FSE}
\end{figure*}

Throughout the paper and this section, we have presented the results for PTS observables with cavities curved out of the bulk systems with $N=1000$ particles. In Fig.~\ref{FSE}, results for configurations with $N=300$ and $N=8000$ are presented for $T=0.101$ and $0.050$. No significant finite-size dependence of the results is observed. 

\begin{table}[htbp]
\begin{tabular}{| l | c | c | c | c | c | c | c | }
\hline
\ \ \ $R$ &&\ $1.2$ &\ $1.5$  &\ $1.8$ &\ $2.1$ &\ $2.4$ &\ $2.7$ \\
\hline
\hline
\ \ \ $n_{\rm ave}$ && 5.16 & 5.67  & 5.72  & 5.18  & 4.44  & 4.02  \\
\hline
\ \ \ $\lambda_{\rm dec}$ && 0.700 & 0.750 & 0.800 & 0.850 & 0.900 & 0.920 \\
\hline
\ \ \ $T_{\rm dec}$ && 0.400 & 0.400  & 0.400  & 0.400  & 0.400  & 0.400 \\
\hline
\ \ \ $s_{\rm eq}$ && 1000 & 1000 & 1000 & 1000 & 1000 & 1000  \\
\hline
\ \ \ $s_{\rm prod}$ && 4000 & 4000 & 4000 & 4000 & 4000 & 4000 \\
\hline
\end{tabular}
\caption{Cavity PTS measurement parameters $T=0.400$, with $100$ cavities}
\label{sample0}
\end{table}
 
\begin{table}[htbp]
\begin{tabular}{| l | c | c | c | c | c | c | c | c | }
\hline
\ \ \ $R$ &&\ $1.5$  &\ $1.8$ &\ $2.1$ &\ $2.4$ &\ $2.7$ &\ $3.0$ &\ $3.3$ \\
\hline
\hline
\ \ \ $n_{\rm ave}$ && 6.28 & 6.35  & 5.96  & 4.98  & 4.86  & 4.95 & 4.39  \\
\hline
\ \ \ $\lambda_{\rm dec}$ && 0.750 & 0.800 & 0.850 & 0.900 & 0.920 & 0.930 & 0.940 \\
\hline
\ \ \ $T_{\rm dec}$ && 0.200 & 0.200  & 0.200  & 0.200  & 0.200  & 0.200 & 0.200 \\
\hline
\ \ \ $s_{\rm eq}$ && 1000 & 1000 & 1000 & 1000 & 1000 & 1000 & 1000  \\
\hline
\ \ \ $s_{\rm prod}$ && 4000 & 4000 & 4000 & 4000 & 4000 & 4000 & 4000 \\
\hline
\end{tabular}
\caption{Cavity PTS measurement parameters $T=0.200$, with $100$ cavities}
\label{sample1}
\end{table}

\begin{table}[htbp]
\begin{tabular}{| l | c | c | c | c | c | c | c | c | }
\hline
\ \ \ $R$ &&\ $1.8$ &\ $2.1$ &\ $2.4$ &\ $2.7$ &\ $3.0$ &\ $3.3$ &\ $3.6$ \\
\hline
\hline
\ \ \ $n_{\rm ave}$ && 6.85 & 6.21  & 5.13  & 5.00  & 5.00  & 4.98 & 4.98  \\
\hline
\ \ \ $\lambda_{\rm dec}$ && 0.800 & 0.850 & 0.900 & 0.920 & 0.930 & 0.940 & 0.945 \\
\hline
\ \ \ $T_{\rm dec}$ && 0.140 & 0.140  & 0.140  & 0.140  & 0.140  & 0.140 & 0.140 \\
\hline
\ \ \ $s_{\rm eq}$ && 1000 & 1000 & 1000 & 1000 & 1000 & 1000 & 1000  \\
\hline
\ \ \ $s_{\rm prod}$ && 4000 & 4000 & 4000 & 4000 & 4000 & 4000 & 4000 \\
\hline
\end{tabular}
\caption{Cavity PTS measurement parameters $T=0.140$, with $100$ cavities}
\label{sample2}
\end{table}

\begin{table}[htbp]
\begin{tabular}{| l | c | c | c | c | c | c | c | c | }
\hline
\ \ \ $R$ &&\ $2.1$ &\ $2.4$ &\ $2.7$ &\ $3.0$ &\ $3.3$ &\ $3.6$ &\ $3.9$ \\
\hline
\hline
\ \ \ $n_{\rm ave}$ && 6.94 & 5.93  & 5.81  & 5.83  & 5.68  & 5.93 & 5.96  \\
\hline
\ \ \ $\lambda_{\rm dec}$ && 0.850 & 0.900 & 0.920 & 0.930 & 0.940 & 0.945 & 0.950 \\
\hline
\ \ \ $T_{\rm dec}$ && 0.125 & 0.125  & 0.125  & 0.125  & 0.125  & 0.125 & 0.125 \\
\hline
\ \ \ $s_{\rm eq}$ && 1000 & 1000 & 1000 & 1000 & 1000 & 1000 & 1000  \\
\hline
\ \ \ $s_{\rm prod}$ && 4000 & 4000 & 4000 & 4000 & 4000 & 4000 & 4000 \\
\hline
\end{tabular}
\caption{Cavity PTS measurement parameters $T=0.101$, with $100$ cavities}
\label{sample3}
\end{table}

\begin{table}[htbp]
\begin{tabular}{| l | c | c | c | c | c | c | c | c | }
\hline
\ \ \ $R$ &&\ $2.4$ &\ $2.7$ &\ $3.0$ &\ $3.3$ &\ $3.6$ &\ $3.9$ &\ $4.2$ \\
\hline
\hline
\ \ \ $n_{\rm ave}$ && 6.90 & 6.72 & 6.93  & 6.94  & 7.03  & 7.15 & 7.92  \\
\hline
\ \ \ $\lambda_{\rm dec}$ && 0.900 & 0.920 & 0.930 & 0.940 & 0.945 & 0.950 & 0.950 \\
\hline
\ \ \ $T_{\rm dec}$ && 0.125 & 0.125  & 0.125  & 0.125  & 0.125  & 0.125 & 0.125 \\
\hline
\ \ \ $s_{\rm eq}$ && 1000 & 1000 & 1000 & 1000 & 1000 & 1000 & 1000  \\
\hline
\ \ \ $s_{\rm prod}$ && 4000 & 4000 & 4000 & 4000 & 4000 & 4000 & 4000 \\
\hline
\end{tabular}
\caption{Cavity PTS measurement parameters $T=0.070$, with $100$ cavities}
\label{sample4}
\end{table}

\begin{table}[htbp]
\begin{tabular}{| l | c | c | c | c | c | c | c | c | c | }
\hline
\ \ \ $R$ &&\ $3.0$ &\ $3.3$ &\ $3.6$ &\ $3.9$ &\ $4.2$ &\ $4.5$ &\ $4.8$ &\ $5.1$\\
\hline
\hline
\ \ \ $n_{\rm ave}$ && 7.91 & 8.01  & 8.41  & 8.97  & 9.43  & 10.04 & 10.79 & 11.27 \\
\hline
\ \ \ $\lambda_{\rm dec}$ && 0.930 & 0.940 & 0.945 & 0.950 & 0.950 & 0.950 & 0.950 & 0.950  \\
\hline
\ \ \ $T_{\rm dec}$ && 0.125 & 0.125  & 0.125  & 0.125  & 0.125  & 0.125 & 0.125 & 0.125 \\
\hline
\ \ \ $s_{\rm eq}$ && 1000 & 1000 & 1000 & 1000 & 1000 & 1000 & 1000 & 1000  \\
\hline
\ \ \ $s_{\rm prod}$ && 4000 & 4000 & 4000 & 4000 & 4000 & 4000 & 4000 & 4000 \\
\hline
\end{tabular}
\caption{Cavity PTS measurement parameters $T=0.050$, with $100$ cavities}
\label{sample5}
\end{table}

\begin{table}[htbp]
\begin{tabular}{| l | c | c | c | c | c | c | c | c | c | c | c | }
\hline
\ \ \ $R$ &&\ $3.3$ &\ $3.6$ &\ $3.9$ &\ $4.2$ &\ $4.5$ &\ $4.8$ &\ $5.1$ &\ $5.4$ &\ $5.7$ &\ $6.0$\\
\hline
\hline
\ \ \ $n_{\rm ave}$ && 8.955 & 9.49  & 10.025  & 10.71  & 11.41 & 12.15 & 12.895 & 13.58 & 14.27 & 15.025 \\
\hline
\ \ \ $\lambda_{\rm dec}$ && 0.940 & 0.945 & 0.950 & 0.950 & 0.950 & 0.950 & 0.950 & 0.950 & 0.950 & 0.950  \\
\hline
\ \ \ $T_{\rm dec}$ && 0.125 & 0.125  & 0.125  & 0.125  & 0.125  & 0.125 & 0.125 & 0.125 & 0.125 & 0.125 \\
\hline
\ \ \ $s_{\rm eq}$ && 1000 & 1000 & 1000 & 1000 & 1000 & 1000 & 1000 & 1000 & 1000 & 1000  \\
\hline
\ \ \ $s_{\rm prod}$ && 4000 & 4000 & 4000 & 4000 & 4000 & 4000 & 4000 & 4000 & 4000 & 4000 \\
\hline
\end{tabular}
\caption{Cavity PTS measurement parameters $T=0.039$, with $200$ cavities}
\label{sample6}
\end{table}

\begin{table}[htbp]
\begin{tabular}{| l | c | c | c | c | c | c | c | c | c | c | c |}
\hline
\ \ \ $R$ &&\ $3.9$ &\ $4.2$ &\ $4.5$ &\ $4.8$ &\ $5.1$ &\ $5.4$ &\ $5.7$ &\ $6.0$ &\ $6.3$ &\ $6.6$\\
\hline
\hline
\ \ \ $n_{\rm ave}$ && 10.38 & 11.13  & 11.975  & 12.815  & 13.51  & 14.30 & 15.02 & 15.80 & 6.61  & 6.915 \\
\hline
\ \ \ $\lambda_{\rm dec}$ && 0.950 & 0.950 & 0.950 & 0.950 & 0.950 & 0.950 & 0.950 & 0.950 & 0.980 & 0.980  \\
\hline
\ \ \ $T_{\rm dec}$ && 0.125  & 0.125  & 0.125  & 0.125 & 0.125 & 0.125 & 0.125 & 0.125 & 0.050 & 0.050 \\
\hline
\ \ \ $s_{\rm eq}$ && 1000 & 1000 & 1000 & 1000 & 1000 & 1000 & 1000 & 1000 & 3000 & 3000   \\
\hline
\ \ \ $s_{\rm prod}$ && 4000 & 4000 & 4000 & 4000 & 4000 & 9000 & 9000 & 9000 & 27000 & 27000 \\
\hline
\end{tabular}
\caption{Cavity PTS measurement parameters $T=0.035$, with $200$ cavities}
\label{sample62}
\end{table}

\begin{table}[htbp]
\begin{tabular}{| l | c | c | c | c | c | c | c | c | c | c | c |}
\hline
\ \ \ $R$ &&\ $3.9$ &\ $4.2$ &\ $4.5$ &\ $4.8$ &\ $5.1$ &\ $5.4$ &\ $5.7$ &\ $6.0$ &\ $6.3$ &\ $6.6$\\
\hline
\hline
\ \ \ $n_{\rm ave}$ && 10.78 & 11.495  & 12.305  & 13.085  & 13.96 & 14.74 & 15.48 & 16.25 & 7.03 & 7.305 \\
\hline
\ \ \ $\lambda_{\rm dec}$ && 0.950 & 0.950 & 0.950 & 0.950 & 0.950 & 0.950 & 0.950 & 0.950 & 0.980 & 0.980  \\
\hline
\ \ \ $T_{\rm dec}$ && 0.125  & 0.125  & 0.125  & 0.125 & 0.125 & 0.125 & 0.125 & 0.125 & 0.050 & 0.050 \\
\hline
\ \ \ $s_{\rm eq}$ && 1000 & 1000 & 1000 & 1000 & 1000 & 1000 & 1000 & 1000 & 3000 & 3000   \\
\hline
\ \ \ $s_{\rm prod}$ && 4000 & 4000 & 4000 & 4000 & 4000 & 9000 & 9000 & 9000 & 27000 & 27000 \\
\hline
\end{tabular}
\caption{Cavity PTS measurement parameters $T=0.033$, with $200$ cavities}
\label{sample64}
\end{table}

\begin{table}[htbp]
\begin{tabular}{| l | c | c | c | c | c | c | c | c | c | c | c |}
\hline
\ \ \ $R$ &&\ $3.9$ &\ $4.2$ &\ $4.5$ &\ $4.8$ &\ $5.1$ &\ $5.4$ &\ $5.7$ &\ $6.0$ &\ $6.3$ &\ $6.6$\\
\hline
\hline
\ \ \ $n_{\rm ave}$ && 10.935 & 11.75 & 12.54  & 13.415 & 14.15 & 14.955 & 15.79 & 16.575 & 7.355 & 7.77  \\
\hline
\ \ \ $\lambda_{\rm dec}$ && 0.950 & 0.950 & 0.950 & 0.950 & 0.950 & 0.950 & 0.950 & 0.950 & 0.980 & 0.980  \\
\hline
\ \ \ $T_{\rm dec}$ && 0.125  & 0.125  & 0.125  & 0.125 & 0.125 & 0.125 & 0.125 & 0.125 & 0.050 & 0.050 \\
\hline
\ \ \ $s_{\rm eq}$ && 1000 & 1000 & 1000 & 1000 & 1000 & 1000 & 1000 & 2000 & 4000 & 4000   \\
\hline
\ \ \ $s_{\rm prod}$ && 4000 & 4000 & 4000 & 4000 & 4000 & 4000 & 9000 & 18000 & 26000 & 26000 \\
\hline
\end{tabular}
\caption{Cavity PTS measurement parameters $T=0.0315$, with $200$ cavities}
\label{sample65}
\end{table}

\begin{table}[htbp]
\begin{tabular}{| l | c | c | c | c | c | c | c | c | c | c | }
\hline
\ \ \ $R$ &&\ $4.2$ &\ $4.5$ &\ $4.8$ &\ $5.1$ &\ $5.4$ &\ $5.7$ &\ $6.0$ &\ $6.3$ &\ $6.6$\\
\hline
\hline
\ \ \ $n_{\rm ave}$ && 12.243 & 13.147  & 14.027  & 14.893  & 15.717  & 16.643 & 17.41 & 8.2767 & 8.6867 \\
\hline
\ \ \ $\lambda_{\rm dec}$ && 0.950 & 0.950 & 0.950 & 0.950 & 0.950 & 0.950 & 0.950 & 0.980 & 0.980  \\
\hline
\ \ \ $T_{\rm dec}$ && 0.125  & 0.125  & 0.125  & 0.125 & 0.125 & 0.125 & 0.125 & 0.050 & 0.050 \\
\hline
\ \ \ $s_{\rm eq}$ && 1000 & 1000 & 1000 & 1000 & 1000 & 1000 & 2000 & 5000  & 8000  \\
\hline
\ \ \ $s_{\rm prod}$ && 4000 & 4000 & 4000 & 4000 & 9000 & 9000 & 18000 & 25000  & 32000\\
\hline
\end{tabular}
\caption{Cavity PTS measurement parameters $T=0.028$, with $300$ cavities}
\label{sample7}
\end{table}

\subsection{Results for hard disks}\label{HD}
\begin{figure*}
\centering
\hspace{-0.1in}\includegraphics[width=0.33\textwidth]{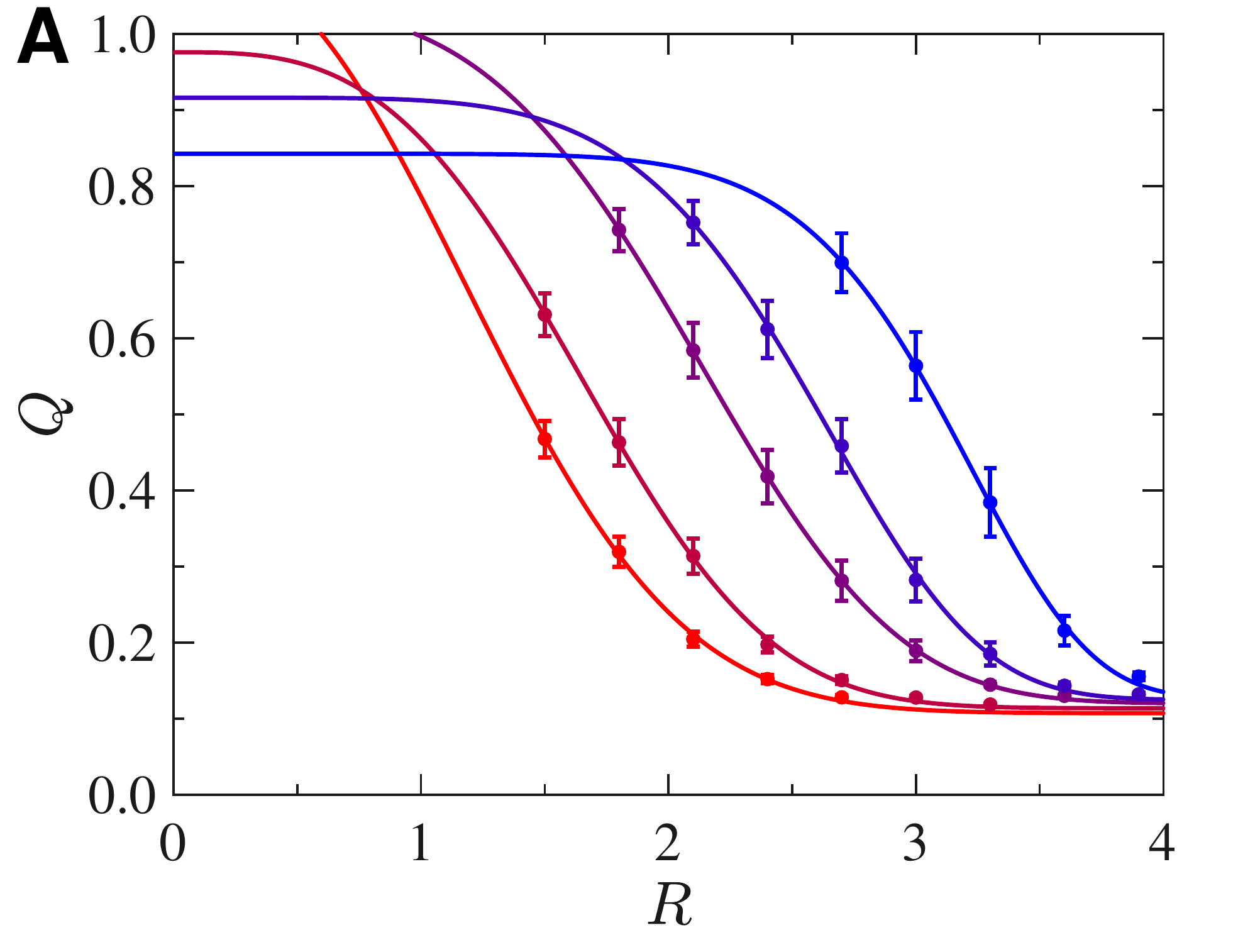}%
\hspace{-0.1in}\includegraphics[width=0.33\textwidth]{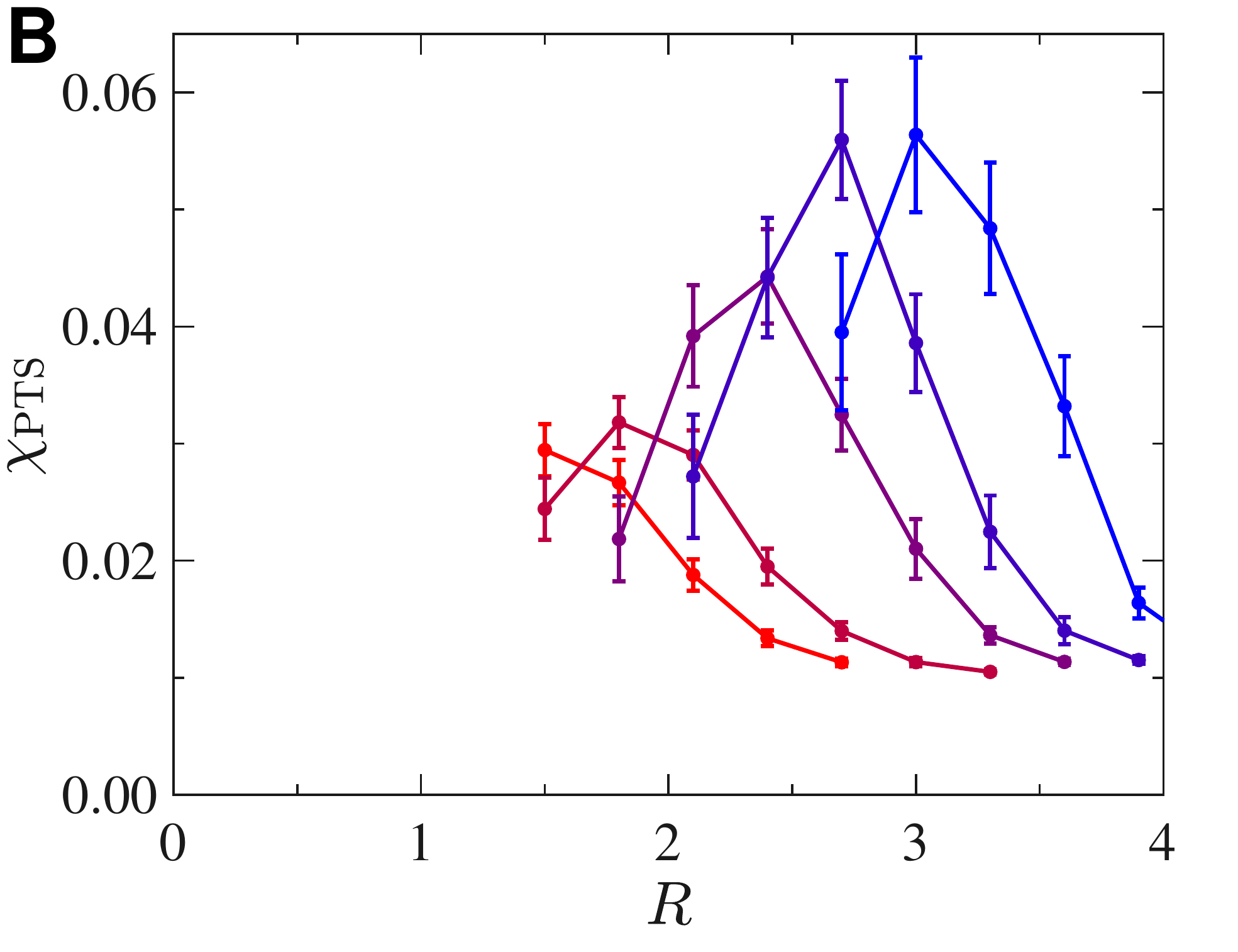}%
\hspace{-0.1in}\includegraphics[width=0.33\textwidth]{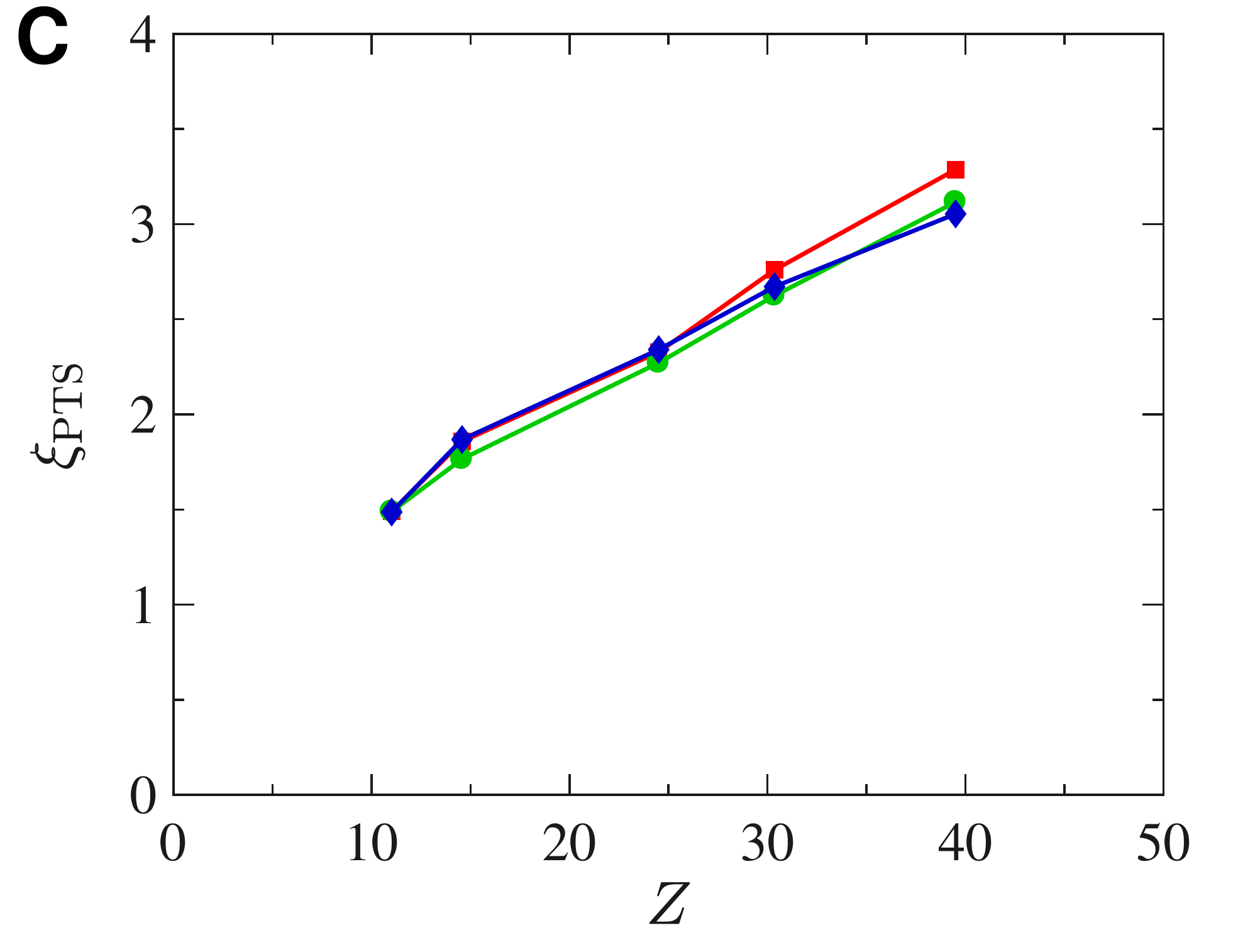}%
\caption{
A Radial decay of the cavity PTS correlation at $\phi = 0.700$, $0.740$, $0.800$, $0.820$, $0.840$ (from red to blue) for hard disks.
Solid lines are fits to a compressed exponential.
B PTS susceptibilities with cavity radius $R$. Solid lines are guides for the eyes.
C PTS lengths $\xi^{\mathrm{fit}}_{\mathrm{PTS}}$ (red-square), $\xi^{\mathrm{th}}_{\mathrm{PTS}}$ (green-circle), and $\xi^{\mathrm{peak}}_{\mathrm{PTS}}$ (blue-diamond) as a function of the reduced pressure $Z$ for hard disks. The clear linear growth of $\xi_{\mathrm{PTS}}$ with $Z$ suggests that $T_{\mathrm{K}}=0$ with RFOT exponent $\theta=1=\frac{d}{2}=d-1$ in $d=2$ spatial dimension.
}
\label{PTSs_HD}
\end{figure*}

Results for hard disks are presented in Fig.~\ref{PTSs_HD}. Most technical details are the same as for the soft-disk case.
The most notable difference concerns the parallel-tempering algorithm, which is adapted from that for $d=3$ hard spheres~\cite{BCCNOY17}, treating the two replicas $a=1$ and $2$ differently from the rest.
Randomized configurations are here prepared by $10^6$
MC sweeps with shrunk particles at $\lambda=0.5$,
and
$\lambda_{\rm dec}$s are chosen appropriately (see Tables \ref{sampleH0}-\ref{sampleH4}) so that at least $96\%$ of all the cavities pass the convergence test
for all packing fractions and radii, with disorder-averaged values converging within $\pm0.01$.
The peak location,  $\xi^{\mathrm{peak}}_{\mathrm{PTS}}$, is here estimated through polynomial extrapolation of three maximal values.

\begin{table}[htbp]
\begin{tabular}{| l | c | c | c | c | c | c | c | }
\hline
\ \ \ $R$ &&\ $1.5$ &\ $1.8$ &\ $2.1$ &\ $2.4$ &\ $2.7$\\
\hline
\hline
\ \ \ $n_{\rm ave}$ && 9.14  & 8.88  & 8.08  & 6.57  & 6.00 \\
\hline
\ \ \ $\lambda_{\rm dec}$ && 0.750 & 0.800 & 0.850 & 0.900 & 0.920  \\
\hline
\ \ \ $t_{\rm rec}$ && 200 & 200 & 200 & 200 & 200  \\
\hline
\ \ \ $s_{\rm eq}$ && 1000 & 1000 & 1000 & 1000 & 1000  \\
\hline
\ \ \ $s_{\rm prod}$ && 4000 & 4000 & 4000 & 4000 & 4000\\
\hline
\end{tabular}
\caption{Cavity PTS measurement parameters $\phi=0.700$, with $100$ cavities
}
\label{sampleH0}
\end{table}

\begin{table}[htbp]
\begin{tabular}{| l | c | c | c | c | c | c | c | c | }
\hline
\ \ \ $R$ &&\  $1.5$ &\ $1.8$ &\ $2.1$ &\ $2.4$ &\ $2.7$ &\ $3.0$ &\ $3.3$\\
\hline
\hline
\ \ \ $n_{\rm ave}$ && 9.45 & 9.44  & 8.44  & 6.94  & 6.49  & 6.32  & 6.02 \\
\hline
\ \ \ $\lambda_{\rm dec}$ && 0.750 & 0.800 & 0.850 & 0.900 & 0.920 & 0.930 & 0.940  \\
\hline
\ \ \ $t_{\rm rec}$ && $10^4$ & $10^4$ & $10^4$ & $10^4$ & $10^4$ & $10^4$ & $10^4$  \\
\hline
\ \ \ $s_{\rm eq}$ && 1000 & 1000 & 1000 & 1000 & 1000 & 1000  & 1000 \\
\hline
\ \ \ $s_{\rm prod}$ && 4000 & 4000 & 4000 & 4000 & 4000 & 4000 & 4000\\
\hline
\end{tabular}
\caption{Cavity PTS measurement parameters $\phi=0.740$, with $100$ cavities}
\label{sampleH1}
\end{table}

\begin{table}[htbp]
\begin{tabular}{| l | c | c | c | c | c | c | c | c | }
\hline
\ \ \ $R$ &&\  $1.8$ &\ $2.1$ &\ $2.4$ &\ $2.7$ &\ $3.0$ &\ $3.3$ &\ $3.6$\\
\hline
\hline
\ \ \ $n_{\rm ave}$ && 10.56 & 9.69  & 8.08  & 8.96  & 7.51  & 7.25  & 7.30 \\
\hline
\ \ \ $\lambda_{\rm dec}$  && 0.800 & 0.850 & 0.900 & 0.900 & 0.930 & 0.940 & 0.945  \\
\hline
\ \ \ $t_{\rm rec}$ && $10^4$ & $10^4$ & $2\cdot10^4$ & $2\cdot10^4$ & $2\cdot 10^4$ & $10^4$ & $10^4$  \\
\hline
\ \ \ $s_{\rm eq}$ && 1000 & 1000 & 1000 & 1000 & 1000 & 1000  & 1000 \\
\hline
\ \ \ $s_{\rm prod}$ && 4000 & 4000 & 4000 & 4000 & 4000 & 4000 & 4000\\
\hline
\end{tabular}
\caption{Cavity PTS measurement parameters $\phi=0.800$, with $100$ cavities}
\label{sampleH2}
\end{table}

\begin{table}[htbp]
\begin{tabular}{| l | c | c | c | c | c | c | c | c | }
\hline
\ \ \ $R$ &&\  $2.1$ &\ $2.4$ &\ $2.7$ &\ $3.0$ &\ $3.3$ &\ $3.6$ &\ $3.9$\\
\hline
\hline
\ \ \ $n_{\rm ave}$ && 10.29 & 8.65  & 8.27  & 8.45  & 8.11  & 8.17  & 8.22 \\
\hline
\ \ \ $\lambda_{\rm dec}$  && 0.850 & 0.900 & 0.920 & 0.930 & 0.940 & 0.945 & 0.950  \\
\hline
\ \ \ $t_{\rm rec}$ && $10^4$ & $10^4$ & $3\cdot10^4$ & $3\cdot10^4$ & $3\cdot 10^4$ & $10^4$ & $10^4$  \\
\hline
\ \ \ $s_{\rm eq}$ && 1000 & 1000 & 1000 & 1000 & 1000 & 1000  & 1000 \\
\hline
\ \ \ $s_{\rm prod}$ && 4000 & 4000 & 4000 & 4000 & 4000 & 4000 & 4000\\
\hline
\end{tabular}
\caption{Cavity PTS measurement parameters $\phi=0.820$, with $100$ cavities}
\label{sampleH3}
\end{table}

\begin{table}[htbp]
\begin{tabular}{| l | c | c | c | c | c | c | c | }
\hline
\ \ \ $R$ &&\  $2.7$ &\ $3.0$ &\ $3.3$ &\ $3.6$ &\ $3.9$ &\ $4.2$\\
\hline
\hline
\ \ \ $n_{\rm ave}$ &&  17.31  & 15.10  & 15.57  & 16.83  & 18.09 & 16.15 \\
\hline
\ \ \ $\lambda_{\rm dec}$  && 0.800 & 0.860 & 0.870 & 0.870 & 0.870 & 0.900  \\
\hline
\ \ \ $t_{\rm rec}$ && $5\cdot10^4$ & $5\cdot10^4$ & $5\cdot 10^4$ & $5\cdot 10^4$ & $3\cdot 10^4$ & $2\cdot 10^4$  \\
\hline
\ \ \ $s_{\rm eq}$ && 1000 & 1000 & 2000 & 2000  & 1000  & 1000 \\
\hline
\ \ \ $s_{\rm prod}$ && 4000 & 4000 & 8000 & 8000 & 4000 & 4000\\
\hline
\end{tabular}
\caption{Cavity PTS measurement parameters $\phi=0.840$, with $100$ cavities}
\label{sampleH4}
\end{table}

\section{Scaling}

\begin{figure}[htbp]
\includegraphics[width=0.5\columnwidth]{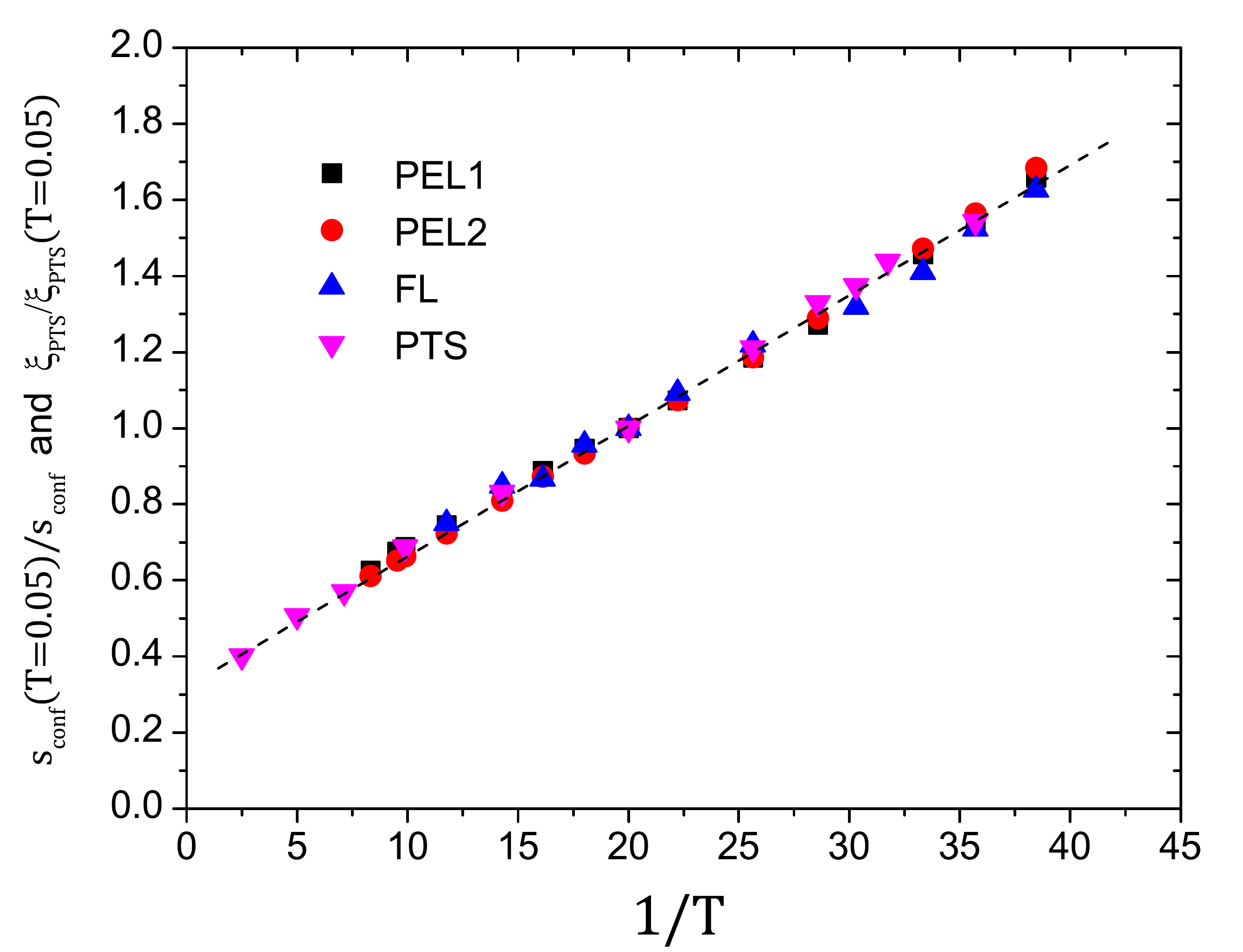}
\caption{
$1/s_{\rm conf}$ and $\xi_{\rm PTS}$ normalized at $T=0.05$.
The dashed line is guide for the eyes.
}
\label{fig:scaling}
\end{figure}  

Here we discuss scaling behaviors of $s_{\rm conf}$ and $\xi_{\rm PTS}$.
Figure~\ref{fig:scaling} shows $1/s_{\rm conf}$ and $\xi_{\rm PTS}$ as a function of the inverse of the temperature, normalized at $T=0.05$.
 We empirically find $\xi_{\rm PTS} \propto 1/s_{\rm conf}=A/T + B$, where $A$ and $B$ are constants, justifying $\theta=1$ chosen in the main text.
This relation also means that $s_{\rm conf}=T/(A+BT)=A^{-1}T-A^{-2}BT^2 + \mathcal{O}(T^3)$, which is consistent with the result of a quadratic extrapolation shown in the main text.
Furthermore, the scaling behavior of $\xi_{\rm PTS}$ can be understood as $\xi_{\rm PTS} \sim |T-T_{\rm K}|^{-\nu}$ with $T_{\rm K}=0$ and $\nu=1$ in $d=2$.

\section{Relationship with recent $d=2$ dynamical studies}


Two-dimensional systems are special in condensed matter physics.
Long-wavelength, Mermin-Wagner density fluctuations then destabilize long-range positional order, and thus finite-temperature crystalline solids cannot exist.
While it has nonetheless long been believed that glassiness in $d=2$ and $d=3$ are essentially the same~\cite{harrowell2006nonlinear}, the putative role of Mermin-Wagner fluctuations was long neglected.
Recent experimental and computational studies of glass-forming liquids have carefully considered the situation~\cite{FS15,shiba2012relationship,shiba2016unveiling,VKCW17,IFKKMK17,tarjus2017glass}.
It is now clear that, in contrast to $d=3$, dynamics in $d=2$ is indeed influenced by the presence of the long-wavelength density fluctuations that enhance the mean-squared displacement of particles and thus seemingly breaks the standard cage picture of glassiness.
It has further been established, however, that such dynamical differences can be eliminated by studying bond-orientational relaxation or by introducing the cage-relative mean-squared displacement, which disentangles the Mermin-Wagner fluctuations from the underlying development of glassiness.
Upon such disentanglement, the cage picture can be recovered in $d=2$ as well.

Here,
we disentangle Mermin-Wagner fluctuations from the measurements of the configurational entropy using approaches in the same spirit as those used in previous dynamical studies.
Even after disentangling effects of these fluctuations, our determination of the configurational entropy and its comparison with $d=3$ results suggest that the glass transition in $d=2$ and $3$ are fundamentally different. In particular, the latter occurs at a finite temperature, whereas the former occurs at zero temperature. Our study thus identifies the lower critical dimension $d_\mathrm{L}=2$ for the long-range amorphous order.

\bibliography{2Dref.bib,SS.bib}

\begin{thebibliography}{59}%
\makeatletter
\providecommand \@ifxundefined [1]{%
 \@ifx{#1\undefined}
}%
\providecommand \@ifnum [1]{%
 \ifnum #1\expandafter \@firstoftwo
 \else \expandafter \@secondoftwo
 \fi
}%
\providecommand \@ifx [1]{%
 \ifx #1\expandafter \@firstoftwo
 \else \expandafter \@secondoftwo
 \fi
}%
\providecommand \natexlab [1]{#1}%
\providecommand \enquote  [1]{``#1''}%
\providecommand \bibnamefont  [1]{#1}%
\providecommand \bibfnamefont [1]{#1}%
\providecommand \citenamefont [1]{#1}%
\providecommand \href@noop [0]{\@secondoftwo}%
\providecommand \href [0]{\begingroup \@sanitize@url \@href}%
\providecommand \@href[1]{\@@startlink{#1}\@@href}%
\providecommand \@@href[1]{\endgroup#1\@@endlink}%
\providecommand \@sanitize@url [0]{\catcode `\\12\catcode `\$12\catcode
  `\&12\catcode `\#12\catcode `\^12\catcode `\_12\catcode `\%12\relax}%
\providecommand \@@startlink[1]{}%
\providecommand \@@endlink[0]{}%
\providecommand \url  [0]{\begingroup\@sanitize@url \@url }%
\providecommand \@url [1]{\endgroup\@href {#1}{\urlprefix }}%
\providecommand \urlprefix  [0]{URL }%
\providecommand \Eprint [0]{\href }%
\providecommand \doibase [0]{http://dx.doi.org/}%
\providecommand \selectlanguage [0]{\@gobble}%
\providecommand \bibinfo  [0]{\@secondoftwo}%
\providecommand \bibfield  [0]{\@secondoftwo}%
\providecommand \translation [1]{[#1]}%
\providecommand \BibitemOpen [0]{}%
\providecommand \bibitemStop [0]{}%
\providecommand \bibitemNoStop [0]{.\EOS\space}%
\providecommand \EOS [0]{\spacefactor3000\relax}%
\providecommand \BibitemShut  [1]{\csname bibitem#1\endcsname}%
\let\auto@bib@innerbib\@empty
\bibitem [{\citenamefont {Chaikin}\ and\ \citenamefont
  {Lubensky}(2000)}]{book1}%
  \BibitemOpen
  \bibfield  {author} {\bibinfo {author} {\bibfnamefont {P.~M.}\ \bibnamefont
  {Chaikin}}\ and\ \bibinfo {author} {\bibfnamefont {T.~C.}\ \bibnamefont
  {Lubensky}},\ }\href@noop {} {\emph {\bibinfo {title} {Principles of
  condensed matter physics}}}\ (\bibinfo  {publisher} {Cambridge University
  Press},\ \bibinfo {year} {2000})\BibitemShut {NoStop}%
\bibitem [{\citenamefont {Privman}(2005)}]{1dbook}%
  \BibitemOpen
  \bibfield  {author} {\bibinfo {author} {\bibfnamefont {V.}~\bibnamefont
  {Privman}},\ }\href@noop {} {\emph {\bibinfo {title} {Nonequilibrium
  statistical mechanics in one dimension}}}\ (\bibinfo  {publisher} {Cambridge
  University Press},\ \bibinfo {year} {2005})\BibitemShut {NoStop}%
\bibitem [{\citenamefont {Goldenfeld}(2018)}]{book2}%
  \BibitemOpen
  \bibfield  {author} {\bibinfo {author} {\bibfnamefont {N.}~\bibnamefont
  {Goldenfeld}},\ }\href@noop {} {\emph {\bibinfo {title} {Lectures on phase
  transitions and the renormalization group}}}\ (\bibinfo  {publisher} {CRC
  Press},\ \bibinfo {year} {2018})\BibitemShut {NoStop}%
\bibitem [{\citenamefont {Berthier}\ and\ \citenamefont
  {Biroli}(2011)}]{rmp11}%
  \BibitemOpen
  \bibfield  {author} {\bibinfo {author} {\bibfnamefont {L.}~\bibnamefont
  {Berthier}}\ and\ \bibinfo {author} {\bibfnamefont {G.}~\bibnamefont
  {Biroli}},\ }\href@noop {} {\bibfield  {journal} {\bibinfo  {journal} {Rev.
  Mod. Phys.}\ }\textbf {\bibinfo {volume} {83}},\ \bibinfo {pages} {587}
  (\bibinfo {year} {2011})}\BibitemShut {NoStop}%
\bibitem [{\citenamefont {Lubchenko}\ and\ \citenamefont
  {Wolynes}(2007)}]{mf1}%
  \BibitemOpen
  \bibfield  {author} {\bibinfo {author} {\bibfnamefont {V.}~\bibnamefont
  {Lubchenko}}\ and\ \bibinfo {author} {\bibfnamefont {P.~G.}\ \bibnamefont
  {Wolynes}},\ }\href@noop {} {\bibfield  {journal} {\bibinfo  {journal} {Annu.
  Rev. Phys. Chem.}\ }\textbf {\bibinfo {volume} {58}},\ \bibinfo {pages} {235}
  (\bibinfo {year} {2007})}\BibitemShut {NoStop}%
\bibitem [{\citenamefont {Charbonneau}\ \emph {et~al.}(2017)\citenamefont
  {Charbonneau}, \citenamefont {Kurchan}, \citenamefont {Parisi}, \citenamefont
  {Urbani},\ and\ \citenamefont {Zamponi}}]{mf2}%
  \BibitemOpen
  \bibfield  {author} {\bibinfo {author} {\bibfnamefont {P.}~\bibnamefont
  {Charbonneau}}, \bibinfo {author} {\bibfnamefont {J.}~\bibnamefont
  {Kurchan}}, \bibinfo {author} {\bibfnamefont {G.}~\bibnamefont {Parisi}},
  \bibinfo {author} {\bibfnamefont {P.}~\bibnamefont {Urbani}}, \ and\ \bibinfo
  {author} {\bibfnamefont {F.}~\bibnamefont {Zamponi}},\ }\href@noop {}
  {\bibfield  {journal} {\bibinfo  {journal} {Annu. Rev. Condens. Matter
  Phys.}\ }\textbf {\bibinfo {volume} {8}},\ \bibinfo {pages} {265} (\bibinfo
  {year} {2017})}\BibitemShut {NoStop}%
\bibitem [{\citenamefont {Kauzmann}(1948)}]{Kauzmann48}%
  \BibitemOpen
  \bibfield  {author} {\bibinfo {author} {\bibfnamefont {W.}~\bibnamefont
  {Kauzmann}},\ }\href@noop {} {\bibfield  {journal} {\bibinfo  {journal}
  {Chem. Rev.}\ }\textbf {\bibinfo {volume} {43}},\ \bibinfo {pages} {219}
  (\bibinfo {year} {1948})}\BibitemShut {NoStop}%
\bibitem [{\citenamefont {Dzero}\ \emph {et~al.}(2005)\citenamefont {Dzero},
  \citenamefont {Schmalian},\ and\ \citenamefont {Wolynes}}]{effort1}%
  \BibitemOpen
  \bibfield  {author} {\bibinfo {author} {\bibfnamefont {M.}~\bibnamefont
  {Dzero}}, \bibinfo {author} {\bibfnamefont {J.}~\bibnamefont {Schmalian}}, \
  and\ \bibinfo {author} {\bibfnamefont {P.~G.}\ \bibnamefont {Wolynes}},\
  }\href@noop {} {\bibfield  {journal} {\bibinfo  {journal} {Phys. Rev. B}\
  }\textbf {\bibinfo {volume} {72}},\ \bibinfo {pages} {100201} (\bibinfo
  {year} {2005})}\BibitemShut {NoStop}%
\bibitem [{\citenamefont {Franz}(2005)}]{effort2}%
  \BibitemOpen
  \bibfield  {author} {\bibinfo {author} {\bibfnamefont {S.}~\bibnamefont
  {Franz}},\ }\href@noop {} {\bibfield  {journal} {\bibinfo  {journal} {Journal
  of Statistical Mechanics: Theory and Experiment}\ }\textbf {\bibinfo {volume}
  {2005}},\ \bibinfo {pages} {P04001} (\bibinfo {year} {2005})}\BibitemShut
  {NoStop}%
\bibitem [{\citenamefont {Angelini}\ and\ \citenamefont
  {Biroli}(2017)}]{effort3}%
  \BibitemOpen
  \bibfield  {author} {\bibinfo {author} {\bibfnamefont {M.~C.}\ \bibnamefont
  {Angelini}}\ and\ \bibinfo {author} {\bibfnamefont {G.}~\bibnamefont
  {Biroli}},\ }\href@noop {} {\bibfield  {journal} {\bibinfo  {journal} {J.
  Stat. Phys.}\ }\textbf {\bibinfo {volume} {167}},\ \bibinfo {pages} {476}
  (\bibinfo {year} {2017})}\BibitemShut {NoStop}%
\bibitem [{\citenamefont {Moore}\ and\ \citenamefont
  {Drossel}(2002)}]{effort5}%
  \BibitemOpen
  \bibfield  {author} {\bibinfo {author} {\bibfnamefont {M.~A.}\ \bibnamefont
  {Moore}}\ and\ \bibinfo {author} {\bibfnamefont {B.}~\bibnamefont
  {Drossel}},\ }\href@noop {} {\bibfield  {journal} {\bibinfo  {journal} {Phys.
  Rev. Lett.}\ }\textbf {\bibinfo {volume} {89}},\ \bibinfo {pages} {217202}
  (\bibinfo {year} {2002})}\BibitemShut {NoStop}%
\bibitem [{\citenamefont {Bouchaud}\ and\ \citenamefont {Biroli}(2004)}]{BB04}%
  \BibitemOpen
  \bibfield  {author} {\bibinfo {author} {\bibfnamefont {J.-P.}\ \bibnamefont
  {Bouchaud}}\ and\ \bibinfo {author} {\bibfnamefont {G.}~\bibnamefont
  {Biroli}},\ }\href@noop {} {\bibfield  {journal} {\bibinfo  {journal} {J.
  Chem. Phys.}\ }\textbf {\bibinfo {volume} {121}},\ \bibinfo {pages} {7347}
  (\bibinfo {year} {2004})}\BibitemShut {NoStop}%
\bibitem [{\citenamefont {Richert}\ and\ \citenamefont
  {Angell}(1998)}]{richert1998dynamics}%
  \BibitemOpen
  \bibfield  {author} {\bibinfo {author} {\bibfnamefont {R.}~\bibnamefont
  {Richert}}\ and\ \bibinfo {author} {\bibfnamefont {C.}~\bibnamefont
  {Angell}},\ }\href@noop {} {\bibfield  {journal} {\bibinfo  {journal} {J.
  Chem. Phys.}\ }\textbf {\bibinfo {volume} {108}},\ \bibinfo {pages} {9016}
  (\bibinfo {year} {1998})}\BibitemShut {NoStop}%
\bibitem [{\citenamefont {Tatsumi}\ \emph {et~al.}(2012)\citenamefont
  {Tatsumi}, \citenamefont {Aso},\ and\ \citenamefont
  {Yamamuro}}]{tatsumi2012thermodynamic}%
  \BibitemOpen
  \bibfield  {author} {\bibinfo {author} {\bibfnamefont {S.}~\bibnamefont
  {Tatsumi}}, \bibinfo {author} {\bibfnamefont {S.}~\bibnamefont {Aso}}, \ and\
  \bibinfo {author} {\bibfnamefont {O.}~\bibnamefont {Yamamuro}},\ }\href@noop
  {} {\bibfield  {journal} {\bibinfo  {journal} {Phys. Rev. Lett.}\ }\textbf
  {\bibinfo {volume} {109}},\ \bibinfo {pages} {045701} (\bibinfo {year}
  {2012})}\BibitemShut {NoStop}%
\bibitem [{\citenamefont {Tarjus}\ \emph {et~al.}(2005)\citenamefont {Tarjus},
  \citenamefont {Kivelson}, \citenamefont {Nussinov},\ and\ \citenamefont
  {Viot}}]{tarjus2005frustration}%
  \BibitemOpen
  \bibfield  {author} {\bibinfo {author} {\bibfnamefont {G.}~\bibnamefont
  {Tarjus}}, \bibinfo {author} {\bibfnamefont {S.~A.}\ \bibnamefont
  {Kivelson}}, \bibinfo {author} {\bibfnamefont {Z.}~\bibnamefont {Nussinov}},
  \ and\ \bibinfo {author} {\bibfnamefont {P.}~\bibnamefont {Viot}},\
  }\href@noop {} {\bibfield  {journal} {\bibinfo  {journal} {J. Phys.: Condens.
  Matter}\ }\textbf {\bibinfo {volume} {17}},\ \bibinfo {pages} {R1143}
  (\bibinfo {year} {2005})}\BibitemShut {NoStop}%
\bibitem [{\citenamefont {Chandler}\ and\ \citenamefont
  {Garrahan}(2010)}]{chandler2010dynamics}%
  \BibitemOpen
  \bibfield  {author} {\bibinfo {author} {\bibfnamefont {D.}~\bibnamefont
  {Chandler}}\ and\ \bibinfo {author} {\bibfnamefont {J.~P.}\ \bibnamefont
  {Garrahan}},\ }\href@noop {} {\bibfield  {journal} {\bibinfo  {journal}
  {Annu. Rev. Phys. Chem.}\ }\textbf {\bibinfo {volume} {61}},\ \bibinfo
  {pages} {191} (\bibinfo {year} {2010})}\BibitemShut {NoStop}%
\bibitem [{\citenamefont {Ninarello}\ \emph {et~al.}(2017)\citenamefont
  {Ninarello}, \citenamefont {Berthier},\ and\ \citenamefont
  {Coslovich}}]{Ninarello2017}%
  \BibitemOpen
  \bibfield  {author} {\bibinfo {author} {\bibfnamefont {A.}~\bibnamefont
  {Ninarello}}, \bibinfo {author} {\bibfnamefont {L.}~\bibnamefont {Berthier}},
  \ and\ \bibinfo {author} {\bibfnamefont {D.}~\bibnamefont {Coslovich}},\
  }\href@noop {} {\bibfield  {journal} {\bibinfo  {journal} {Phys. Rev. X}\
  }\textbf {\bibinfo {volume} {7}},\ \bibinfo {pages} {021039} (\bibinfo {year}
  {2017})}\BibitemShut {NoStop}%
\bibitem [{\citenamefont {Berthier}\ \emph {et~al.}(2017)\citenamefont
  {Berthier}, \citenamefont {Charbonneau}, \citenamefont {Coslovich},
  \citenamefont {Ninarello}, \citenamefont {Ozawa},\ and\ \citenamefont
  {Yaida}}]{BCCNOY17}%
  \BibitemOpen
  \bibfield  {author} {\bibinfo {author} {\bibfnamefont {L.}~\bibnamefont
  {Berthier}}, \bibinfo {author} {\bibfnamefont {P.}~\bibnamefont
  {Charbonneau}}, \bibinfo {author} {\bibfnamefont {D.}~\bibnamefont
  {Coslovich}}, \bibinfo {author} {\bibfnamefont {A.}~\bibnamefont
  {Ninarello}}, \bibinfo {author} {\bibfnamefont {M.}~\bibnamefont {Ozawa}}, \
  and\ \bibinfo {author} {\bibfnamefont {S.}~\bibnamefont {Yaida}},\
  }\href@noop {} {\bibfield  {journal} {\bibinfo  {journal} {Proc. Natl. Acad.
  Sci. U. S. A.}\ }\textbf {\bibinfo {volume} {114}},\ \bibinfo {pages} {11356}
  (\bibinfo {year} {2017})}\BibitemShut {NoStop}%
\bibitem [{\citenamefont {Flenner}\ and\ \citenamefont {Szamel}(2015)}]{FS15}%
  \BibitemOpen
  \bibfield  {author} {\bibinfo {author} {\bibfnamefont {E.}~\bibnamefont
  {Flenner}}\ and\ \bibinfo {author} {\bibfnamefont {G.}~\bibnamefont
  {Szamel}},\ }\href@noop {} {\bibfield  {journal} {\bibinfo  {journal} {Nat.
  Commun.}\ }\textbf {\bibinfo {volume} {6}} (\bibinfo {year}
  {2015})}\BibitemShut {NoStop}%
\bibitem [{\citenamefont {Vivek}\ \emph {et~al.}(2017)\citenamefont {Vivek},
  \citenamefont {Kelleher}, \citenamefont {Chaikin},\ and\ \citenamefont
  {Weeks}}]{VKCW17}%
  \BibitemOpen
  \bibfield  {author} {\bibinfo {author} {\bibfnamefont {S.}~\bibnamefont
  {Vivek}}, \bibinfo {author} {\bibfnamefont {C.~P.}\ \bibnamefont {Kelleher}},
  \bibinfo {author} {\bibfnamefont {P.~M.}\ \bibnamefont {Chaikin}}, \ and\
  \bibinfo {author} {\bibfnamefont {E.~R.}\ \bibnamefont {Weeks}},\ }\href@noop
  {} {\bibfield  {journal} {\bibinfo  {journal} {Proc. Natl. Acad. Sci. U. S.
  A.}\ }\textbf {\bibinfo {volume} {114}},\ \bibinfo {pages} {1850} (\bibinfo
  {year} {2017})}\BibitemShut {NoStop}%
\bibitem [{\citenamefont {Illing}\ \emph {et~al.}(2017)\citenamefont {Illing},
  \citenamefont {Fritschi}, \citenamefont {Kaiser}, \citenamefont {Klix},
  \citenamefont {Maret},\ and\ \citenamefont {Keim}}]{IFKKMK17}%
  \BibitemOpen
  \bibfield  {author} {\bibinfo {author} {\bibfnamefont {B.}~\bibnamefont
  {Illing}}, \bibinfo {author} {\bibfnamefont {S.}~\bibnamefont {Fritschi}},
  \bibinfo {author} {\bibfnamefont {H.}~\bibnamefont {Kaiser}}, \bibinfo
  {author} {\bibfnamefont {C.~L.}\ \bibnamefont {Klix}}, \bibinfo {author}
  {\bibfnamefont {G.}~\bibnamefont {Maret}}, \ and\ \bibinfo {author}
  {\bibfnamefont {P.}~\bibnamefont {Keim}},\ }\href@noop {} {\bibfield
  {journal} {\bibinfo  {journal} {Proc. Natl. Acad. Sci. U. S. A.}\ }\textbf
  {\bibinfo {volume} {114}},\ \bibinfo {pages} {1856} (\bibinfo {year}
  {2017})}\BibitemShut {NoStop}%
\bibitem [{\citenamefont {Sciortino}\ \emph {et~al.}(1999)\citenamefont
  {Sciortino}, \citenamefont {Kob},\ and\ \citenamefont {Tartaglia}}]{Kob}%
  \BibitemOpen
  \bibfield  {author} {\bibinfo {author} {\bibfnamefont {F.}~\bibnamefont
  {Sciortino}}, \bibinfo {author} {\bibfnamefont {W.}~\bibnamefont {Kob}}, \
  and\ \bibinfo {author} {\bibfnamefont {P.}~\bibnamefont {Tartaglia}},\
  }\href@noop {} {\bibfield  {journal} {\bibinfo  {journal} {Phys. Rev. Lett.}\
  }\textbf {\bibinfo {volume} {83}},\ \bibinfo {pages} {3214} (\bibinfo {year}
  {1999})}\BibitemShut {NoStop}%
\bibitem [{\citenamefont {Ozawa}\ and\ \citenamefont
  {Berthier}(2017)}]{misaki2017}%
  \BibitemOpen
  \bibfield  {author} {\bibinfo {author} {\bibfnamefont {M.}~\bibnamefont
  {Ozawa}}\ and\ \bibinfo {author} {\bibfnamefont {L.}~\bibnamefont
  {Berthier}},\ }\href@noop {} {\bibfield  {journal} {\bibinfo  {journal} {J.
  Chem. Phys.}\ }\textbf {\bibinfo {volume} {146}},\ \bibinfo {pages} {014502}
  (\bibinfo {year} {2017})}\BibitemShut {NoStop}%
\bibitem [{\citenamefont {Sengupta}\ \emph {et~al.}(2012)\citenamefont
  {Sengupta}, \citenamefont {Karmakar}, \citenamefont {Dasgupta},\ and\
  \citenamefont {Sastry}}]{2dsastry}%
  \BibitemOpen
  \bibfield  {author} {\bibinfo {author} {\bibfnamefont {S.}~\bibnamefont
  {Sengupta}}, \bibinfo {author} {\bibfnamefont {S.}~\bibnamefont {Karmakar}},
  \bibinfo {author} {\bibfnamefont {C.}~\bibnamefont {Dasgupta}}, \ and\
  \bibinfo {author} {\bibfnamefont {S.}~\bibnamefont {Sastry}},\ }\href@noop {}
  {\bibfield  {journal} {\bibinfo  {journal} {Phys. Rev. Lett.}\ }\textbf
  {\bibinfo {volume} {109}},\ \bibinfo {pages} {095705} (\bibinfo {year}
  {2012})}\BibitemShut {NoStop}%
\bibitem [{\citenamefont {Frenkel}\ and\ \citenamefont {Ladd}(1984)}]{Frenkel}%
  \BibitemOpen
  \bibfield  {author} {\bibinfo {author} {\bibfnamefont {D.}~\bibnamefont
  {Frenkel}}\ and\ \bibinfo {author} {\bibfnamefont {A.~J.~C.}\ \bibnamefont
  {Ladd}},\ }\href@noop {} {\bibfield  {journal} {\bibinfo  {journal} {J. Chem.
  Phys.}\ }\textbf {\bibinfo {volume} {81}},\ \bibinfo {pages} {3188} (\bibinfo
  {year} {1984})}\BibitemShut {NoStop}%
\bibitem [{\citenamefont {Ozawa}\ \emph
  {et~al.}(2018{\natexlab{a}})\citenamefont {Ozawa}, \citenamefont {Parisi},\
  and\ \citenamefont {Berthier}}]{misaki2018}%
  \BibitemOpen
  \bibfield  {author} {\bibinfo {author} {\bibfnamefont {M.}~\bibnamefont
  {Ozawa}}, \bibinfo {author} {\bibfnamefont {G.}~\bibnamefont {Parisi}}, \
  and\ \bibinfo {author} {\bibfnamefont {L.}~\bibnamefont {Berthier}},\
  }\href@noop {} {\bibfield  {journal} {\bibinfo  {journal} {arXiv:1805.06017}\
  } (\bibinfo {year} {2018}{\natexlab{a}})}\BibitemShut {NoStop}%
\bibitem [{\citenamefont {Berthier}\ and\ \citenamefont
  {Coslovich}(2014)}]{cosloentropy}%
  \BibitemOpen
  \bibfield  {author} {\bibinfo {author} {\bibfnamefont {L.}~\bibnamefont
  {Berthier}}\ and\ \bibinfo {author} {\bibfnamefont {D.}~\bibnamefont
  {Coslovich}},\ }\href@noop {} {\bibfield  {journal} {\bibinfo  {journal}
  {Proc. Natl. Acad. Sci. U. S. A.}\ }\textbf {\bibinfo {volume} {111}},\
  \bibinfo {pages} {11668} (\bibinfo {year} {2014})}\BibitemShut {NoStop}%
\bibitem [{\citenamefont {Franz}\ and\ \citenamefont {Parisi}(1997)}]{FP}%
  \BibitemOpen
  \bibfield  {author} {\bibinfo {author} {\bibfnamefont {S.}~\bibnamefont
  {Franz}}\ and\ \bibinfo {author} {\bibfnamefont {G.}~\bibnamefont {Parisi}},\
  }\href@noop {} {\bibfield  {journal} {\bibinfo  {journal} {Phys. Rev. Lett.}\
  }\textbf {\bibinfo {volume} {79}},\ \bibinfo {pages} {2486} (\bibinfo {year}
  {1997})}\BibitemShut {NoStop}%
\bibitem [{\citenamefont {Biroli}\ \emph {et~al.}(2008)\citenamefont {Biroli},
  \citenamefont {Bouchaud}, \citenamefont {Cavagna}, \citenamefont {Grigera},\
  and\ \citenamefont {Verrocchio}}]{BBCGV08}%
  \BibitemOpen
  \bibfield  {author} {\bibinfo {author} {\bibfnamefont {G.}~\bibnamefont
  {Biroli}}, \bibinfo {author} {\bibfnamefont {J.-P.}\ \bibnamefont
  {Bouchaud}}, \bibinfo {author} {\bibfnamefont {A.}~\bibnamefont {Cavagna}},
  \bibinfo {author} {\bibfnamefont {T.~S.}\ \bibnamefont {Grigera}}, \ and\
  \bibinfo {author} {\bibfnamefont {P.}~\bibnamefont {Verrocchio}},\
  }\href@noop {} {\bibfield  {journal} {\bibinfo  {journal} {Nat. Phys.}\
  }\textbf {\bibinfo {volume} {4}},\ \bibinfo {pages} {771} (\bibinfo {year}
  {2008})}\BibitemShut {NoStop}%
\bibitem [{\citenamefont {Berthier}\ \emph
  {et~al.}(2016{\natexlab{a}})\citenamefont {Berthier}, \citenamefont
  {Charbonneau},\ and\ \citenamefont {Yaida}}]{BCY16}%
  \BibitemOpen
  \bibfield  {author} {\bibinfo {author} {\bibfnamefont {L.}~\bibnamefont
  {Berthier}}, \bibinfo {author} {\bibfnamefont {P.}~\bibnamefont
  {Charbonneau}}, \ and\ \bibinfo {author} {\bibfnamefont {S.}~\bibnamefont
  {Yaida}},\ }\href@noop {} {\bibfield  {journal} {\bibinfo  {journal} {J.
  Chem. Phys.}\ }\textbf {\bibinfo {volume} {144}},\ \bibinfo {pages} {024501}
  (\bibinfo {year} {2016}{\natexlab{a}})}\BibitemShut {NoStop}%
\bibitem [{\citenamefont {Kirkpatrick}\ \emph {et~al.}(1989)\citenamefont
  {Kirkpatrick}, \citenamefont {Thirumalai},\ and\ \citenamefont
  {Wolynes}}]{KTW89}%
  \BibitemOpen
  \bibfield  {author} {\bibinfo {author} {\bibfnamefont {T.~R.}\ \bibnamefont
  {Kirkpatrick}}, \bibinfo {author} {\bibfnamefont {D.}~\bibnamefont
  {Thirumalai}}, \ and\ \bibinfo {author} {\bibfnamefont {P.~G.}\ \bibnamefont
  {Wolynes}},\ }\href@noop {} {\bibfield  {journal} {\bibinfo  {journal} {Phys.
  Rev. A}\ }\textbf {\bibinfo {volume} {40}},\ \bibinfo {pages} {1045}
  (\bibinfo {year} {1989})}\BibitemShut {NoStop}%
\bibitem [{\citenamefont {Stillinger}(1988)}]{stillinger1988supercooled}%
  \BibitemOpen
  \bibfield  {author} {\bibinfo {author} {\bibfnamefont {F.~H.}\ \bibnamefont
  {Stillinger}},\ }\href@noop {} {\bibfield  {journal} {\bibinfo  {journal} {J.
  Chem. Phys.}\ }\textbf {\bibinfo {volume} {88}},\ \bibinfo {pages} {7818}
  (\bibinfo {year} {1988})}\BibitemShut {NoStop}%
\bibitem [{\citenamefont {Debenedetti}\ \emph {et~al.}(2003)\citenamefont
  {Debenedetti}, \citenamefont {Stillinger},\ and\ \citenamefont
  {Shell}}]{debenedetti2003model}%
  \BibitemOpen
  \bibfield  {author} {\bibinfo {author} {\bibfnamefont {P.~G.}\ \bibnamefont
  {Debenedetti}}, \bibinfo {author} {\bibfnamefont {F.~H.}\ \bibnamefont
  {Stillinger}}, \ and\ \bibinfo {author} {\bibfnamefont {M.~S.}\ \bibnamefont
  {Shell}},\ }\href@noop {} {\bibfield  {journal} {\bibinfo  {journal} {J.
  Phys. Chem. B}\ }\textbf {\bibinfo {volume} {107}},\ \bibinfo {pages} {14434}
  (\bibinfo {year} {2003})}\BibitemShut {NoStop}%
\bibitem [{\citenamefont {Kob}\ and\ \citenamefont {Berthier}(2013)}]{pinning}%
  \BibitemOpen
  \bibfield  {author} {\bibinfo {author} {\bibfnamefont {W.}~\bibnamefont
  {Kob}}\ and\ \bibinfo {author} {\bibfnamefont {L.}~\bibnamefont {Berthier}},\
  }\href@noop {} {\bibfield  {journal} {\bibinfo  {journal} {Phys. Rev. Lett.}\
  }\textbf {\bibinfo {volume} {110}},\ \bibinfo {pages} {245702} (\bibinfo
  {year} {2013})}\BibitemShut {NoStop}%
\bibitem [{\citenamefont {Berthier}\ \emph
  {et~al.}(2016{\natexlab{b}})\citenamefont {Berthier}, \citenamefont
  {Coslovich}, \citenamefont {Ninarello},\ and\ \citenamefont
  {Ozawa}}]{BCNO16}%
  \BibitemOpen
  \bibfield  {author} {\bibinfo {author} {\bibfnamefont {L.}~\bibnamefont
  {Berthier}}, \bibinfo {author} {\bibfnamefont {D.}~\bibnamefont {Coslovich}},
  \bibinfo {author} {\bibfnamefont {A.}~\bibnamefont {Ninarello}}, \ and\
  \bibinfo {author} {\bibfnamefont {M.}~\bibnamefont {Ozawa}},\ }\href@noop {}
  {\bibfield  {journal} {\bibinfo  {journal} {Phys. Rev. Lett.}\ }\textbf
  {\bibinfo {volume} {116}},\ \bibinfo {pages} {238002} (\bibinfo {year}
  {2016}{\natexlab{b}})}\BibitemShut {NoStop}%
\bibitem [{\citenamefont {Santos}\ \emph {et~al.}(2002)\citenamefont {Santos},
  \citenamefont {Yuste},\ and\ \citenamefont {Lopez~de
  Haro}}]{santos2002contact}%
  \BibitemOpen
  \bibfield  {author} {\bibinfo {author} {\bibfnamefont {A.}~\bibnamefont
  {Santos}}, \bibinfo {author} {\bibfnamefont {S.~B.}\ \bibnamefont {Yuste}}, \
  and\ \bibinfo {author} {\bibfnamefont {M.}~\bibnamefont {Lopez~de Haro}},\
  }\href@noop {} {\bibfield  {journal} {\bibinfo  {journal} {J. Chem. Phys.}\
  }\textbf {\bibinfo {volume} {117}},\ \bibinfo {pages} {5785} (\bibinfo {year}
  {2002})}\BibitemShut {NoStop}%
\bibitem [{\citenamefont {Royall}\ and\ \citenamefont
  {Williams}(2015)}]{Royall2015}%
  \BibitemOpen
  \bibfield  {author} {\bibinfo {author} {\bibfnamefont {C.~P.}\ \bibnamefont
  {Royall}}\ and\ \bibinfo {author} {\bibfnamefont {S.~R.}\ \bibnamefont
  {Williams}},\ }\href@noop {} {\bibfield  {journal} {\bibinfo  {journal}
  {Phys. Rep.}\ }\textbf {\bibinfo {volume} {560}},\ \bibinfo {pages} {1}
  (\bibinfo {year} {2015})}\BibitemShut {NoStop}%
\bibitem [{\citenamefont {Russo}\ and\ \citenamefont {Tanaka}(2015)}]{RT15}%
  \BibitemOpen
  \bibfield  {author} {\bibinfo {author} {\bibfnamefont {J.}~\bibnamefont
  {Russo}}\ and\ \bibinfo {author} {\bibfnamefont {H.}~\bibnamefont {Tanaka}},\
  }\href@noop {} {\bibfield  {journal} {\bibinfo  {journal} {Proc. Natl. Acad.
  Sci. U. S. A.}\ }\textbf {\bibinfo {volume} {112}},\ \bibinfo {pages} {6920}
  (\bibinfo {year} {2015})}\BibitemShut {NoStop}%
\bibitem [{\citenamefont {G{\"o}tze}(2008)}]{Goetze2008}%
  \BibitemOpen
  \bibfield  {author} {\bibinfo {author} {\bibfnamefont {W.}~\bibnamefont
  {G{\"o}tze}},\ }\href@noop {} {\emph {\bibinfo {title} {Complex dynamics of
  glass-forming liquids: A mode-coupling theory}}},\ Vol.\ \bibinfo {volume}
  {143}\ (\bibinfo  {publisher} {OUP Oxford},\ \bibinfo {year}
  {2008})\BibitemShut {NoStop}%
\bibitem [{\citenamefont {Ediger}\ \emph {et~al.}(1996)\citenamefont {Ediger},
  \citenamefont {Angell},\ and\ \citenamefont {Nagel}}]{ANGELL96}%
  \BibitemOpen
  \bibfield  {author} {\bibinfo {author} {\bibfnamefont {M.~D.}\ \bibnamefont
  {Ediger}}, \bibinfo {author} {\bibfnamefont {C.~A.}\ \bibnamefont {Angell}},
  \ and\ \bibinfo {author} {\bibfnamefont {S.~R.}\ \bibnamefont {Nagel}},\
  }\href@noop {} {\bibfield  {journal} {\bibinfo  {journal} {J. Phys. Chem.}\
  }\textbf {\bibinfo {volume} {100}},\ \bibinfo {pages} {13200} (\bibinfo
  {year} {1996})}\BibitemShut {NoStop}%
\bibitem [{\citenamefont {Elmatad}\ \emph {et~al.}(2009)\citenamefont
  {Elmatad}, \citenamefont {Chandler},\ and\ \citenamefont
  {Garrahan}}]{Elmatad2009}%
  \BibitemOpen
  \bibfield  {author} {\bibinfo {author} {\bibfnamefont {Y.~S.}\ \bibnamefont
  {Elmatad}}, \bibinfo {author} {\bibfnamefont {D.}~\bibnamefont {Chandler}}, \
  and\ \bibinfo {author} {\bibfnamefont {J.~P.}\ \bibnamefont {Garrahan}},\
  }\href@noop {} {\bibfield  {journal} {\bibinfo  {journal} {J. Phys. Chem. B}\
  }\textbf {\bibinfo {volume} {113}},\ \bibinfo {pages} {5563} (\bibinfo {year}
  {2009})}\BibitemShut {NoStop}%
\bibitem [{\citenamefont {Hecksher}\ \emph {et~al.}(2008)\citenamefont
  {Hecksher}, \citenamefont {Nielsen}, \citenamefont {Olsen},\ and\
  \citenamefont {Dyre}}]{Hecksher2008}%
  \BibitemOpen
  \bibfield  {author} {\bibinfo {author} {\bibfnamefont {T.}~\bibnamefont
  {Hecksher}}, \bibinfo {author} {\bibfnamefont {A.~I.}\ \bibnamefont
  {Nielsen}}, \bibinfo {author} {\bibfnamefont {N.~B.}\ \bibnamefont {Olsen}},
  \ and\ \bibinfo {author} {\bibfnamefont {J.~C.}\ \bibnamefont {Dyre}},\
  }\href@noop {} {\bibfield  {journal} {\bibinfo  {journal} {Nat. Phys.}\
  }\textbf {\bibinfo {volume} {4}},\ \bibinfo {pages} {737} (\bibinfo {year}
  {2008})}\BibitemShut {NoStop}%
\bibitem [{\citenamefont {Coluzzi}\ \emph {et~al.}(2000)\citenamefont
  {Coluzzi}, \citenamefont {Parisi},\ and\ \citenamefont
  {Verrocchio}}]{coluzzi2000lennard}%
  \BibitemOpen
  \bibfield  {author} {\bibinfo {author} {\bibfnamefont {B.}~\bibnamefont
  {Coluzzi}}, \bibinfo {author} {\bibfnamefont {G.}~\bibnamefont {Parisi}}, \
  and\ \bibinfo {author} {\bibfnamefont {P.}~\bibnamefont {Verrocchio}},\
  }\href@noop {} {\bibfield  {journal} {\bibinfo  {journal} {J. Chem. Phys.}\
  }\textbf {\bibinfo {volume} {112}},\ \bibinfo {pages} {2933} (\bibinfo {year}
  {2000})}\BibitemShut {NoStop}%
\bibitem [{\citenamefont {Ozawa}\ \emph {et~al.}(2015)\citenamefont {Ozawa},
  \citenamefont {Kob}, \citenamefont {Ikeda},\ and\ \citenamefont
  {Miyazaki}}]{ozawa2015equilibrium}%
  \BibitemOpen
  \bibfield  {author} {\bibinfo {author} {\bibfnamefont {M.}~\bibnamefont
  {Ozawa}}, \bibinfo {author} {\bibfnamefont {W.}~\bibnamefont {Kob}}, \bibinfo
  {author} {\bibfnamefont {A.}~\bibnamefont {Ikeda}}, \ and\ \bibinfo {author}
  {\bibfnamefont {K.}~\bibnamefont {Miyazaki}},\ }\href@noop {} {\bibfield
  {journal} {\bibinfo  {journal} {Proc. Nat. Acad. Sci., U.S.A.}\ }\textbf
  {\bibinfo {volume} {112}},\ \bibinfo {pages} {6914} (\bibinfo {year}
  {2015})}\BibitemShut {NoStop}%
\bibitem [{\citenamefont {Coluzzi}\ \emph {et~al.}(1999)\citenamefont
  {Coluzzi}, \citenamefont {M{\'e}zard}, \citenamefont {Parisi},\ and\
  \citenamefont {Verrocchio}}]{coluzzi1999thermodynamics}%
  \BibitemOpen
  \bibfield  {author} {\bibinfo {author} {\bibfnamefont {B.}~\bibnamefont
  {Coluzzi}}, \bibinfo {author} {\bibfnamefont {M.}~\bibnamefont {M{\'e}zard}},
  \bibinfo {author} {\bibfnamefont {G.}~\bibnamefont {Parisi}}, \ and\ \bibinfo
  {author} {\bibfnamefont {P.}~\bibnamefont {Verrocchio}},\ }\href@noop {}
  {\bibfield  {journal} {\bibinfo  {journal} {J. Chem. Phys.}\ }\textbf
  {\bibinfo {volume} {111}},\ \bibinfo {pages} {9039} (\bibinfo {year}
  {1999})}\BibitemShut {NoStop}%
\bibitem [{\citenamefont {Sastry}(2000)}]{sastry2000evaluation}%
  \BibitemOpen
  \bibfield  {author} {\bibinfo {author} {\bibfnamefont {S.}~\bibnamefont
  {Sastry}},\ }\href@noop {} {\bibfield  {journal} {\bibinfo  {journal} {J.
  Phys. Condens. Matter}\ }\textbf {\bibinfo {volume} {12}},\ \bibinfo {pages}
  {6515} (\bibinfo {year} {2000})}\BibitemShut {NoStop}%
\bibitem [{\citenamefont {Angelani}\ and\ \citenamefont
  {Foffi}(2007)}]{angelani2007configurational}%
  \BibitemOpen
  \bibfield  {author} {\bibinfo {author} {\bibfnamefont {L.}~\bibnamefont
  {Angelani}}\ and\ \bibinfo {author} {\bibfnamefont {G.}~\bibnamefont
  {Foffi}},\ }\href@noop {} {\bibfield  {journal} {\bibinfo  {journal} {J.
  Phys. Condens. Matter}\ }\textbf {\bibinfo {volume} {19}},\ \bibinfo {pages}
  {256207} (\bibinfo {year} {2007})}\BibitemShut {NoStop}%
\bibitem [{\citenamefont {Ozawa}\ \emph
  {et~al.}(2018{\natexlab{b}})\citenamefont {Ozawa}, \citenamefont {Ikeda},
  \citenamefont {Miyazaki},\ and\ \citenamefont {Kob}}]{ozawa2018ideal}%
  \BibitemOpen
  \bibfield  {author} {\bibinfo {author} {\bibfnamefont {M.}~\bibnamefont
  {Ozawa}}, \bibinfo {author} {\bibfnamefont {A.}~\bibnamefont {Ikeda}},
  \bibinfo {author} {\bibfnamefont {K.}~\bibnamefont {Miyazaki}}, \ and\
  \bibinfo {author} {\bibfnamefont {W.}~\bibnamefont {Kob}},\ }\href@noop {}
  {\bibfield  {journal} {\bibinfo  {journal} {arXiv preprint arXiv:1804.02324}\
  } (\bibinfo {year} {2018}{\natexlab{b}})}\BibitemShut {NoStop}%
\bibitem [{\citenamefont {Shiba}\ \emph {et~al.}(2018)\citenamefont {Shiba},
  \citenamefont {Keim},\ and\ \citenamefont {Kawasaki}}]{shiba2018isolating}%
  \BibitemOpen
  \bibfield  {author} {\bibinfo {author} {\bibfnamefont {H.}~\bibnamefont
  {Shiba}}, \bibinfo {author} {\bibfnamefont {P.}~\bibnamefont {Keim}}, \ and\
  \bibinfo {author} {\bibfnamefont {T.}~\bibnamefont {Kawasaki}},\ }\href@noop
  {} {\bibfield  {journal} {\bibinfo  {journal} {J. Phys. Condens. Matter}\
  }\textbf {\bibinfo {volume} {30}},\ \bibinfo {pages} {094004} (\bibinfo
  {year} {2018})}\BibitemShut {NoStop}%
\bibitem [{\citenamefont {Sciortino}(2005)}]{sciortino2005potential}%
  \BibitemOpen
  \bibfield  {author} {\bibinfo {author} {\bibfnamefont {F.}~\bibnamefont
  {Sciortino}},\ }\href@noop {} {\bibfield  {journal} {\bibinfo  {journal} {J.
  Stat. Mech.}\ }\textbf {\bibinfo {volume} {2005}},\ \bibinfo {pages} {P05015}
  (\bibinfo {year} {2005})}\BibitemShut {NoStop}%
\bibitem [{\citenamefont {Charbonneau}\ \emph {et~al.}(2016)\citenamefont
  {Charbonneau}, \citenamefont {Dyer}, \citenamefont {Lee},\ and\ \citenamefont
  {Yaida}}]{CDLY16}%
  \BibitemOpen
  \bibfield  {author} {\bibinfo {author} {\bibfnamefont {P.}~\bibnamefont
  {Charbonneau}}, \bibinfo {author} {\bibfnamefont {E.}~\bibnamefont {Dyer}},
  \bibinfo {author} {\bibfnamefont {J.}~\bibnamefont {Lee}}, \ and\ \bibinfo
  {author} {\bibfnamefont {S.}~\bibnamefont {Yaida}},\ }\href@noop {}
  {\bibfield  {journal} {\bibinfo  {journal} {J. Stat. Mech. Theory Exp.}\
  }\textbf {\bibinfo {volume} {2016}},\ \bibinfo {pages} {074004} (\bibinfo
  {year} {2016})}\BibitemShut {NoStop}%
\bibitem [{\citenamefont {Yaida}\ \emph {et~al.}(2016)\citenamefont {Yaida},
  \citenamefont {Berthier}, \citenamefont {Charbonneau},\ and\ \citenamefont
  {Tarjus}}]{YBCT16}%
  \BibitemOpen
  \bibfield  {author} {\bibinfo {author} {\bibfnamefont {S.}~\bibnamefont
  {Yaida}}, \bibinfo {author} {\bibfnamefont {L.}~\bibnamefont {Berthier}},
  \bibinfo {author} {\bibfnamefont {P.}~\bibnamefont {Charbonneau}}, \ and\
  \bibinfo {author} {\bibfnamefont {G.}~\bibnamefont {Tarjus}},\ }\href@noop {}
  {\bibfield  {journal} {\bibinfo  {journal} {Phys. Rev. E}\ }\textbf {\bibinfo
  {volume} {94}},\ \bibinfo {pages} {032605} (\bibinfo {year}
  {2016})}\BibitemShut {NoStop}%
\bibitem [{\citenamefont {Frenkel}\ and\ \citenamefont {Smit}(2001)}]{FS01}%
  \BibitemOpen
  \bibfield  {author} {\bibinfo {author} {\bibfnamefont {D.}~\bibnamefont
  {Frenkel}}\ and\ \bibinfo {author} {\bibfnamefont {B.}~\bibnamefont {Smit}},\
  }\href@noop {} {\emph {\bibinfo {title} {Understanding Molecular
  Simulation}}}\ (\bibinfo  {publisher} {Academic Press, New York, ed. 2.},\
  \bibinfo {year} {2001})\BibitemShut {NoStop}%
\bibitem [{\citenamefont {Fukunishi}\ \emph {et~al.}(2002)\citenamefont
  {Fukunishi}, \citenamefont {Watanabe},\ and\ \citenamefont {Takada}}]{FWT02}%
  \BibitemOpen
  \bibfield  {author} {\bibinfo {author} {\bibfnamefont {H.}~\bibnamefont
  {Fukunishi}}, \bibinfo {author} {\bibfnamefont {O.}~\bibnamefont {Watanabe}},
  \ and\ \bibinfo {author} {\bibfnamefont {S.}~\bibnamefont {Takada}},\
  }\href@noop {} {\bibfield  {journal} {\bibinfo  {journal} {J. Chem. Phys.}\
  }\textbf {\bibinfo {volume} {116}},\ \bibinfo {pages} {9058} (\bibinfo {year}
  {2002})}\BibitemShut {NoStop}%
\bibitem [{\citenamefont {Cavagna}\ \emph {et~al.}(2012)\citenamefont
  {Cavagna}, \citenamefont {Grigera},\ and\ \citenamefont
  {Verrocchio}}]{BICtest12}%
  \BibitemOpen
  \bibfield  {author} {\bibinfo {author} {\bibfnamefont {A.}~\bibnamefont
  {Cavagna}}, \bibinfo {author} {\bibfnamefont {T.~S.}\ \bibnamefont
  {Grigera}}, \ and\ \bibinfo {author} {\bibfnamefont {P.}~\bibnamefont
  {Verrocchio}},\ }\href@noop {} {\bibfield  {journal} {\bibinfo  {journal} {J.
  Chem. Phys.}\ }\textbf {\bibinfo {volume} {136}},\ \bibinfo {pages} {204502}
  (\bibinfo {year} {2012})}\BibitemShut {NoStop}%
\bibitem [{\citenamefont {Harrowell}(2006)}]{harrowell2006nonlinear}%
  \BibitemOpen
  \bibfield  {author} {\bibinfo {author} {\bibfnamefont {P.}~\bibnamefont
  {Harrowell}},\ }\href@noop {} {\bibfield  {journal} {\bibinfo  {journal}
  {Nat. Phys.}\ }\textbf {\bibinfo {volume} {2}},\ \bibinfo {pages} {157}
  (\bibinfo {year} {2006})}\BibitemShut {NoStop}%
\bibitem [{\citenamefont {Shiba}\ \emph {et~al.}(2012)\citenamefont {Shiba},
  \citenamefont {Kawasaki},\ and\ \citenamefont
  {Onuki}}]{shiba2012relationship}%
  \BibitemOpen
  \bibfield  {author} {\bibinfo {author} {\bibfnamefont {H.}~\bibnamefont
  {Shiba}}, \bibinfo {author} {\bibfnamefont {T.}~\bibnamefont {Kawasaki}}, \
  and\ \bibinfo {author} {\bibfnamefont {A.}~\bibnamefont {Onuki}},\
  }\href@noop {} {\bibfield  {journal} {\bibinfo  {journal} {Phys. Rev. E}\
  }\textbf {\bibinfo {volume} {86}},\ \bibinfo {pages} {041504} (\bibinfo
  {year} {2012})}\BibitemShut {NoStop}%
\bibitem [{\citenamefont {Shiba}\ \emph {et~al.}(2016)\citenamefont {Shiba},
  \citenamefont {Yamada}, \citenamefont {Kawasaki},\ and\ \citenamefont
  {Kim}}]{shiba2016unveiling}%
  \BibitemOpen
  \bibfield  {author} {\bibinfo {author} {\bibfnamefont {H.}~\bibnamefont
  {Shiba}}, \bibinfo {author} {\bibfnamefont {Y.}~\bibnamefont {Yamada}},
  \bibinfo {author} {\bibfnamefont {T.}~\bibnamefont {Kawasaki}}, \ and\
  \bibinfo {author} {\bibfnamefont {K.}~\bibnamefont {Kim}},\ }\href@noop {}
  {\bibfield  {journal} {\bibinfo  {journal} {Phys. Rev. Lett.}\ }\textbf
  {\bibinfo {volume} {117}},\ \bibinfo {pages} {245701} (\bibinfo {year}
  {2016})}\BibitemShut {NoStop}%
\bibitem [{\citenamefont {Tarjus}(2017)}]{tarjus2017glass}%
  \BibitemOpen
  \bibfield  {author} {\bibinfo {author} {\bibfnamefont {G.}~\bibnamefont
  {Tarjus}},\ }\href@noop {} {\bibfield  {journal} {\bibinfo  {journal} {Proc.
  Nat. Acad. Sci., U.S.A.}\ ,\ \bibinfo {pages} {201700193}} (\bibinfo {year}
  {2017})}\BibitemShut {NoStop}%
\end{thebibliography}%

\end{document}